\documentclass[preprint,floats,floatfix,tightenlines]{revtex4}
 \usepackage{graphicx}

\usepackage{amssymb}

\begin{document}

\message{present finished studies - distinguish info of internal use vs.info of use to general reader}



\title{Event Reconstruction and Data Acquisition for the RICE Experiment at the South Pole}

\author{I. Kravchenko}
\address{Massachusetts Institute of Technology Laboratory for
Nuclear Science, Cambridge, MA  02139}
\author{A. Hase, D. Besson, S. Graham, Z. Kessler}
\address{University of Kansas Dept. of Physics and Astronomy, Lawrence KS
66045-2151}
\author{J. Ledford, K. Ratzlaff}
\address{University of Kansas Instrumentation Design Laboratory, Lawrence KS 66045-2151}
\author{Xin-Hua Bai, Allan Baker, Philip Braughton, Christina Hammock, Michael Offenbacher, Mark Noske, Nicolas Hart-Michel}\address{Amundsen-Scott South Pole Station}


\begin{abstract}
The RICE experiment seeks to measure ultra-high energy neutrinos
($E_\nu>10^{16}$ eV) by detection of the radio wavelength
Cherenkov radiation produced by neutrino-ice collisions within Antarctic
ice. An array of 16 dipole
antennas, buried at depths of
100-400 m, and sensitive over the 100-500 MHz frequency
range, has been continuously taking data for the
last seven years. 
We herein describe the design and performance of the RICE experiment's
event trigger and data acquisition system, highlighting elements not
covered in previous publications. 
\end{abstract}

\maketitle


\section{Introduction\label{label:intro}}
The RICE experiment seeks measurement of ultra-high energy
(``UHE''; $E_\nu>10^{15}$ eV)
neutrinos interacting in Antarctic ice, by measurement of
the radiowavelength Cherenkov radiation resulting from
the collision of a neutrino with an ice molecule.
Previous publications described initial limits on the incident
neutrino flux\citep{rice03a}, calibration procedures\citep{rice03b}, 
ice properties' measurements\citep{KravchenkoRFre}, and successor
analyses of micro-black hole production\citep{dougshahidBH06},
gamma-ray burst production of UHE neutrinos\citep{grb06}, and
tightened limits on the diffuse neutrino flux\citep{rice06}.

\subsection{Signal Strength}
\begin{figure}[htpb]
\centerline{\includegraphics[width=9cm]{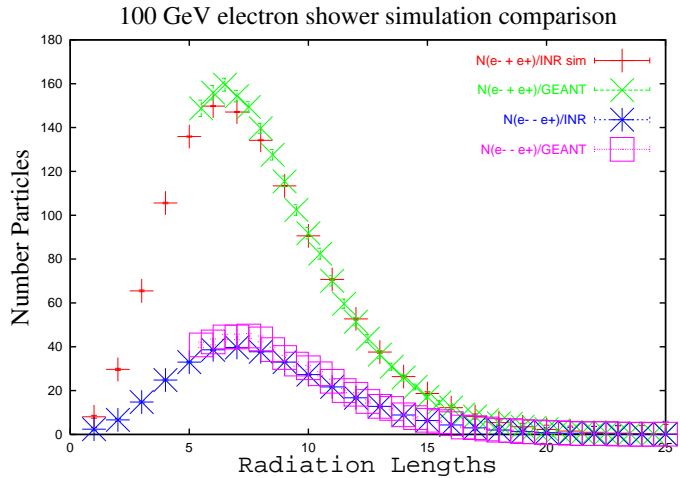}}
\caption{Comparison of Institute of Nuclear Research simulation of
100 GeV shower characteristics\citep{SIMEX} vs. GEANT-4 simulations run
at University of Kansas\citep{GEANTsims}.}
\label{fig:shower-comparison}
\end{figure}

The expected coherent radio pulse from a neutrino-initiated
electromagnetic shower in ice was initially derived from the
Monte Carlo simulation work of Zas, Halzen and Stanev\citep{ZHS92} and
more recently explored with other studies, including
a variety of simulations building on the ZHS code\citep{Jaime}, 
GEANT simulations\citep{GEANTsims}, and an independent time-domain
only simulation code\citep{SIMEX}.
All simulations model the cascade as it develops
in a semi-infinite medium (ice) and calculate the number of
atomic electrons swept into the forward-moving cascade as well
as the number of shower positrons depleted through annihilation
with atomic electrons. 
From that, one can determine the expected 
net Cherenkov electric field
signal strength at an arbitrary observation point ${\vec R}$ by summing
the Cherenkov electric field vectors ${\vec E}({\vec R},\omega)$ for
each particle participating in the forward-moving shower. 
Although simplest in the Fraunoffer approximation, full calculations
have also been carried out for receivers in the Fresnel zone\citep{Fresnel}.
All simulations
give the same qualitative conclusion - at large distances, the signal 
at the antenna
is a symmetric pulse, approximately 1-2 ns wide in the time domain. The
power spectrum as a function of frequency is
approximately linearly rising with
frequency, as expected in the long-wavelength limit. In that limit, the
excess negative charge in the shower front can be treated as a single
(``coherent'') charge emitting Cherenkov radiation with an
energy per photon: $E=\hbar\omega$, i.e., $E(\omega)\sim\omega$.
Figure \ref{fig:shower-comparison}
shows the status of 
our current calculations.
For perfect signal transmission (no cable signal losses),
the signal induced in a 50-Ohm
antenna (on the Cherenkov cone)
due to a 1 PeV neutrino initiating 
a shower at R=1 km from an antenna is $\sim$20$\mu$V$\sqrt{B}$, with
$B$ the system bandwidth in GHz. This is comparable to
the 300 K thermal noise over that bandwidth
in the same antenna, 
prior to amplification.
Due to the finite experimental 
bandwidth, the time domain
signal is broadened ($\Delta t\sim 1/B$). Finite bandwidth is introduced by: a) finite response of the
antenna, b) cable losses as a function of frequency, which
tend to attenuate the high-frequency signal components, c) 
bandpass filters,
which, in the case of RICE, 
remove $f<$200 MHz to suppress large 
radiofrequency (RF) noise generated by the AMANDA(/IceCube) photo-tubes. 
The resulting pulse is 
therefore stretched in the time domain to $\sim$5 ns. 


\section{Hardware and Signal Path and Experimental Layout}
\begin{figure}[htpb]
\centerline{\includegraphics[width=9cm]{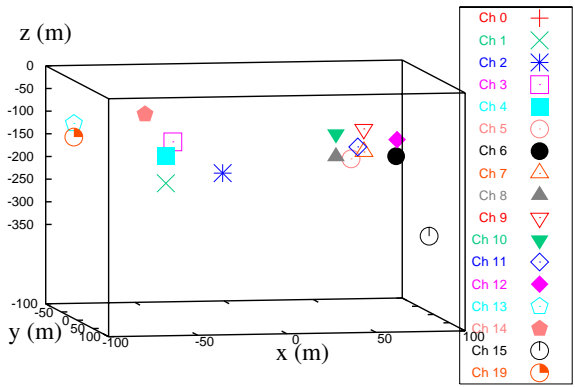}}
\caption{RICE antenna deployment geometry.}
\label{fig:RICE-3dview.eps}
\end{figure}

The sensitive detector elements, 
radio receivers, are submerged at depths
of several hundred meters close to the
Geographic South Pole. 
The 
(Martin A. Pomerantz Observatory) MAPO building houses hardware for several experiments, including the RICE and AMANDA surface electronics, and is centered at (x$\sim$40m, y$\sim -30$m) on the surface. The AMANDA array is located approximately 600 m (AMANDA-A) to 2400 m (AMANDA-B) below the RICE array in the ice; the South Pole Air Shower Experiment (SPASE) is located on the surface at ($x\sim-400$m, $y\sim$0m). 
The coordinate system conforms to the convention used by the AMANDA experiment; grid North is defined by the Greenwich Meridian and coincides with the +y-direction in the Figure. 
The geometry
of the deployed antennas is presented in Figure \ref{fig:RICE-3dview.eps}
and also presented in Table \ref{tab:Rxlocations}.
\begin{table}[hptb]\begin{center}\begin{tabular}{c|c|c|c|c}Channel number & x- (m) & y- (m) & z- (m) & time delay (ns) \\ \hline 0 & 4.8 &  102.8 &  -166 &  1336 \\ 1 & -56.3 &  34.2 &  -213 &  1416 \\ 2 & -32.1 &  77.4 &  -176 & 1293 \\ 3 & -61.4 &  85.3 &  -103 & 1230 \\ 4 & -56.3 & 34.2 & -152 & 1166 \\ 5 & 47.7 & 33.8 & -166 & 1181 \\ 6 & 78.0 & 13.8 & -170 & 944 \\ 7 & 64.1 & -18.3 & -171 & 939 \\ 8 & 43.9 & 7.3 & -171 & 946 \\ 9 & 64.1 & -18.3 & -120 & 809 \\ 10 & 43.9 & 7.3 & -120 & 672 \\ 11 & 67.5 & -39.5 & -168 & 952 \\ 12 & 66.3 & 74.7 & -110 & 1051 \\ 13 & -95.1 & -38.3 & -105 & 116 \\ 14 & -46.7 & -86.6 & -105 & 1051 \\ 15 & 95.2 & 12.7 & -347 & 1984  \\ 19 & -95.1 & -38.3 & -135 & 1276 \\ \hline \end{tabular} \caption{\label{tab:Rxlocations}Location of RICE radio receivers. We have adopted the coordinate system convention used by the AMANDA collaboration. 
}
\end{center}\end{table}

\begin{figure}[htpb]
\centerline{\includegraphics[width=9cm]{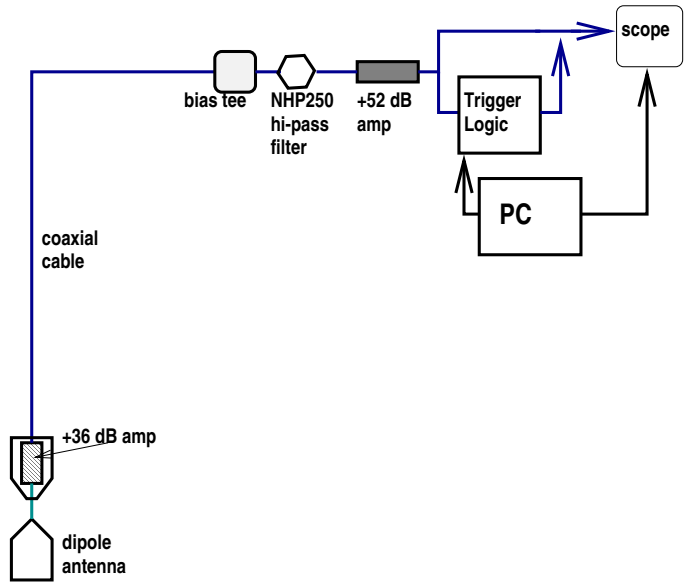}}
\caption{Block diagram, showing primary experimental components.}
\label{fig:RICE-block-diagram.xfig.eps}
\end{figure}

A block diagram of the experiment, showing the signal
path from in-ice to the surface electronics, is shown in Figure 
\ref{fig:RICE-block-diagram.xfig.eps}.\footnote{With the exception of the
RICE dipole antennas, all components are 50-ohm.} We now discuss in 
greater detail experimental 
components not previously discussed in other publications\citep{rice03a,rice03b,rice06}. 

\subsection{Antenna Calibration and Transfer Function}

\begin{figure}\vspace{12cm}\includegraphics{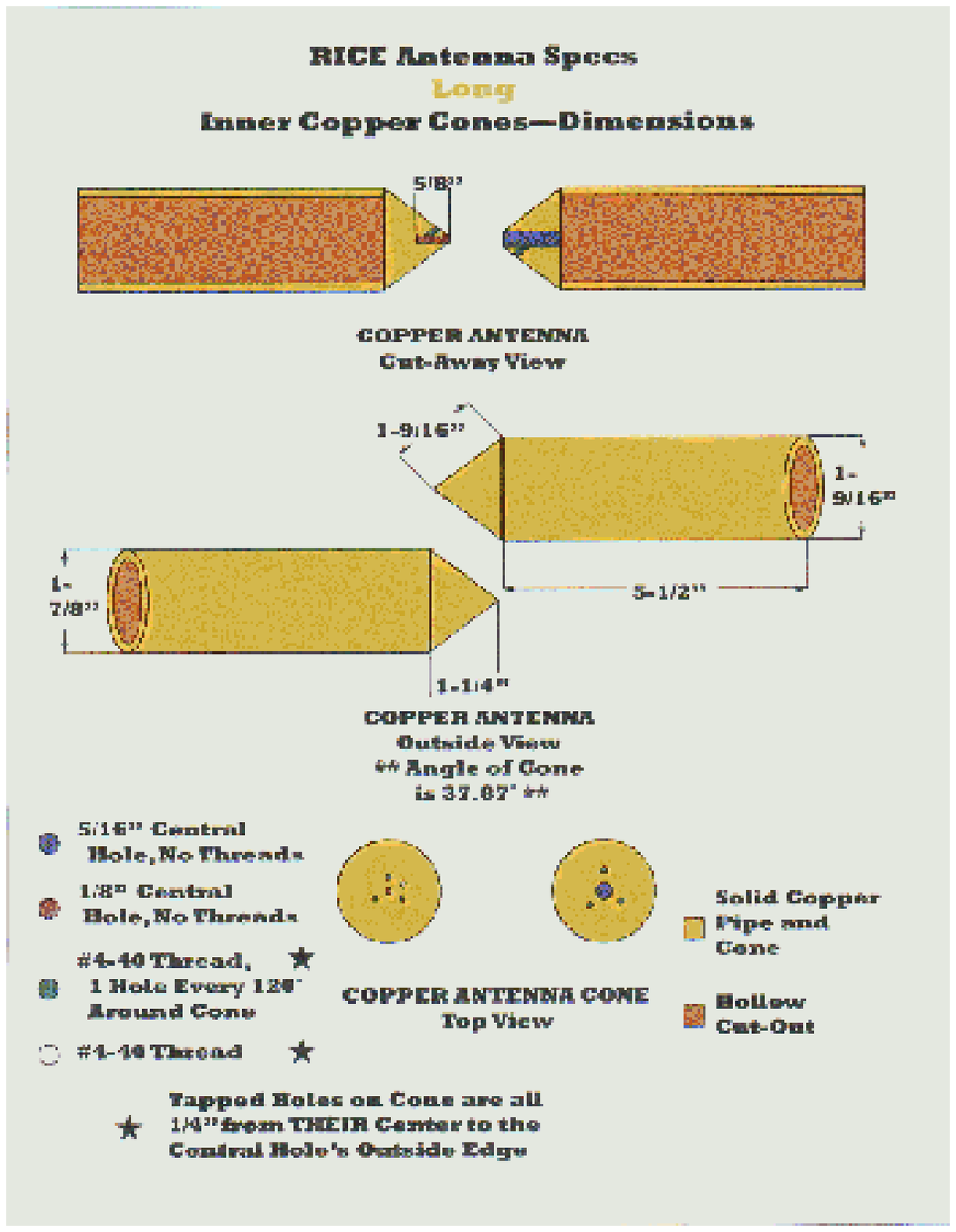}\caption{RICE fat dipole antennas.} \label{fig:superdb3}\end{figure}

The RICE antennas are ``standard'' fat dipole antennas, tuned to 
a center frequency of approximately 450 MHz in air. A sketch of 
the antenna construction is shown in Figure \ref{fig:superdb3}.

\subsubsection{Antenna Calibration Procedure}
We sought to determine
a method of antenna calibration, 
for purposes of predicting signal shape (i.e. voltage amplitude in the time 
domain) resulting from broadband electromagnetic waves incident on a 
receiving RICE antenna. This is applicable for radio frequency detectors in 
situations where the incident field due to events under investigation is 
predicted by other models, and sensitivity of the detectors to these events 
must be determined, as is the case for RICE.
The complex transmission equation presented here is 
analogous to the Friis Transmission Formula which
is familiar in antenna literature. However, the former, unlike the latter, 
retains phase information required for predicting signal shape in the time 
domain. 
What is 
needed is the complex \textit{transfer function} $T$, defined as:
\begin{equation}
\label{eq1}
V=TE
\end{equation}
where $V$ is the voltage output at the antenna terminals, and E is the field 
incident on the antenna. $T$ depends upon the antenna's 
\textit{complex effective height}\citep{heff-refs} 
$h_{eff}$, its
impedance, and the impedance of the load to which the 
antenna is connected. Of these the \textit{complex effective height} 
is difficult to measure directly, but 
can be calculated using a \textit{complex transmission equation}, which is analogous to the Friis Transmission 
Formula familiar in antenna literature. The latter relates 
input power delivered at the input terminals of a transmitting antenna to 
the power received at the output terminals of a receiving antenna. Other 
common parameters characterizing the performance of antenna, (e.g. gain), 
likewise are defined in terms of power, discarding phase information. 

We use a pair of antennas, one
transmitting, 
the other receiving, to extract $h_{eff}$.
As the transmission coefficient for
a single antenna is readily measured with a network 
analyzer, then if the complex effective height of one antenna is 
known, the other may be calculated. Alternatively the \textit{complex effective height} of two identical 
antennas may be determined.

\subsubsection{Transfer Function}
\label{sec:transfer}
The transfer function relates the output 
voltage of the antenna to the incident field. Decomposing the field into 
plane waves in the frequency domain, we can re-cast our definition
for $T$ as:
\begin{equation}
\label{eq2}
V_\omega =T_{\omega \theta \phi \rho } E_{\omega \theta \phi \rho } 
\end{equation}
where $\omega$ is angular frequency, $(\theta ,\phi )$ give the polar
and azimuthal orientation 
of the line of propagation of $E$, and $\rho$ gives the polarization of 
$E$ with respect to the plane made by the axis $\theta=0$ and the line of 
propagation of $E$ 

The output signal, the time domain $V_t$, is given simply by the
Fourier transform of $V_\omega$. I.e., the Fourier transform, 
$T_t$ of $T_\omega$ is simply a Green's function for the antenna, 
characterizing its response to incident waves along a given line of 
propagation, $(\theta ,\phi )$, with a given polarization, $\rho $.

The transfer function is determined by the antenna's impedance, the load 
impedance, and the complex effective height of the antenna. The receiving 
antenna, along with a load (e.g., cable and DAQ electronics) of impedance 
$Z^{L}$, may be modeled by an \textit{equivalent circuit} in which the antenna is replaced by a 
Thevenin generator with an \textit{EMF} 
or \textit{open circuit voltage,} $V^{\mathrm{oc}}$, induced by the incident 
field, with a series impedance, $Z^{\mathrm{rx}}$.
Thus the 
voltage, $V_{rx}$, at the output of the antenna terminals, and
dropped across the load is given by
\begin{equation}
\label{eq3}
V^{\mathrm{rx}}=\frac{V^{\mathrm{oc}}Z^{\mathrm{L}}}{Z{^{\mathrm{L}}}+Z^{\mathrm{rx}}}
\end{equation}
where the functional dependence of all variables on the frequency, $\omega$, 
is implicit.


\subsubsection{Complex Effective Height}
\label{sec:complex}
The complex effective height, in the case of a receiving antenna, relates 
the incident field to the open circuit voltage of the antenna.
\begin{equation}
\label{eq4}
V^{\mathrm{oc}}=HE.
\end{equation}
Combining the
above equations, we can fully 
characterize the transfer function:
\begin{equation}
\label{eq5}
T=\frac{HZ^{\mathrm{L}}}{Z^{\mathrm{L}}+Z^{\mathrm{rx}}}.
\end{equation}
The complex effective height can also be defined for a transmitting antenna, 
relating the far field to the current, $I_{tx} $, at the input terminals of 
the antenna.
\begin{equation}
\label{eq6}
E_{\mathrm{far~field}} =-H^{\mathrm{tx}}I^{\mathrm{tx}}\frac{ie^{{i\omega r} 
\mathord{\left/ {\vphantom {{i\omega r} c}} \right. 
\kern-\nulldelimiterspace} c}Z_0 \omega }{2\pi rc}
\end{equation}
where $r$ is the distance, $Z_0$ is the characteristic impedance of free 
space (377$\Omega$), 
and $c$ is the speed of light. The modulus of $H^{\mathrm{tx}}$ can be 
interpreted as the length of a uniform linear current segment, centered at 
the antenna, aligned parallel to the orientation of $E_{\mathrm{far~field}}$, 
with current $I^{\mathrm{tx}}$, which induces the same far field as the 
antenna in a given direction $(\theta ,\phi)$ and polarization $\rho$. 
The phase angle of $H^{\mathrm{tx}}$
then gives the phase of $E_{\mathrm{far~field}}$
due to the antenna relative to the $E_{\mathrm{far~field}}$ that 
would be induced by such a uniform current segment.

There is no immediately obvious relationship between the transmitting and 
receiving definitions of complex effective height. However, by 
the Reciprocity Theorem for antennas, the two definitions 
are in fact equivalent. That is,
\begin{equation}
\label{eq7}
H_{\omega \theta \phi \rho }^{\mathrm{rx}} =H_{\omega \theta \phi \rho 
}^{\mathrm{tx}} .
\end{equation}
This fact is of particular importance for calibrating a pair of identical 
antennas of unknown complex effective height. For the remainder of this
discussion, we assume reciprocity to be valid. (The accuracy of our
full-system gain, ex post facto, validates this assumption, to some degree.)

\subsubsection{Complex Transmission Equation}
\label{sec:mylabel2}
\begin{figure}
\centerline{\includegraphics[width=8cm,angle=0]{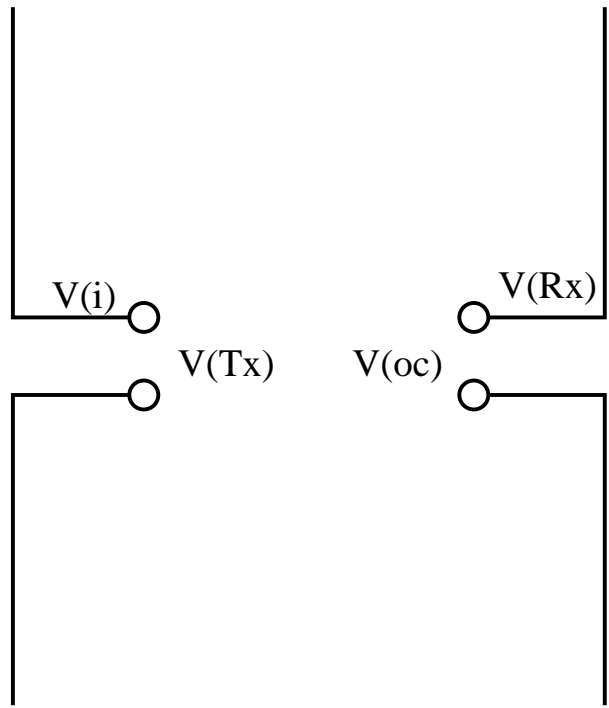}}
\caption{Schematic of voltage variables used in equivalent height
calculation.}
\label{fig:tx-rx_equiv_circ}
\end{figure}

\begin{figure}
\centerline{\includegraphics[width=10cm,angle=0]{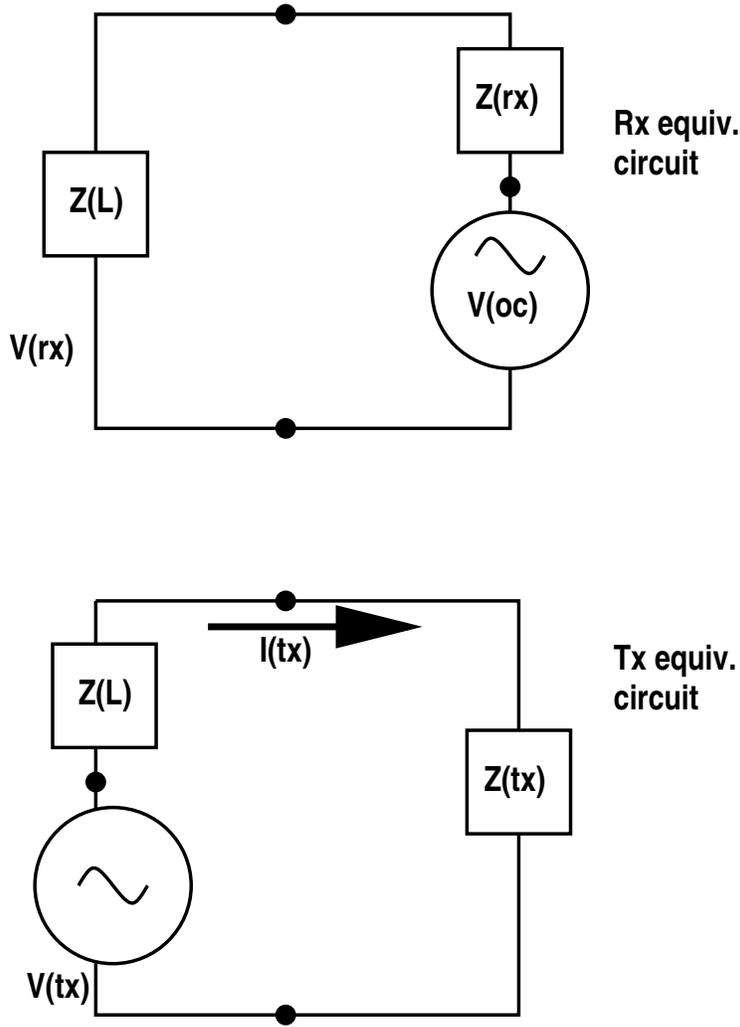}}
\caption{Equivalent Circuit for a Transmitting or Receiving Antenna.}
\label{fig:rx_equiv_circ}
\end{figure}

Although
it is difficult to measure $E$ in the space near an antenna,
the complex transmission coefficient 
between a transmitting and receiving antenna is readily measured by a 
network analyzer. What we wish to construct is an equation that relates the 
complex effective heights, $H^\alpha$ and $H^\beta$, of a pair of antenna 
$\alpha$ and $\beta$, to their complex transmission coefficient, 
$t^{\alpha \beta } $. The complex transmission coefficient is the ratio of 
the voltage incident on the input terminals of the transmitting antenna, 
$V^{\mathrm{i}}$, to the voltage output at the terminals of the receiving 
antenna, $V^{\mathrm{rx}}$. Figure \ref{fig:tx-rx_equiv_circ} also shows
the voltages at the output terminals of the transmitting antenna
($V^{\mathrm{tx}}$) and the input to the receiving antenna
($V^{\mathrm{oc}}$).
Note that due to reciprocity, it does not matter 
which antenna is the transmitter and which is the receiver. That is,
\begin{equation}
\label{eq8}
t^{\alpha \beta }=t^{\beta \alpha }=\frac{V^{\mathrm{rx}}}{V^{\mathrm{i}}}
\end{equation}
Before we can construct our complex transmission equation, we must first 
relate $V^{\mathrm{i}}$ to $I^{\mathrm{tx}}$. First consider the 
equivalent circuit for a transmitting antenna. In the equivalent circuit the 
antenna is represented by its impedance, $Z^{\mathrm{tx}}$ in series with a 
function generator consisting of an ideal AC voltage source, 
$V^{\mathrm{tx}}$ and a series impedance, $Z^L$ (see Fig. 
\ref{fig:rx_equiv_circ}). The current 
at the input terminals of the antenna is then given by
\begin{equation}
\label{eq9}
I^{\mathrm{tx}}=\frac{V^{\mathrm{tx}}}{Z^{\mathrm{L}}+Z^{\mathrm{tx}}}.
\end{equation}
At first it may seem that the incident voltage, $V^{\mathrm{i}}$, and the open 
circuit voltage $V^{\mathrm{tx}}$ of the transmitter's function generator would 
be identical. However, the complex transmission coefficient is defined in 
terms of a voltage wave incident on the interface between the function 
generator and the transmitting antenna, which gives rise to a reflected and 
transmitted wave.  If we replace the antenna impedance with a 
zero or infinite impedance (close or open circuit), there will be total 
reflection, and the magnitude of the reflected wave must equal that of the 
incident wave, so $V^{\mathrm{tx}}=2V^{\mathrm{i}}$. Thus, in this limit,
\message{WHY IS THIS FACTOR OF TWO GENERAL?}
\begin{equation}
\label{eq10}
I^{\mathrm{tx}}=\frac{2V^{\mathrm{i}}}{Z^{\mathrm{L}}+Z^{\mathrm{tx}}}.
\end{equation}
Now we can combine equations above to arrive at the 
complex transmission equation:
\begin{equation} 
\label{eq11}
t^{\alpha \beta 
}=-H^{\mathrm{tx}}H^{\mathrm{rx}}\frac{2Z^{\mathrm{L}}}{(Z^{\mathrm{L}}+Z^{\mathrm{tx}})(Z^{\mathrm{L}}+Z^{\mathrm{rx}})}\frac{ie^{{
i\omega r} \mathord{\left/ {\vphantom {{i\omega r} c}} \right. 
\kern-\nulldelimiterspace} c}Z_0 \omega }{2\pi rc},
\end{equation}
where $r$ is the distance between the antennas, $Z_0$ is the characteristic 
impedance of free space, and $c$ is the speed of light. Note that we have 
treated the impedance of the devices at either end as identical, which need 
not generally be the case, although it typically will be the case if a 
network analyzer is being used to measure $t^{\alpha \beta }$. In the case 
where two identical antennas are used, each oriented in the same manner with 
respect to the other, the complex transmission equation becomes
\begin{equation}
\label{eq12}
t^{\alpha \alpha }=-(H^\alpha )^{^2}\frac{2Z^{\mathrm{L}}}{(Z^{\mathrm{L}}+Z^\alpha 
)^2}\frac{ie^{{i\omega r} \mathord{\left/ {\vphantom {{i\omega r} c}} 
\right. \kern-\nulldelimiterspace} c}Z_0 \omega }{2\pi rc}.
\end{equation}
This formalism therefore
makes it possible to calibrate a pair of identical antennas. One caveat 
here is that when solving for $H^\alpha $, there is an 
ambiguity introduced when taking the square root. We resolve this ambiguity 
by requiring the phase angle of $H_\omega ^\alpha$ to be a continuous 
function of $\omega$, and also require that the Fourier transform, 
$H_t^\alpha $, of $H_\omega ^\alpha $, be causal in the time domain, as it 
is the Green's function for the antenna's open circuit response to the 
incident wave.

\subsubsection{Measurement of Complex Effective Height}
\label{sec:measurement}
Measurement of the complex effective height of an 
antenna is simply a matter of measuring the antenna's impedance, then 
measuring the complex transmission coefficients between pairs of antennas, 
one of which has known effective height in at least one orientation, or 
measuring the complex transmission coefficient between a pair of identical 
antennas which are each oriented in the same manner with respect to the 
other.

If the environment in which the transmission measurements are taken is not 
ideal, there may be significant errors introduced by reflections in the 
environment. Care must be taken to eliminate the error due to reflections by 
taking multiple transmission measurements and randomizing the effects of 
reflections. This is done by moving the pair of antennas in tandem, 
maintaining the same distance and relative orientation, and taking a 
transmission measurement at each location. In this way the line-of-sight 
path between the antennas is preserved, while indirect paths due to 
reflections will be changed with each repositioning of the antenna pair. The 
range of motion for this procedure should be several times the largest 
wavelength for which accurate measurements are desired.

\subsubsection{The RICE Receivers}
The signal path from each antenna to digital oscilloscope consists of the 
antenna itself, a high pass filter, an amplifier, several hundred meters of 
coaxial cable, another amplifier, splitter, and finally a digital 
oscilloscope. We refer to this signal path in its entirety as a `receiver' 
to distinguish it from the antenna proper, which is simply the first 
component of the receiver. All components of the receiver, with the 
exception of the antenna, are impedance matched to 50 ohms to eliminate 
reflections. We can directly measure a complex transmission coefficient for 
each component. The product of these with the transfer function of the 
antenna provides the transfer function for the entire receiver.

Because the receivers are embedded in ice, 
the index of refraction changes significantly. As a 
first approximation, we treat the response of the antenna to wavelengths as 
identical in both media (the validity of this assumption is discussed
later in this document). That is,
\begin{equation}
\label{eq13}
T_\lambda ^{\mathrm{ice}} =T_\lambda ^{\mathrm{air}} ,
\end{equation}
where $\lambda ={2\pi c} \mathord{\left/ {\vphantom {{2\pi c} {n\omega }}} 
\right. \kern-\nulldelimiterspace} {n\omega }$, where $c$ is the speed of 
light and $n$ is the index of refraction of the surrounding medium (air or 
ice), and we have exchanged the functional dependence of $T$ on $\omega$ for 
a dependence on $\lambda$. 


\subsubsection{Measurement of RICE Dipole Transfer Function}
To utilize the complex transmission equation, the impedances of the two
antennas involved must also be known. Measurement of the impedance of the
antennas was also performed using the network analyzer, based on a
reflection measurement. The network analyzer measures a reflection
coefficient for reflected waves returning via its transmitting port, and the
complex impedance of the connected device is directly derivable
from the complex reflection
coefficient. Note that the impedance
measured is, in effect, the lumped impedance of whatever is connected to the
analyzer, including effects of its environment. In measuring the antenna
impedances, we have assumed that the environment produces negligible
reflections returning to the antenna. We have confirmed this by measuring
impedance in various locations and found only small variation
($\sim$5 in the VSWR), indicating
that reflections from objects in the vicinity of the antenna have little
effect on the impedance measurement.

\subsubsection{Complex Effective Height Measurement Scheme}
\begin{figure}[htpb]
\centerline{\includegraphics[width=10cm,angle=0]{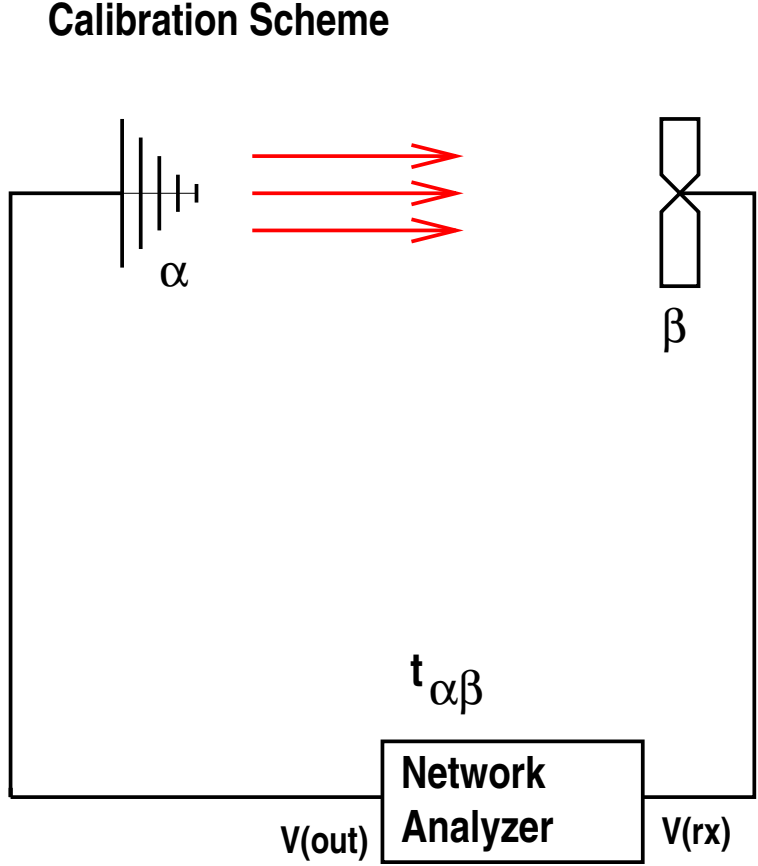}}
\caption{Geometry of effective height measurements for RICE antennas.}
\label{fig:calibration_scheme.ps}
\end{figure}

The complex effective height of the RICE dipoles was measured using the
scheme illustrated in Fig. \ref{fig:calibration_scheme.ps}.
A pair of antennas, one transmitting and one
receiving, are 
connected via coaxial cable to a network analyzer, which measures
the transmission coefficient (ratio of V$^{out}$ incident on the input
terminals of the transmitting antenna to V$^{rx}$ output at the terminals of
the receiving antenna), as
depicted in Fig. \ref{fig:calibration_scheme.ps}. 
The analyzer is calibrated to compensate for the
effect of the cables. The measurements were carried out on the roof of
the KU Physics Department (Malott Hall) in Lawrence, KS. 
Given the transmission
coefficient and the impedance of both antennsa, the complex effective height
is determined using the complex transmission formula.
                                                                                
The complex effective height of two identical Yagi antennas was first
measured in this manner. A calibrated Yagi antenna
was then used as the standard to
measure the complex effective height of two different RICE dipoles. In both
cases, the complex transmission equation is used to determine the effective
height from the transmission coefficients.
                                                                                
These measurements were taken at a range of 9.38 meters, which corresponds
to $r$ in the complex transmission formula. Error in this value has the effect
of translating the response of the antenna in time. However, this bias will
be identical for all antennas, and as we are ultimately only interested in
the relative timing of signal arrival at various antennas, the inclusion of
a constant bias in the timing of signals across all antennas will have no
effect on our results. We caution that, for wavelengths comparable to the
separation distance itself, unprobed
Fresnel zone effects may become significant.

\subsubsection{Polarization and Incident Angles}
The effective height was measured for an incident wave aligned parallel to
the axis of the RICE dipole antenna. That is, we have specifically measured
$H_{\omega ,\frac{\pi }{2},\varphi ,0} $. Due to the symmetry of the RICE
dipole design, we do not expect significant variation dependent on $\varphi$. 
For incident angles and polarizations other than this optimal alignment,
it has been assumed that $H$ is proportional to $\cos (\alpha)$, where
$\alpha$ is the angle between the field vector and the antenna axis
($\alpha$ is a function of $\rho$ and $\theta)$. (Additional measurements
suggested that although the $\cos (\alpha)$ relationship is valid for 
off-vertical polarization angles, it may not be a good approximation for
off-horizontal incident angles [i.e., $\theta \ne \pi /2]$. For example,
particularly at higher frequencies, the response can actually be greater for
$\theta =\pi /4$, than for $\theta =\pi /2$, suggesting a multilobed antenna
pattern at the higher frequencies. We estimate that this is a $\sim$10\% 
effect.)

\subsection{Results}
\begin{figure}[htpb]
\centerline{\includegraphics[width=11cm]{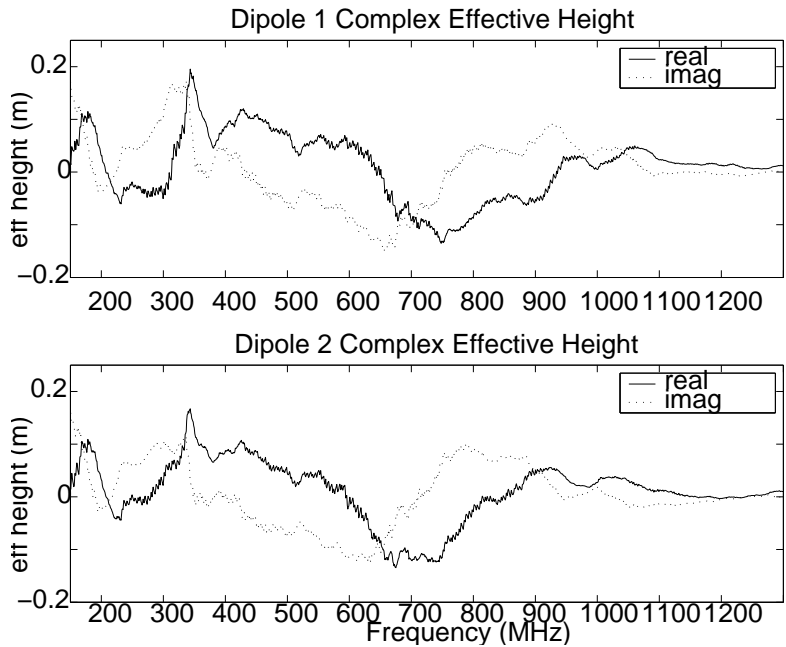}}
\caption{Real and Imaginary parts of
RICE complex effective height, measured for two dipole antennas.}
\label{fig:rice_dipole_H.eps}
\end{figure}

\begin{figure}[htpb]
\centerline{\includegraphics[width=11cm]{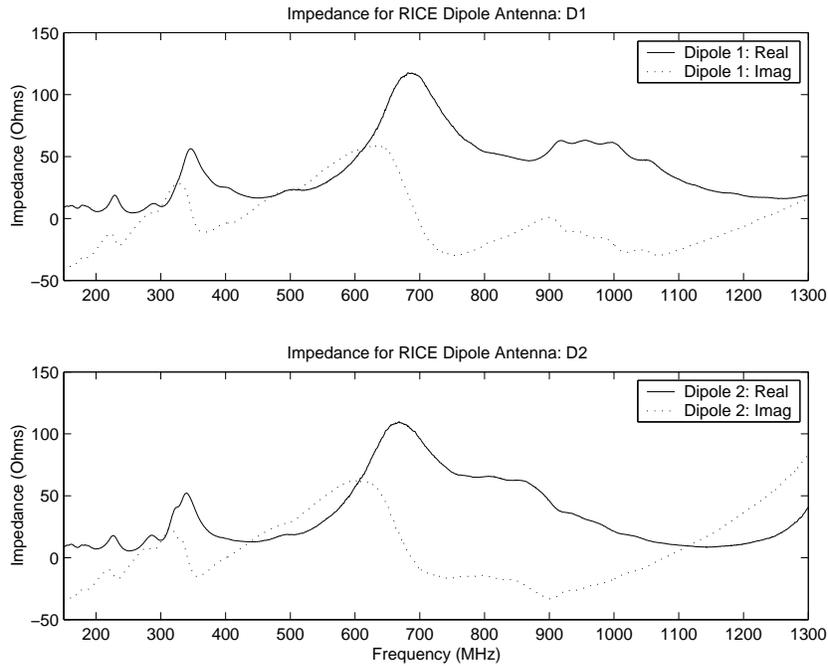}}
\caption{Real and Imaginary parts of complex impedance, measured for
two RICE dipoles.}
\label{fig:rice_dipole_Z.eps}
\end{figure}

Fig. \ref{fig:rice_dipole_H.eps}
shows the complex effective height 
two different dipoles, indicating
the degree of consistency between the two antennas. 
Fig. \ref{fig:rice_dipole_Z.eps} displays the complex impedance for the
same two dipoles.
\begin{figure}[htpb]
\centerline{\includegraphics[width=11cm]{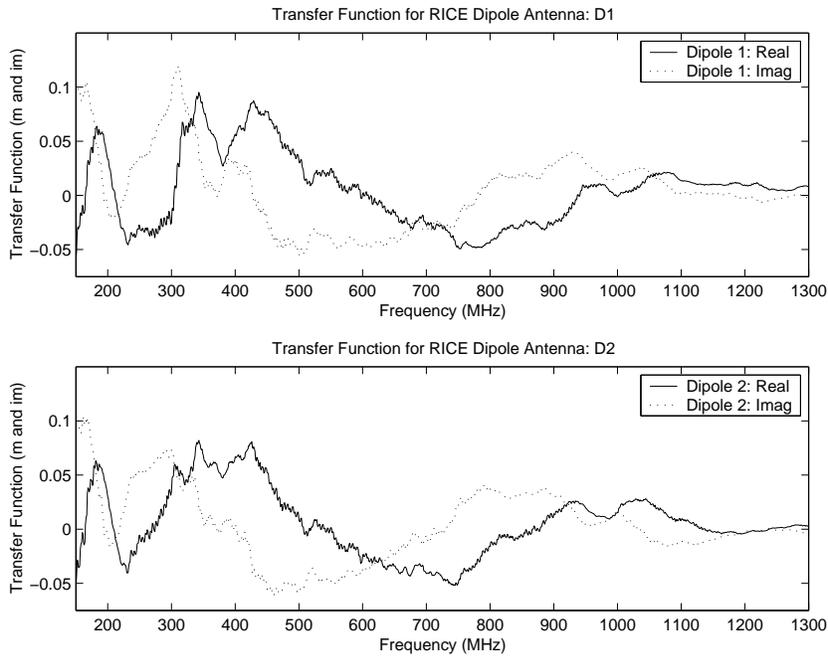}}
\caption{Real and Imaginary parts of transfer function, measured for
two RICE dipole antennas coupled to 50$\Omega$ coaxial cable.}
\label{fig:rice_dipole_T.eps}
\end{figure}
\begin{figure}[htpb]
\centerline{\includegraphics[width=9cm]{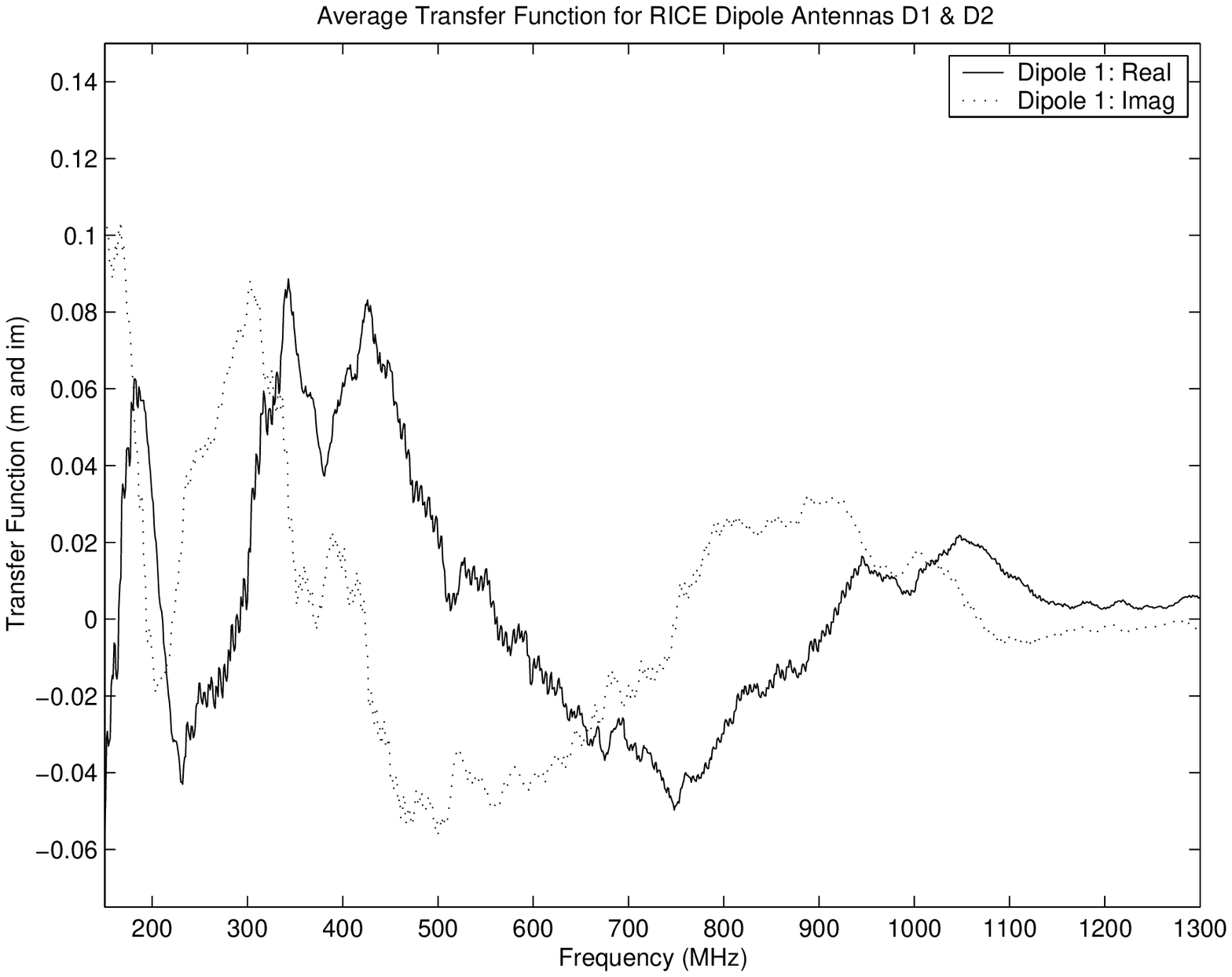}}
\caption{Average transfer function used in Monte Carlo simulations and
calculation of neutrino sensitivity.}
\label{fig:rice_dipole_Tavg.eps}
\end{figure}
Fig. \ref{fig:rice_dipole_T.eps}
shows the
transfer functions for these antennas computed from the impedance and
complex effective heights, and Fig. \ref{fig:rice_dipole_Tavg.eps}
shows the average transfer function, which is implemented in the
RICE Monte Carlo simulation (`radiomc').
                                                                                
\subsection{Systematic errors in antenna calibration}
It is difficult to obtain an exact value for the amount of error
in these measurements for the complex effective heights, impedance, and
ultimately the transfer functions of the RICE dipoles. The network analyzer
itself performs averaging over several repeated measurements of the
transmission or reflection coefficients, which reduce 
the statistical uncertainty in those
measurements to less than $\mathcal{O}$(1{\%}). However,
considerable error may be introduced due to reflections in the environment.
The amount of
error introduced into any single measurement is highly frequency dependent,
presumably as a result of whether reflections along different paths tend to
reinforce or cancel each other in a given band of frequencies. Of course
which frequencies are most affected will also change with the location of
the pair of the antennas.
                                                                                
Other random error can be introduced simply by jostling of the antennas as
they are repositioned. The distance between antennas can only be maintained
to within about 0.1 meter, and their relative orientation to within about 5
degrees. However variation among identical measurements is on the order of
two or three percent -- much smaller than the error introduced by
reflections.
                                                                                
Another source of uncertainty in the calibration of the RICE dipoles is
variation among the antennas themselves. We have extensively tested
the pair of dipoles mentioned previously. Comparing
several antennas, raw reflection coefficient
measurements show variations which are typically $\le$5\%.

\subsection{Scaling of RICE Dipoles to Ice}
Our current assumption is that the transfer
function is identical for identical wavelengths. This is equivalent to
the statement that the impedance of the environment, having index of
refraction n, is given by $\zeta/n$, where $\zeta$ is the impedance of
free-space ($\sqrt{\mu_0\epsilon_0}$=377 $\Omega$).
That is,
\[
T_\lambda ^{\mathrm{ice}} =T_\lambda ^{\mathrm{air}} .
\]
Intuitively, if we imagine the antenna as a 
resonant device analogous to a resonant pipe, the above scaling
seems reasonable - if the wavelength through air changes, but the
size of the pipe stays the same, then the resoonant frequency
will migrate proportionately. 

\begin{figure}
\centerline{\includegraphics[width=11cm,angle=0]{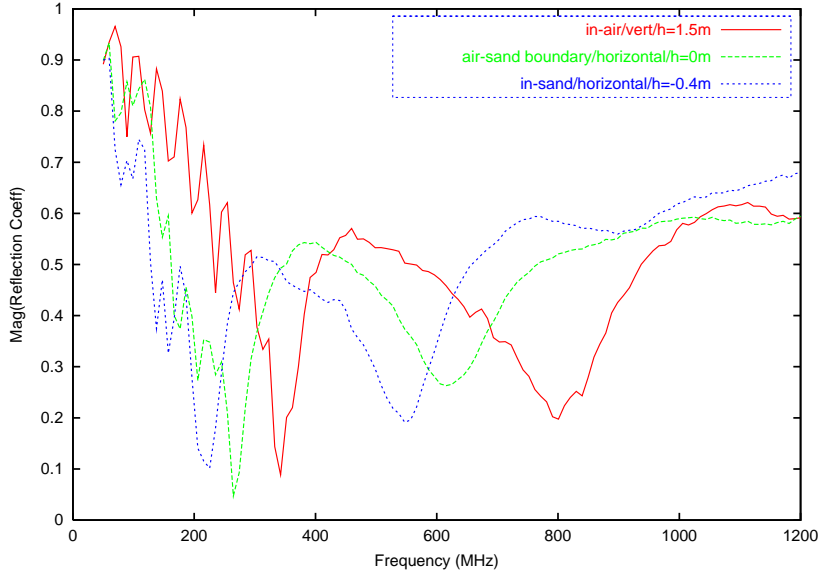}}
\caption{Comparison of antenna VSWR, measured in-sand compared 
to in-air, as a function of frequency. Note the downward shift in frequency
of the peak antenna response.}
\label{refl-coeff-sandVair}
\end{figure}

To qualitatively 
probe environment-dependent antenna response effects, 
three measurements of the complex
reflection coefficient (``S11'') of an antenna in proximity to a
large (3 meters across and 1 meter deep) sandbox. The
voltage standing-wave ratio (VSWR) was then calculated from the
complex reflection coefficient. In the first
measurement, a standard RICE dipole was oriented vertically,
about 1.5 m above
the sandbox. For the second, the dipole was lying horizontally on the
surface; \message{I neglected to do a vertical, half-in/half-out msrmnt}
for the third, the dipole
was dropped about 30 cm into the sand, still oriented horizontally. 
As indicated in Fig. \ref{refl-coeff-sandVair},
the peak response migrates from poles at $\sim$350 MHz in air to 250
MHz at the air-sand interface to 215 MHz when buried 30 cm,
more or less consistent with the scaling
expected by n (expected to be $\sim\sqrt{1.6}$ by $h_{eff}\sim\sqrt{GZ/Z_0}$
with the impedance of free space $Z_0\to Z_0/n$ in a medium of
index of refraction $n$).\footnote{This, 
of course, does not constitute a measurement of the complex effective height.}
In our simulations, we determine the expected in-ice response by
re-coupling the antenna complex impedance, in medium, to the
50-Ohm coaxial cable.

\subsection{Coaxial Cables \label{subsect:cable}}
Three types of coaxial cables, all having
similar attenuation characteristics, connect the in-ice RICE 
hardware to DAQ electronics in MAPO. With our current detector
geometry, we require approximately 200-400 m of cable to reach
the DAQ logic in MAPO from the under-ice antennas.
A summary of the main cable types and their 
loss specifications\footnote{As quoted by the manufacturer; lab tests
verified these loss values to within 2-3\%} is specified
in Table \ref{tab:cables}.

\begin{table}[htpb]
\centering
\begin{tabular}{|c|c|c|} \hline
Type 
& $v_{\rm propagation}/c$ & Power Attenuation(/100') @ 150 MHz/450 MHz \\
\hline
LMR-500 
& 0.86 & 1.22 dB/2.17 dB \\
LMR-600 
& 0.87 & 0.964 dB/1.72 dB \\
Cablewave FLC12-50J 
& 0.88 & 0.845 dB/1.51 dB \\
Andrews LDF4-50A 
& 0.88 & 0.73 dB/1.41 dB  \\
Andrews LDF5-50A 
& 0.88 & 0.46 dB/0.83 dB  \\ \hline
\end{tabular}
\caption{RICE coaxial cable specifications}
\label{tab:cables}
\end{table}

\subsubsection{\label{subsubsect:cabledispersion}Cable Dispersion}
\begin{figure}[htpb]
\centerline{\includegraphics[width=9cm,angle=-90]{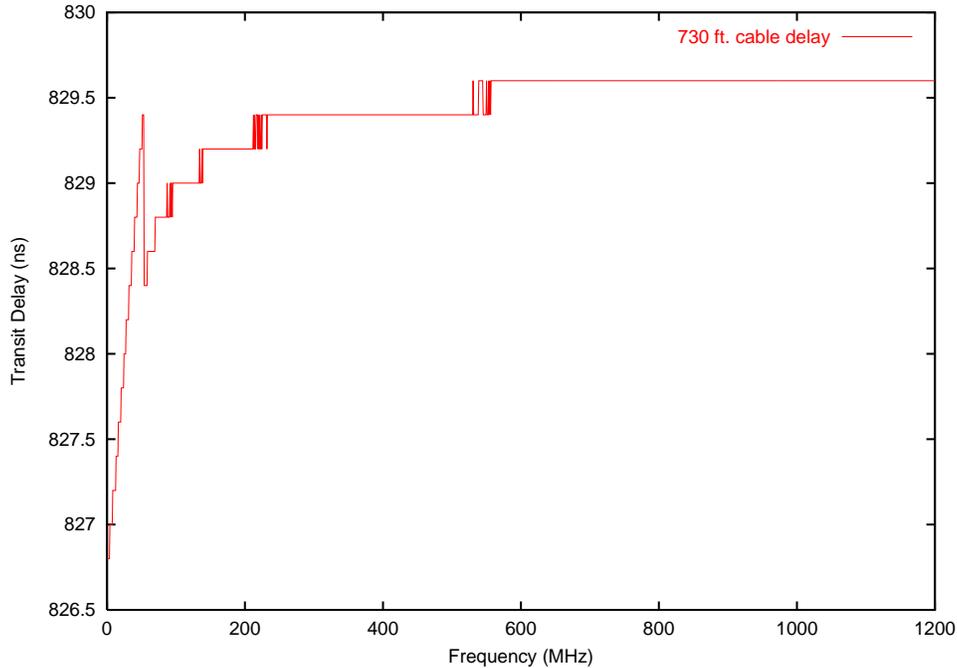}}
\caption{Measured propagation time, as a function of frequency, 
through $\sim$730 feet
of LCF-50J Cablewave cable only.}
\label{fig:cable-delay-lcf-50j.ps}
\end{figure}

\begin{figure}
\centerline{\includegraphics[width=11cm,angle=0]{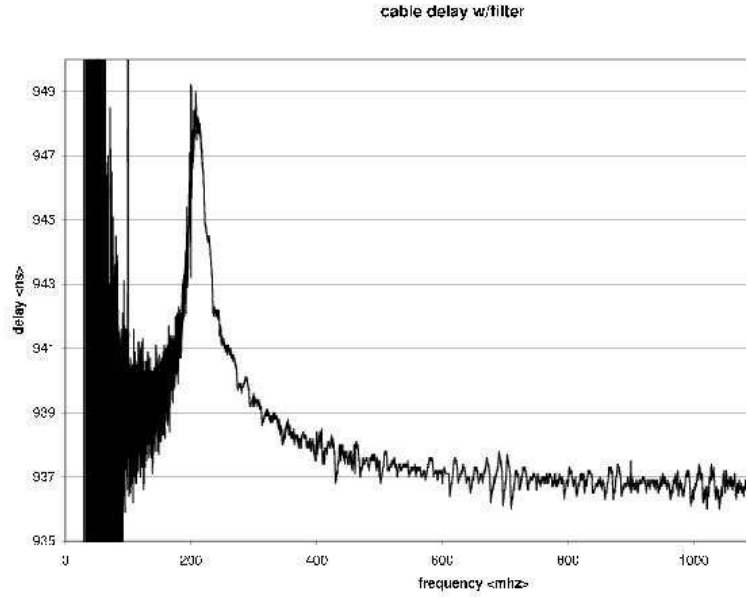}}
\caption{Measured propagation time, as a function of frequency, 
through $\sim$840 feet 
of LMR-600 cable, with high-pass filter connected at one end and terminated 
into 50$\Omega$. Effect of the high-pass filter is evident for frequencies
less than 300 MHz.}
\label{fig:cable+filter-delay.ps}
\end{figure}

In addition to the attenuation characteristics given in Table \ref{tab:cables},
dispersive effects must also be quantified.
Due to increasing
signal absorption with frequency, sub-ns duration signals will be
``spread'' in the time domain as they propagate through our cables.
We have directly quantified cable dispersive effects in the frequency
domain by measuring the signal propagation time through 800 foot lengths
of the LMR-600 and
LCF-50J cables, as a function of frequency
(Figures 
\ref{fig:cable-delay-lcf-50j.ps} and \ref{fig:cable+filter-delay.ps}).
In the RICE passband ($>$200 MHz), the cable is observed to be largely
non-dispersive, with phase differences less than 0.25 rad. However, through
the passband, the high-pass filter has considerable dispersive effects
in the regime 200-300 MHz. Such effects are explicitly incorporated into
our simulation.

\subsubsection{Cable Damage during deployment}
\begin{figure}[htpb]
\centerline{\includegraphics[width=9cm]{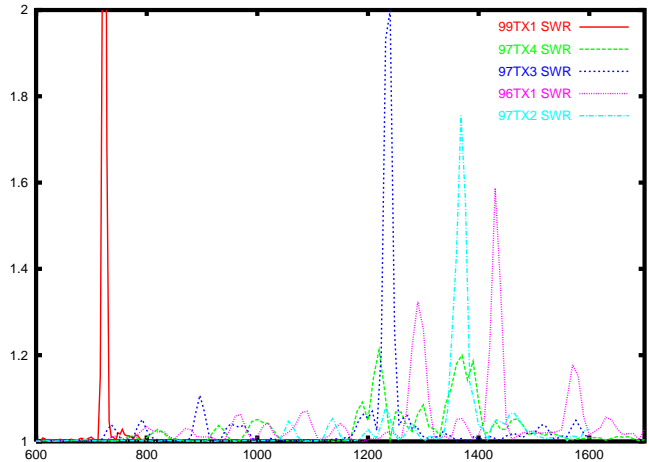}}
\caption{Standing Wave Ratio, all transmitters. Horizontal axis shows time
delay (ns) relative to output signal from network analyzer.}
\label{fig:AllTx-SWR.eps}
\end{figure}

Following deployment, there is the possibility that the cables are
damaged or kinked after freeze-in. 
Since both our neutrino signal strength, as well as the 
overall gain calibration depends on the cable signal loss, it is
important that any damage to the cable during deployment be
minimized.
Figure \ref{fig:AllTx-SWR.eps} shows
the standing-wave-ratio (SWR) obtained for 5 transmitters, frozen into
the ice. We note the presence of large returns in 97Tx4, indicating the
possibility of damage during freeze-in.

\subsubsection{Cable cross-talk}
Although the RICE coaxial cable should be well-shielded, 
cross-talk effects may lead to spurious hits recorded in our data.
We have, therefore, searched for 
cross-talk between antennas in the same hole, 
knowing the spatial difference between
any pair of antennas, as well as the signal propagation velocity through
the cable. 
For example, channels 7 and 9 are in the same hole and are
separated by approximately 50 m (similarly 
for channels 8 and 10). Consider
two possibilities: a) the deep receiver is hit first, sending a signal
up through the cable, which cross-talks to the shallower receiver, within
a time scale of 1-2 ns or so. In this model, the time difference between
the two signals (original plus induced through cross-talk)
received at the surface is very small (the same 1-2 ns, assuming $v_{cable,1}$
=$v_{cable,2}$ and the cable lengths are the same), 
b) the shallow receiver is hit, a cross-talk signal propagates both up and
down the cable attached to the deep receiver. At the surface, we would
measure an induced signal in the deep receiver delayed, relative to the
former case, by the total extra cable transit time for the deep antenna.
Note that $\sim$0 ns observed time difference would also correspond
to signals induced in the surface electronics.\footnote{Such
events are observed periodically during typical data-taking.}
\message{DBT: USE THE NO-ANTENNA DATA TO QUANTIFY SURFACE-GENERATED NOISE - cannot find these data 09-21-06}

\message{DBT: ADD WAVEFORMS FROM TRANSMITTER EVENTS FOR TWO RECEIVERS IN SAME HOLE. WHY??}

\subsubsection{Cross-talk due to AMANDA (and RICE) cables}

\begin{figure}[htpb]
\centerline{\includegraphics[width=4cm]{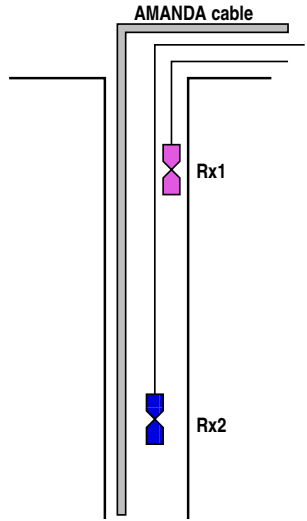}}
\caption{Geometry of possible cross-talk effects.}
\label{fig:Xtalk.xfig.eps}
\end{figure}

\begin{figure}[htpb]
\centerline{\includegraphics[width=9cm]{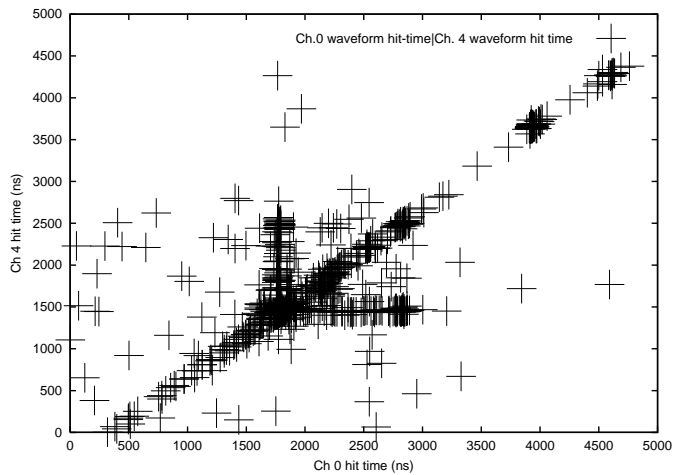}}
\caption{Hit time registered in channel 1 (x-) vs. hit
time registered by hit in channel 4 (y-), for general events. Note that
when either channel is the 4th hit of the 4-hit coincidence, its
time is set to $\sim 1.5\mu$sec, corresponding to the vertical and
horizontal off-diagonal bands.}
\label{fig:Xtalk-ch0-ch4}
\end{figure}

We considered two possible cases - a) cross-talk in holes with AMANDA cables, as well as b) cross-talk in the ``dry'' holes drilled specifically
for RICE. An illustration of the cross-talk geometry is shown
in Fig. \ref{fig:Xtalk.xfig.eps}.
Channel 1 is 50 m deeper than channel 4 in an AMANDA hole,
corresponding to an additional cable time delay of $\sim$190 ns if
cross-talk pickup is in the RICE cable itself, and $\sim$280 ns
if cross-talk pickup is in the AMANDA cabls. Given
the full cable-length 
time delays (1416 ns, and 1166 ns, respectively), 
and knowing the cable lengths $l'$ and $l$, we would
expect the channel 1 signal to arrive at the DAQ either
440 ns, or 530 ns after channel 4
signals if all the channel 1 signals were due to cross-talk effects. 
For channel 7 and channel 9, we expect channel 7 to arrive 380 ns after
channel 9 for complete cross-talk pickup in channel 9. 
Such bands are not observed in data
(Fig. \ref{fig:Xtalk-ch0-ch4}).\footnote{The
in-ice signal propagation velocity can also be derived from plots such as
these; this will be discussed later in this document when we describe
our attempt to observe cosmic ray muons.}

\subsection{Double pulses}
\begin{figure}[htpb]
\centerline{\includegraphics[width=9cm]{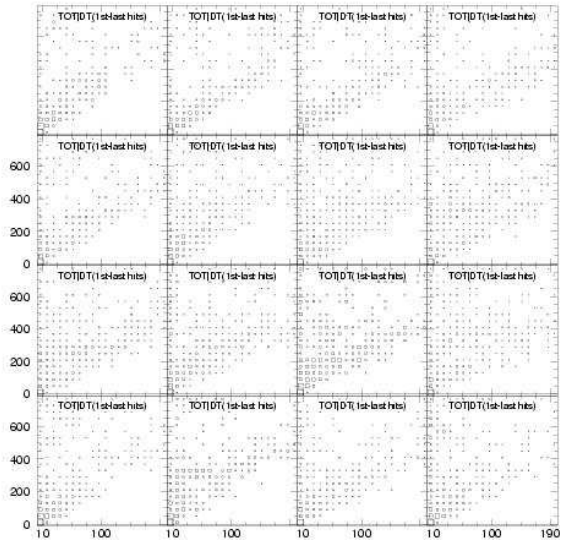}}
\caption{Time over threshold in each channel (horizontal) vs.
time difference between last and first $6\sigma$ excursion, in a waveform.
``Double pulses'', in which a waveform is repeated multiple times with a fixed time delay between the first and next pulse, should appear as single bin peaks in these plots.}
\label{fig:dtVtot}
\end{figure}

Visual scanning of our event sample 
shows a large fraction of events which contain double pulses.
(It is, in fact, the presence of such double pulses which gave rise
to some concern that triggered-sampling digitizers may miss 
pre-trigger or post-trigger double pulses which would otherwise
disqualify a hit as 'valid'.)
Algorithms
which target, and reject such 
double-pulse events, would also be potentially very 
worthwhile in our efforts to minimize backgrounds. We have searched for
double pulses 
in two ways: first, by looking at the time-over-threshold distribution,
scatter plotted against the time
difference between the time-of-first-$6\sigma$-excursion minus the
time-of-last-$6\sigma$-excursion (Figure \ref{fig:dtVtot}), to 
determine if our backgrounds are characterized by a simple series of
high-amplitude signals, having fixed temporal extent. In this model,
the pieces of the waveform before and after this background 
are quiescent. Although Figure \ref{fig:dtVtot} does, indeed,
show a clear correlation between these two quantities, the fact that the
diagonal band in each channel is relatively thick, and not thin, indicates
that the temporal extent of the receiver hit has large scatter.

\begin{figure}[htpb]
\centerline{\includegraphics[width=9cm]{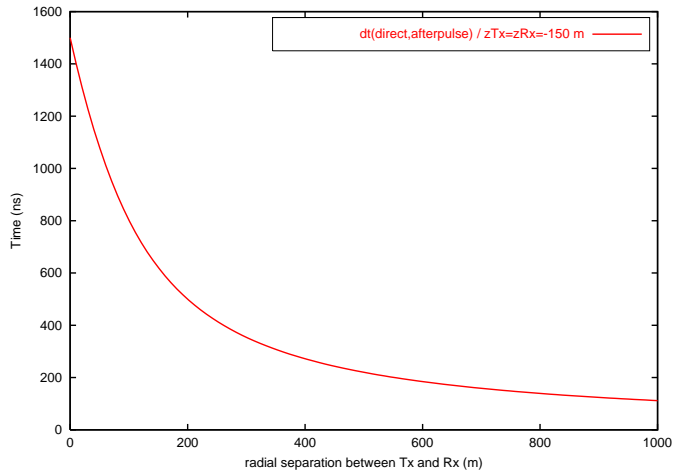}}
\caption{Expected time difference between direct and after pulse (simulation),
given indicated transmitter/receiver geometry.}
\label{fig:deltat-direct-afterpulse}
\end{figure}

\begin{figure}[htpb]
\centerline{\includegraphics[width=9cm]{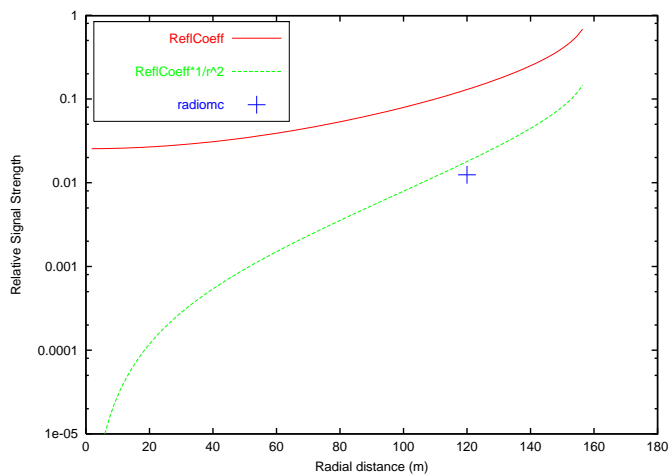}}
\caption{Amplitude of
``Reflected'' second pulse, based on calculated
Fresnel reflection coefficient, compared with ray-tracing code currently used
in RICE signal Monte Carlo simulation.}
\label{fig:ReflPulse}
\end{figure}

We expect real double pulses due to signals which traverse a ``direct'' path
from source to measurement point, followed by the signal reflected off of the 
air-surface interface 
and back down to the antenna. By contrast, noise generated
on the surface can travel down the cable, and ``reflect'' off an
antenna back up to the DAQ.
Figure \ref{fig:deltat-direct-afterpulse} shows the expected
time delay between the direct and the surface-reflected signals 
(in nanoseconds) for the case where both transmitter and receiver
are at a depth of --150 m, for a given radial separation between
transmitter and receiver. Note that the current double pulse cut 
in the reconstruction code is
set to 800 ns, which should cut out very few events resulting from
true physical reflections at the air-ice interface.
The magnitude of the afterpulse is determined by the additional pathlength
(to first order, the signal power $\propto 1/r^2$) and the reflection
coefficient of the signal at the surface (determined by the appropriate
Fresnel coefficients). 
For the same geometry as assumed in the previous figure, and ignoring
any defocusing of the signal as it reflects off the uneven top surface
(this is certainly not the correct case), Figure \ref{fig:ReflPulse} gives
the signal strength of the reflected pulse compared to the initial signal
strength, based on an analytic calculation just using the 
Fresnel Reflection 
Coefficient expected at the surface interface
(red, assuming straight line
trajectories through the firn), then including $1/r^2$ spreading of field lines
(green), and compared with one calculation of the current Monte Carlo
simulation (point), which includes the differential bending of the
ray as it traverses the firn. Note also that these plots assume a
spherical source geometry -- in the rare instance of a 
neutrino induced signal, for which a RICE receiver is
hit at the Cherenkov angle, the afterpulse would be diminished by 
the cone angular width factor $\sim exp(-(\theta-\theta_c)^2/(2*(d\theta)^2)$,
with $d\theta\sim$0.04 radians at 500 MHz. 
\message{I dont get this - this seems to be just showing that the MC has the right factors in it}

\subsubsection{Time Structure of After Pulses}
Observations indicate that
the after-pulse in a double pulse event
occurs typically anywhere from 500 ns -- 2000 ns after the initial pulse
in an event, and $\sim$10\%--50\% in amplitude of the initial
pulse (for events that saturate, the amplitude could, in principle, be
measured simply from the time-over-threshold of the signal, modulo
amplifier saturation effects). It was also
noticed that after-pulses occur, to some extent, in
transmitter data, as well. In several cases, we observe that the
time delay between the hit-time of the primary pulse in an event and the
after-pulse is of order twice the cable delay between the surface and
the transmitter that was broadcasting the signal\footnote{This may, in fact, be
the most reliable way of calibrating this transit time, and may give one
the t0 of a transmitter signal}, consistent with a picture where
the source emits a signal which is partially transmitted and
partially reflected back to the generator, then subsequently re-reflected
down to the transmitter. In fact, laboratory tests performed at KU indicate
that the situation is more complicated than that -- not only do we observe
clearly the (generator$\to$Tx$\to$generator$\to$Tx) reflection, but the
(Rx$\to$scope$\to$Rx$\to$scope) reflection is also evident,\footnote{The 
reflection at the scope obviously depends on the scope's input impedance.} albeit at a
lower amplitude than the former reflection

\begin{figure}[htpb]
\centerline{\includegraphics[width=11cm]{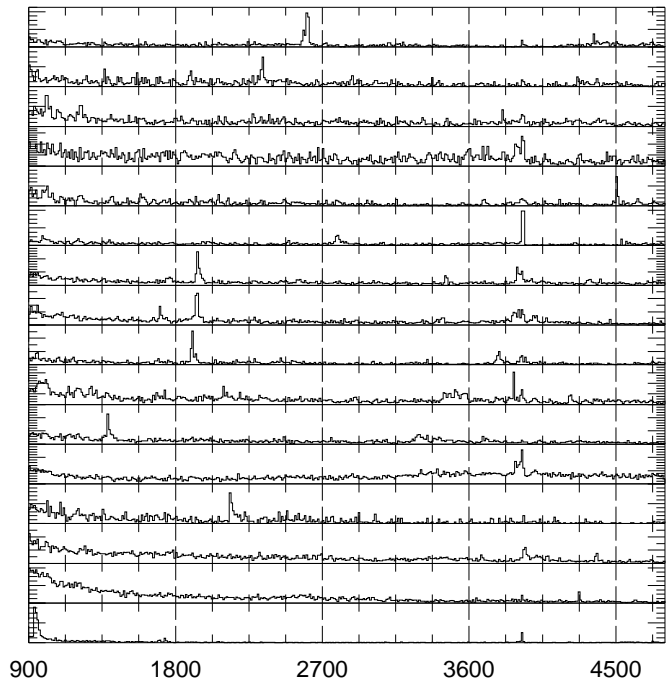}}
\caption{Distribution of time difference between time of largest 
amplitude hit in a 
given channel relative to time of second 
largest amplitude hit, requiring that the largest
hit be at least 800 ns separated in time from the 2nd hit (general events).
Each panel represents an oscilloscope waveform channel, beginning with
channel 0 (top) to channel 15 (bottom). Horizontal axis is time in
ns.} 
\label{fig:doublepulse_general_dt500ns}
\end{figure}

\begin{figure}[htpb]
\centerline{\includegraphics[width=11cm]{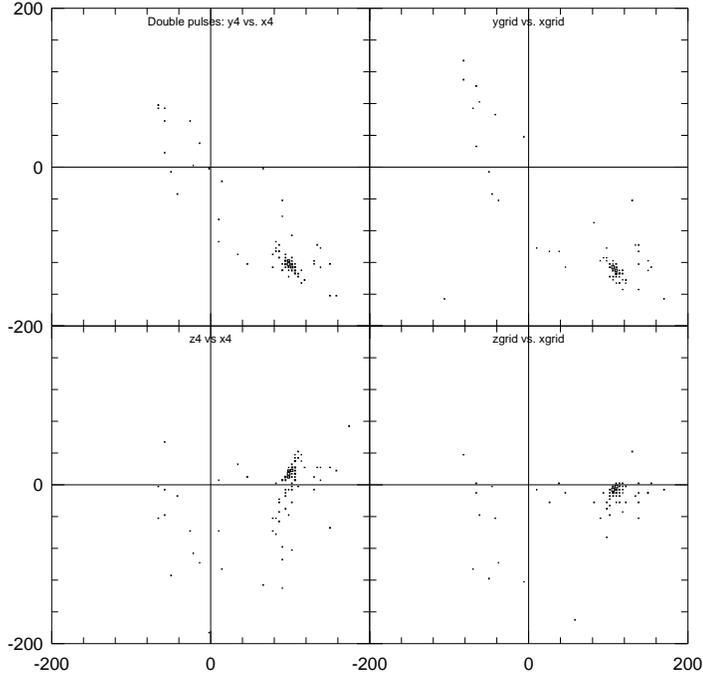}}
\caption{Vertex distributions, for events which show clear $Rx$
reflections.}
\label{fig:dblpls-vtxdist}
\end{figure}

We have looked in the 2000 data for evidence of double pulses,
using the variable t800nXX, where ``t800n'' indicates that
we are examining the first large-amplitude excursion
at least 800 ns following the first hit in a waveform,
and ``XX'' designates a channel number.
Figure
\ref{fig:doublepulse_general_dt500ns}
shows the distribution in the time difference
between t800nXX and the time of the maximum voltage in a given waveform,
for general event triggers.
One observes the expected
time difference that would correspond to a reflection off the surface
electronics
itself and back down to the in-ice receiver 
(i.e. a time difference which is twice the listed time delay
for that channel) clearly in channels 6, 7, 8 , 10 and 12. Further 
investigation shows that the appearance of double pulses is strongly 
correlated for channels 6, 7, and 8 (i.e., when one channel shows a
receiver reflection at the exptected time delay, the other two channels
do, as well). 
In addition, such events show a strong correlation with
the vertex location (x,y,z) = (100,--140,0), for both vertexing algorithms,
as shown in Figure \ref{fig:dblpls-vtxdist}. Although this location
does not obviously correspond to a feature on the surface, it is 
important to remember that ray tracing distortions complicate surface
source locations.

\message{TBD: 
COMMENT ON THIS SOURCE LOCATION - USE RADIOMC TO GENERATE A MAP GIVING
TRUE VS. RECONSTRUCTED VERTEX, SHOW THE MAP AND COMMENT ON PUTATIVE SOURCE
LOCATION.}

\begin{figure}[htpb]
\centerline{\includegraphics[width=9cm]{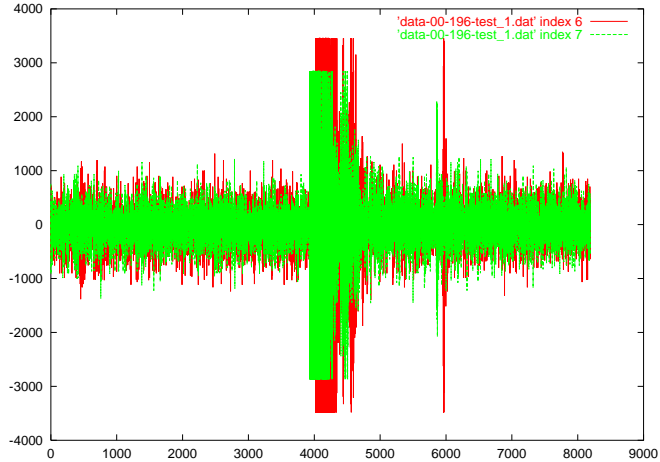}}
\caption{Typical waveforms for events which show ``$Rx$''-type
reflections, for two channels. (Horizontal axis is time in nanoseconds;
vertical scale is arbitrary.) Note that the time difference between first-
and second-pulses are comparable for these two channels (channel 6 and 7).}
\label{fig:dblpls-evt}
\end{figure}

As an example of what the time domain 
waveforms look like for such double pulse
events, Figure
\ref{fig:dblpls-evt} 
shows the waveforms observed for event 8, day 196, of the year 2000 data
(channels correspond to the `index' indicated in these plots).

\subsubsection{Discriminator Performance}
\begin{figure}[htpb]
\centerline{\includegraphics[width=5cm,angle=-90]{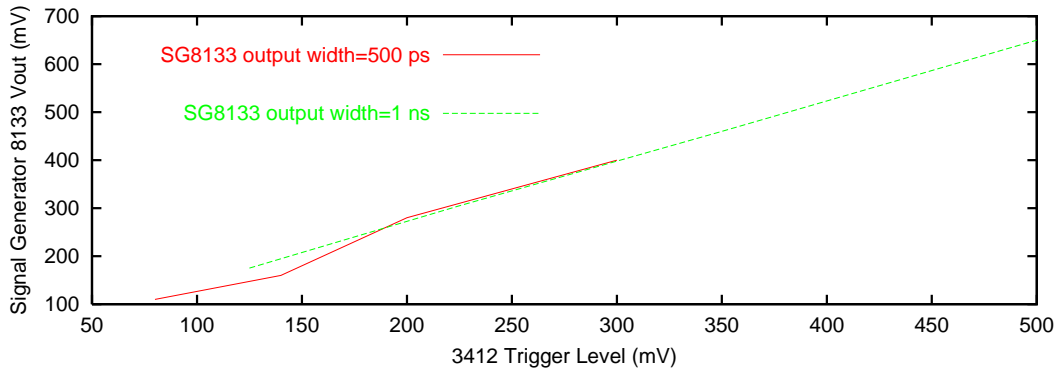}}
\caption{LeCroy 3412E Discriminator efficiency study. The lines 
correspond to the 3412 Trigger level required for 100\% efficiency.
Units are millivolts.}
\label{fig:3412-efficiency}
\end{figure}

After the cables, the next element in the signal path is
the LeCroy 3412 discriminator. Since the neutrino signal has an
intrinsic bandwidth of $\sim$1 GHz, the discriminator
must be highly efficient for ns-duration RF pulses.
A series of studies were conducted to quantify the efficiency of the
LeCroy 3412. A fast signal generator (HP8133A), capable of
generating 500 ps width signals was used to pulse a 3412 module
in the lab; the signal
was split and monitored with an HP54542 digital oscilloscope. The nominal
threshold of the 3412 is varied by turning a set screw on the front panel
of the 3412 module; a voltage probe allows to read back the corresponding
threshold value. 
As shown in Figure \ref{fig:3412-efficiency}, the 3412
efficiency turn-on is not sharp; rather, we measure the 3412 set
voltages corresponding to initial response, up to 100\%
efficiency. Down to a signal width (FWHM) of 500 ps, the discriminator
performs as expected (to within 10\%), and also out-performs the 
internal trigger of the HP54542. 

We have also determined, in the laboratory, that the 3412 discriminator is
capable of triggering on a wide range of amplitudes and signal strengths.
A test was run for which the 3412 threshold was set to 
0.9$\times V_{generator}$, where $V_{generator}$ is the output signal
voltage as recorded by an HP8133 signal generator. Figure 
\ref{fig:3412-efficiency} shows the voltage levels recorded for both the
generator, and also as read off one of the RICE HP54542	digital oscilloscopes,
for which the trigger efficiency exceeded 90\%. 
We observe good efficiency
for signals down to 0.5 ns.

\subsection{Amplifiers}
RICE voltage signals are typically boosted by a factor of $10^4$ before
waveforms are recorded. 
Such a large gain requires daily monitoring.
We now describe amplifier gain measurements and calibration.
\subsubsection{Gain Drift\label{sect:gain-calibration}}
\begin{figure}[htpb]
\centerline{\includegraphics[width=9cm]{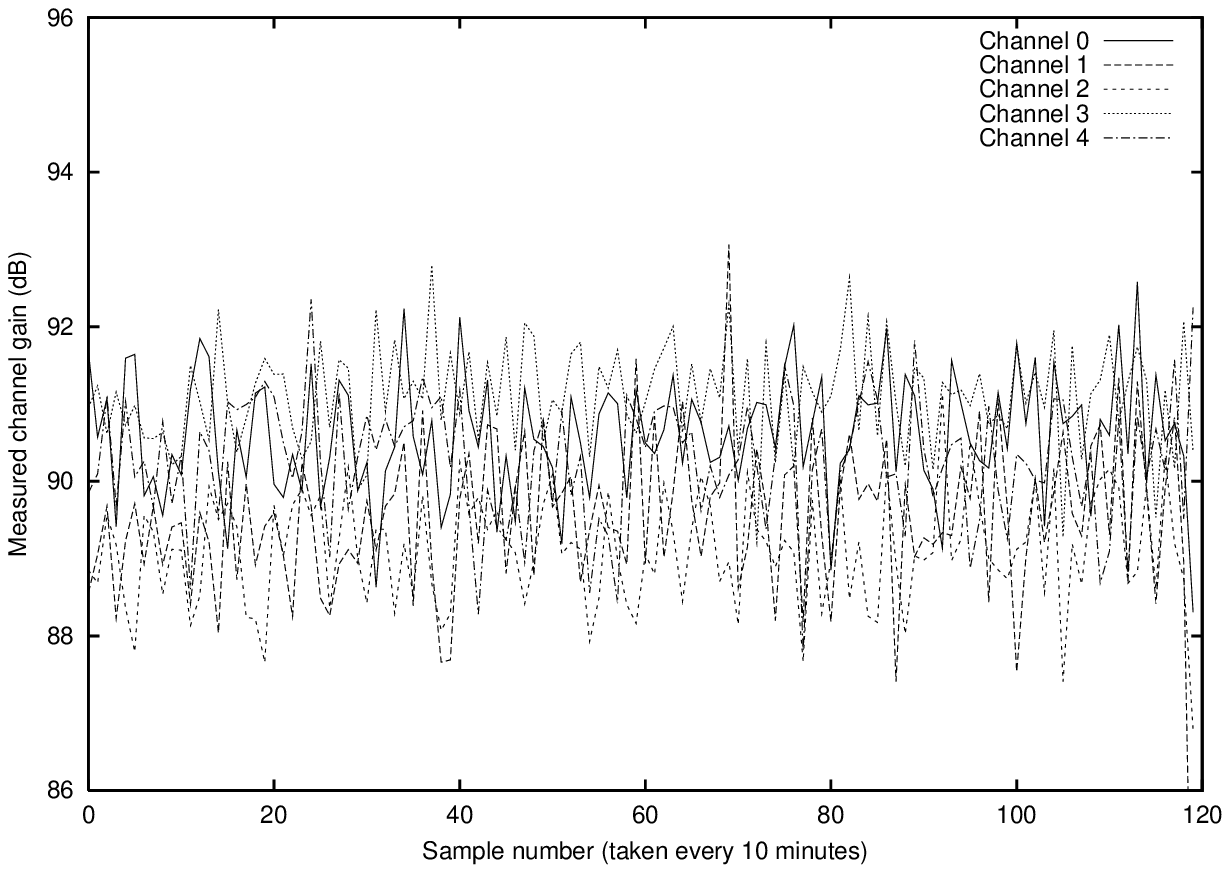}}
\caption{Calculated gains, based on normalizing to thermal noise, 
for ``unbiased'' triggers,
taken from April 15, 2001 data for five (random) channels.}
\label{fig:amp-gains.eps}
\end{figure}

\begin{figure}[htpb]
\centerline{\includegraphics[width=9cm]{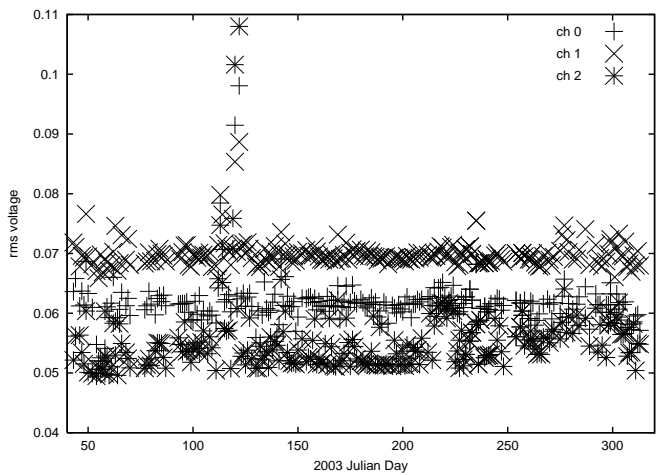}}
\caption{Voltage rms in three RICE channels, monitored over a full-year
period; approximately one entry per day.}
\label{fig:amp-rms-2003.eps}
\end{figure}

As described previously\citep{rice03b,rice06},
the amplifier gain is calculated on-line, simply by measuring the rms
of the voltage distributions recorded in `unbiased' events, and,
after correcting for the known cable losses in each channel,
relating the noise power in a bandwidth B to the voltage
$V$ by $V^2/Z=kTB$, with $Z$ the characteristic impedance of the
system. Figure \ref{fig:amp-gains.eps} shows the monitored amplifier gains,
over a 20-hour period. Figure
\ref{fig:amp-rms-2003.eps} shows the monitored gains, over
a longer time baseline.
We estimate the statistical error from each measurement
to be of order 1 dB. Some `spikes' are visible, which contribute to our
overall systematic error.\footnote{As re-iterated below,
we remind that, in our previous publication, we estimated an overall
system power gain uncertainty of 6 dB for a full circuit (signal generator$\to$transmitter~antenna$\to$receiver antenna$\to$surface data acquisition).}

\subsubsection{Amplifier Saturation}
\begin{figure}[htpb]
\centerline{\includegraphics[width=9cm]{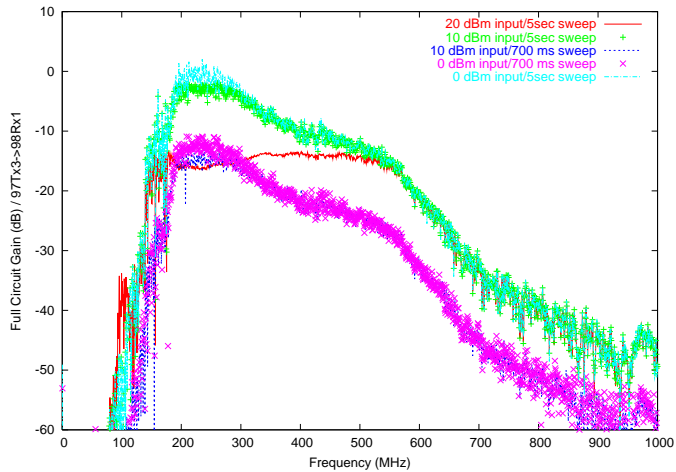}}
\caption{Calculated full-circuit gain
(signal generator$\to$transmitter~antenna$\to$receiver antenna$\to$surface data acquisition), as a function of network analyzer 
sweep time. We observe saturation effects at fast sweep times coupled with
large input power.}
\label{fig:NWA-amp-saturation-studies}
\end{figure}

At large enough signal input amplitudes, most amplifiers will
saturate (roughly, when the output 
voltage is of order 
0.1$\times V_{bias}^{plateau}$, with $V_{bias}^{plateau}$ 
the bias voltage required for the 
amplifier to reach it's maximum gain). 
By sweeping at either very high power, or fast sweep time, we can probe
possible amplifier saturation effects.
This is illustrated in
Figure \ref{fig:NWA-amp-saturation-studies}. We expect linearity, however,
for the short-duration excitations induced by neutrino events.

\section{Online Software}
Data are collected online using a custom-written LabView code
(running in the Windows environment) which 
interfaces to the CAMAC hardware and also sets the parameters for the
digital oscilloscopes. Runs are initiated daily by the Raytheon Polar
Services winter-over (for which we are eternally grateful);
data are also archived to CD on a daily basis, as well.

\subsection{Trigger scheme}
\label{sec:trigger}

There are three  possible conditions that can produce a RICE trigger:
\begin{enumerate}
     \item The main RICE trigger is a ``self-trigger'', which is formed if
$\ge$N underground antenna receivers fire over threshold within the 1.25
microsecond gate. At present we use a 4-fold coincidence
criterion.  These are our 
primary physics events (``general'' trigger events).

   \item Random noise triggers, or so called Unbiased events, are
triggers forced by the DAQ periodically to sample the noise environment.

    \item An AMANDA-B or SPASE coincidence trigger (aka
``external'' trigger) fires if at least
one underground antenna is hit within $\pm$1.25us from the trigger
signal received from the ``big'' AMANDA-B or SPASE. The ``big'' AMANDA-B
trigger signal corresponds to a
30-fold AMANDA multiplicity trigger.\footnote{We thank Steve Barwick for implementing this large trigger.}

\end{enumerate}

\begin{figure}
\vspace{22cm}
  \includegraphics{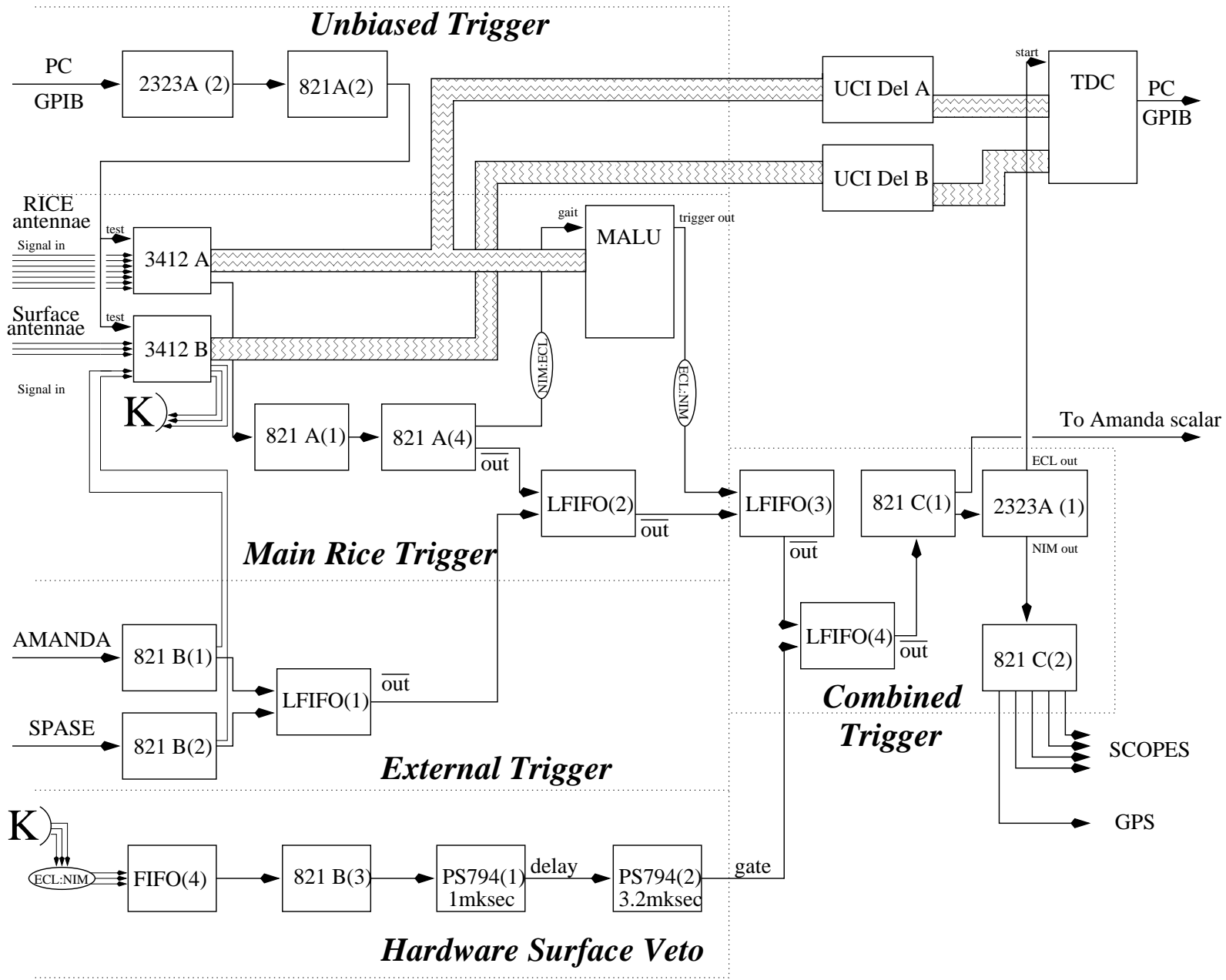}
        \caption{ Functional diagram of the DAQ electronics.
\label{function}} 
\end{figure}

A functional diagram of the DAQ and trigger is presented in Fig. \ref{function}.
  The figure is divided into logical sections by a dotted line.
First, consider the {\bf Main Rice Trigger}. Analog data
from antennas arrive at the CAMAC discriminators 3412E and (B).
If any channel exceeds the discriminator threshold, 
a NIM pulse appears at the output of this channel. 
The ``Current Sum Output''
for all hit
channels of the 3412E is converted into a 1.25$\mu$s long NIM pulse 
by means of two discriminators 821 A(1) and 821 A(4). 
Transformed into an ECL signal, this
then opens the gate of the MALU 4532 module.
While the gate is open, the MALU counts rising edges 
arriving at its inputs 0-15 from the 3412E (only one hit
per channel is counted). If the number of pulses
seen by the MALU is greater than or equal to some preset value N
(by default, 4), a general trigger pulse is produced by this module.
  
The {\bf Noise Events Trigger} produces Unbiased events.
The chain for this trigger starts at the Gate 2323A (2).
At appropriate moments, this Gate 
is triggered by the DAQ executable from the main PC
through the GPIB interface. Two copies of this signal created using
the 821 A(2) module are connected to the ``Test'' inputs of both
discriminators 3412E and B. A pulse appearing at the ``Test'' input 
produces hits in all channels of the discriminators' outputs.
These hits then propagate through the remainder of the DAQ
as with the Main Rice Trigger. 
Afterwards, the events develop as described
above for the Main Rice Trigger.

The remaining trigger line is the
{\bf External Trigger}, that is, either the AMANDA-B or
SPASE coincidence triggers. First, we ensure that the AMANDA and SPASE
triggers are standard NIM signals using the discriminators 821 B(1)
and 821 B(2). Next, we form a logical OR of AMANDA and SPASE
by means of the Logic FIFO (1) module. Then, we want to 
impose the logical 
condition $A.and.B=C$ where A is detection of AMANDA or SPASE
and B is at least one underground RICE; while C is the result. Since
our LFIFO 
has only OR logic in it, we therefore
use the fact that $A.and.B=C$ is logically equivalent
to $\bar{A}.or.\bar{B}=\bar{C}$, which is then employed in the
FIFO (2) logic. I.e., we take the .OR. of 
the inverted AMANDA.or.SPASE logic signal
and the inverted ``at least 1 underground hit'' logic signal. 
We then take an inverted output,
corresponding to our desired ``C'', External Trigger.

These three chains form the input for the {\bf Combined Trigger}.
The Main RICE trigger is ORed with the External Trigger in the
Logic FIFO (3). A {\bf Veto} can be invoked with
the following logic: $A.and.\bar{B}=C$, where A is
a valid trigger, B is the veto pulse and C is the final trigger.
In the absence of an .AND. module, we use the fact that
$A.and.\bar{B}=C$ is equivalent to $\bar{A}.or.B=\bar{C}$.
Thus, we combine the inverted result of LFIFO(3) with the plain
output of the veto branch in Logic FIFO(4) and take the inverted
output of this combination. 
This output is finally run through 821 C(1) to
obtain a constant duration logic pulse. The result
is the ``raw'' RICE trigger. One copy of this is used to
monitor the raw trigger rates, connected to an external scalar, e.g.
The other copy triggers the gate unit 2323A (1); the output
of it is the final RICE trigger. The last unit is needed
as it latches the system and preserves the event until it
is read out.

  The discriminator 821 C(2) produces 5 copies of the final trigger
that latch the scopes and freeze the event time in the GPS module.
The TDC also receives its ``Start'' from the final RICE trigger,
while data inputs come from the discriminators 3412E and 3412
delayed by 1.5 $\mu$s in the Delay modules A and B.

  Once triggered, the system ignores any other possible valid
triggers and retains data in the TDC and the scopes until cleared
and enabled again.

\subsubsection{Software Online Trigger Veto }
\label{sec:veto}

\begin{figure}[htpb]\centerline{\includegraphics[width=9cm]{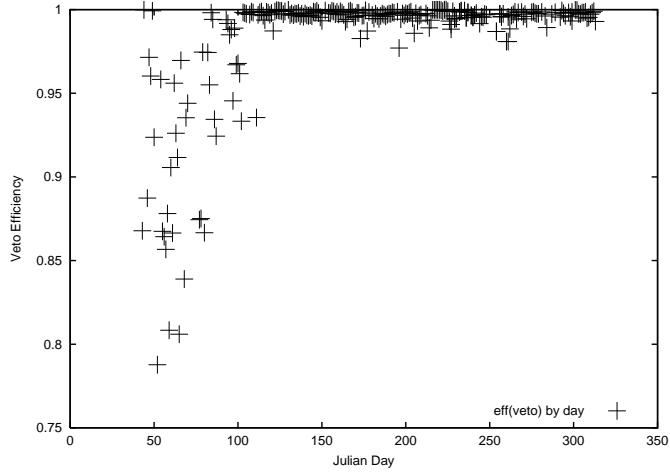}}\caption{Calculated surface-source veto efficiency (software), as a function of Julian date (2002).(This is NOT the efficiency for passing neutrino events.)}\label{fig:Veto-eff-2003.eps}\end{figure}

The second (software) level
vetoes time-of-hit patterns for `general' 
triggers (as registered by on-line DAQ TDC's) 
which are characteristic of surface-generated noise. These were
determined by simulating sources located at known South Pole 
science stations in the vicinity of the RICE experiment, including
MAPO, the ASTRO, and the SPASE experiments. The resulting hit times were
stored in a ``library'', which was then compared with an ensemble of
expected antenna hit times from simulated neutrino events.
This online filter works reasonably well - typical
efficiencies of this filter range from 96\%--99\%
(based simply on the ratio of general triggers to rejected veto
triggers, assuming that all the general triggers are, in fact, 
surface in origin). The efficiency, as a function of
Julian day, for 2003, is shown in Figure \ref{fig:Veto-eff-2003.eps}.

\subsection{Deadtime Determination}
\subsubsection{Online Live Time monitoring}
 
\begin{figure}
\vspace{10cm}
\includegraphics{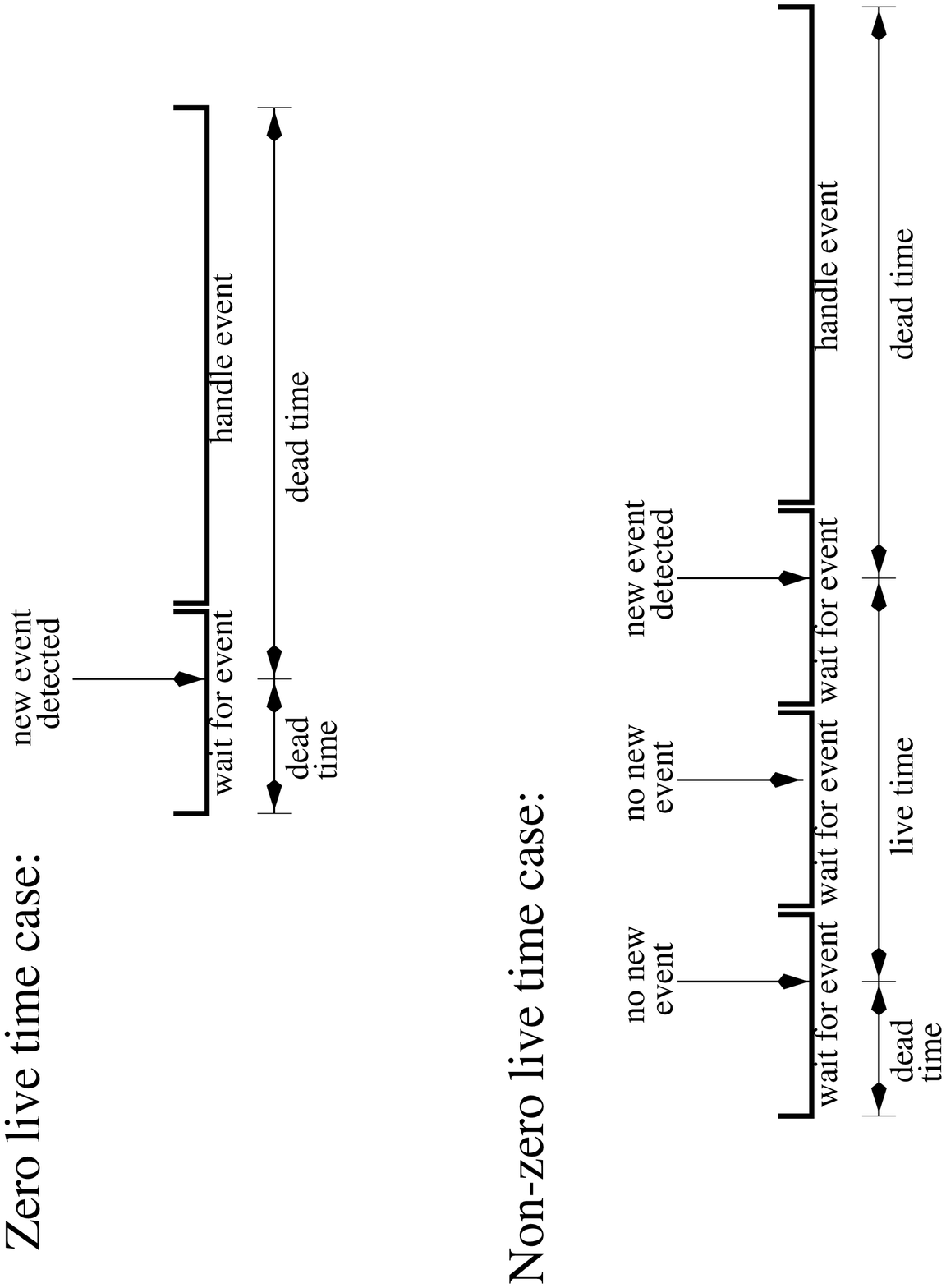}
\caption{Live/dead time definition}
 \label{fig:livefrac}
\end{figure}

     The Live Time of the DAQ is calculated on-line
during data taking.
Each event cycle can be roughly broken into two pieces:
waiting for event and event processing, as shown in 
Fig. \ref{fig:livefrac}.
When the new event arrives, the DAQ does not respond instantly,
but with a certain delay. In the process of waiting, the
executable stays within a ``while'' loop, checking for a
trigger every cycle as well as performing some other
actions. One waiting cycle takes approx. 4-5ms, about 1.2ms
of which is actual checking for the trigger occurrence. The 
definition of the dead/live time is illustrated in 
Fig. \ref{fig:livefrac}.
Out of the full time that one event takes, the first ``waiting
for event'' cycle and full ``handling event'' part is counted
as dead time, while the extra (above 1) ``waiting for event''
cycles are counted as live time. The live fraction is then defined
as Live Time divided by the total event time for this event.
The total live fraction for the run can now be calculated
as the total live time for all events detected divided by the total
run time.
In addition, the experiment is essentially `shut down' when the
South Pole Station uplink (at 303 MHz) to the LES-9 satellite is
active (approximately 5 hours out of every day prior to 2005, when
LES was disabled by NASA).

A fair amount of effort was invested in attempting to
assess the dead time intrinsic
in forming the event trigger, and also
associated with making the software veto decision described above,
as well as the maximum rates that the DAQ is capable of.
  There are two areas that have been investigated: the maximum rate at which 
DAQ can process events and decide whether to accept or reject them, 
and dead time incurred when a good event 
is being saved. 

\subsubsection{ Maximum trigger rate}

  First, we measure the maximum DAQ veto rate and veto dead time.
A four-fold copy of signals coming out of a signal generator
is sent into the 3412E. The time spacing between the signals is set 
(using appropriate delays from the PS 794 unit) so
that the event should constitute a four-fold coincidence, but the timing
pattern should be consistent with a `veto' event. The frequency of
the Signal Generator output is then varied, and the trigger rate through 
the DAQ recorded. If the dead time were zero, then the trigger rate 
through the DAQ would exactly track the input signal generator rate. 
The results of this test, for data taken in 2000, 
are given in Table \ref{tab:maxrate}.
Subsequent improvements to the DAQ and the CAMAC interface
gave approximately four-fold reductions in dead time.

\begin{table}[htpb]
\begin{center}
\begin{tabular}{|c|c|c|c|c|}
\hline
  Generator & \multicolumn{2}{c|}{Fast mode} 
                         & \multicolumn{2}{c|}{Slow mode} \\
  frequency, Hz & Event Rate, Hz & Live fraction 
                         & Event Rate, Hz & Live fraction \\
\hline
  10   &  10.1  & 0.918  &  9.9 & 0.800 \\
  20   &  20.3  & 0.838  & 19.7 & 0.580 \\
  30   &  30.2  & 0.758  & 30.2 & 0.294 \\
  40   &  39.8  & 0.679  & 39.9 & 0.157 \\
  50   &  50.4  & 0.599  & 40.1 & 0.165 \\
  60   &  60.5  & 0.517  & -    &  - \\
  70   &  69.1  & 0.440  & 34.6 & 0.228 \\
  80   &  79.4  & 0.357  & - & - \\
  90   &  89.3  & 0.279  & - & - \\
 100   &  99.4  & 0.194  & 44.2 & 0.092 \\
 105   & 105.3  & 0.155  & - & - \\
 110   & 109.9  & 0.117  & - & - \\
 115   & 114.2  & 0.078  & - & - \\
 120   & 113.7  & 0.084  & - & - \\
 125   & 112.0  & 0.093  & - & - \\
 130   & 105.1  & 0.148  & - & - \\
 135   &  93.2  & 0.256  & - & - \\
 140   &  78.0  & 0.368  & - & - \\
 150   &  80.8  & 0.346  & 43.2 & 0.044 \\
 160   &  79.3  & 0.356  & - & - \\
 170   &  84.3  & 0.316  & - & - \\
 200   &  99.9  & 0.194  & - & - \\
 300   & 100.9  & 0.188  & - & - \\
 400   & 100.8  & 0.188  & - & - \\
 500   & 118.3  & 0.037  & - & - \\
 1000  & 123.4  & 0.000  & - & - \\
 5000  & 124.9  & 0.000  & 48.3 & 0.000 \\
\hline
\end{tabular}
\end{center}
\caption{Veto rate as a function of the input frequency of the
simulated signal. Slow mode is different from fast mode by the presence
of the LabView Event Display window. Unmeasured quantities are indicated by a dash.}
 \label{tab:maxrate}
\end{table}

From the table, we draw the following conclusions:
\begin{itemize}
   \item Saturation of the rate happens at about 125 Hz 
indicating a dead time of 0.008 seconds.

   \item The event rate is independent of how many scopes/channels are in the 
system, as expected (the numbers in the table correspond to runs
with 16 channels used; the runs with a single channels active 
give the same result).
   \item The event rate does not rise 
monotonically but goes up and down before 
reaching saturation. This is likely to be caused by the interference 
between the cycles with different time constants: the frequency of the 
generator and DAQ polling rate. Naturally, this does not affect the dead 
time of the system which should be calculated at the rate saturation point.
\end{itemize}

\subsubsection{Calculation of correlation between trigger rate and deadtime
(general events)}
\begin{figure}[htpb]
\centerline{\includegraphics[width=9cm]{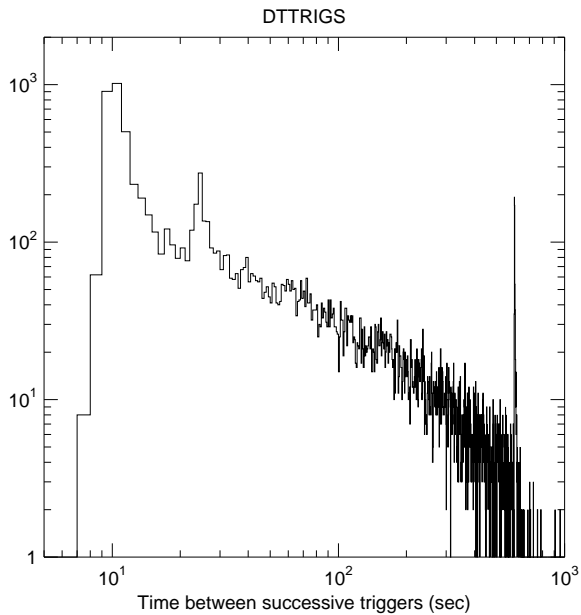}}
\caption{Time between successive triggers, all event types, January,
2005 data.}
\label{fig:dttrigs}
\end{figure}

First, we examine the time between
successive triggers (``DTTRIG'', Figure \ref{fig:dttrigs}). 
Three peaks
are evident - at 10 seconds (the time required to read out
general events), 20 seconds (the time required to read out 
an unbiased event), and t=600 seconds (the time between 
successive unbiased events, if there are no other intervening
triggers). Of the 14136 events in this plot, 10546 of them
correspond to DTTRIG$>$20 seconds. Alternately, 
during this data-taking period $>$70\%
of our triggers are not occurring in 
``general-trigger-saturation'' mode. Given that these data
were taken during IceCube drilling and the ambient noise levels
are large, this is not surprising.

\begin{figure}[htpb]
\centerline{\includegraphics[width=9cm]{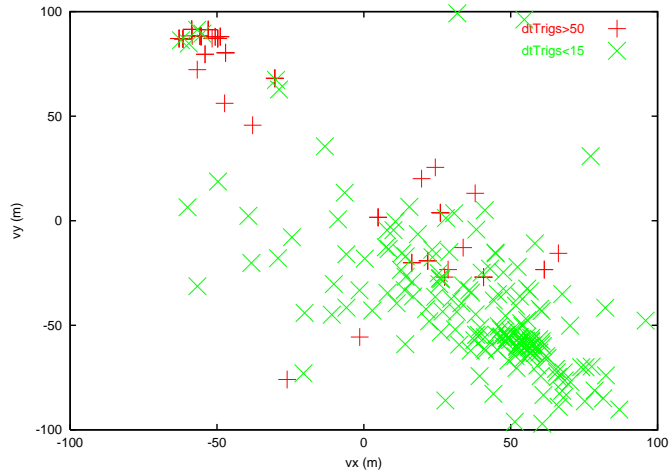}}
\caption{Vertices, 2002 data, divided into subsamples of large
time-since-last-trigger, and small time-since-last-trigger. The
MAPO building is centered
at approximately (x=--50,y=--50) in the Figure.}
\label{fig:dttrigsA}
\end{figure}

Figure \ref{fig:dttrigs} suggests considering whether
surface-originating events can also be obviously characterized by
their time-since-last-trigger. Events which are saturating the DAQ,
such that triggers are being registered as fast as the DAQ can record them,
are likely to be of a single origin. 
For cases of small time-since-last-trigger, we expect the
vertices to cluster at the surface, since such times are clearly
dominated by background.
Figure \ref{fig:dttrigsA} shows the reconstructed vertices
according to time-since-last-trigger. The small time-since-last-trigger
events, corresponding to small livetime, cluster around the known
location of the MAPO building.

\subsection{HSV board}
\begin{figure}
\centerline{\includegraphics[width=9cm]{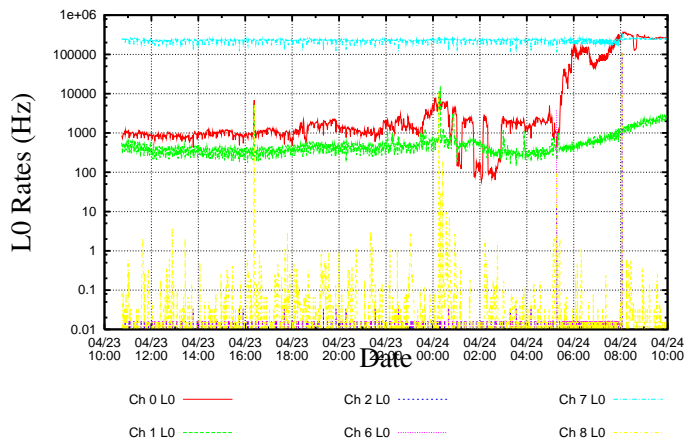}}
\caption{Raw (``Level 0'') HSV singles rates.}
\label{fig:HSV_singles_06115.eps}
\end{figure}

\begin{figure}
\centerline{\includegraphics[width=9cm]{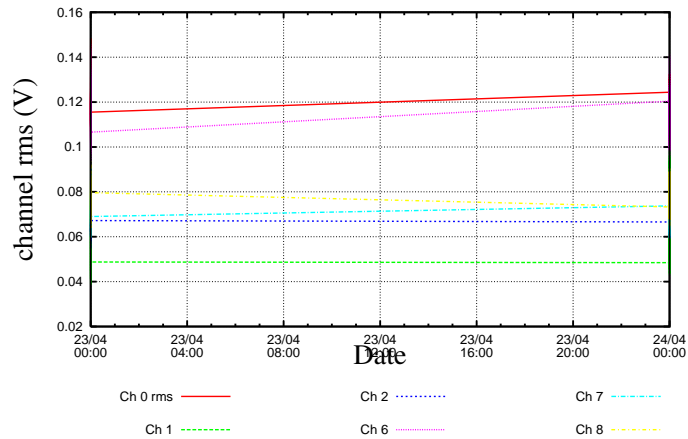}}
\caption{Corresponding rms voltages of channels plotted in previous figure.
Color scheme identical to previous plot.}
\label{fig:rms_06115}
\end{figure}

As discussed elsewhere, an additional Hardware Surface Veto 
CAMAC board was integrated into the RICE Data Acquisition system beginning in January, 2005\citep{rice06}. For hits separated by at least 20 ns, this board has been measured to provide a veto of potential surface-generated noise at a rate greater than 200 KHz.
This board also allows us the capability of measuring singles rates in the RICE antennas (Fig. \ref{fig:HSV_singles_06115.eps} and also
Fig. \ref{fig:rms_06115}, which shows the possible correlation of the
singles rate with large rms values for each channel). We
estimate that the HSV board enhances our neutrino
sensitivity by $\sim$30-40\%.

\subsubsection{TDC performance and stability}
\begin{figure}
\centerline{\includegraphics[width=9cm]{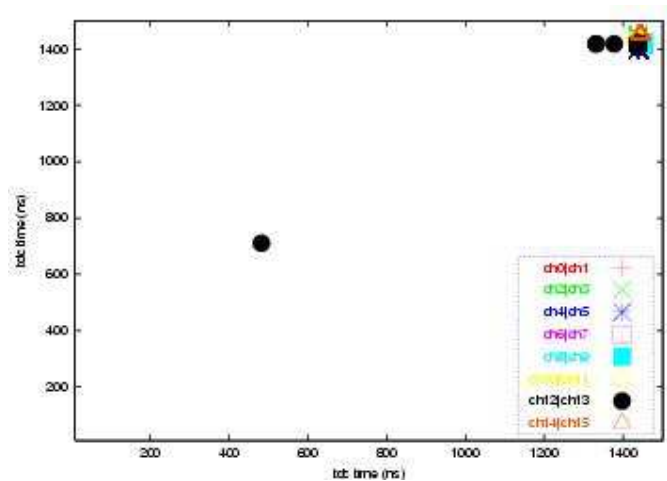}}
\caption{Raw TDC times for the indicated channels, unbiased events.
Number of times each channel deviates
from 1440 in unbiased evts indicates jitter (bin at (1450,1450)
contains 837 events).}
\label{fig:tdc1Vtdc2_unbiased.ps}
\end{figure}

We know that the TDC hit time in unbiased events should roughly correspond
to the signal propagation time through the full DAQ (dominated by the
delay board), or approximately 1.4 microseconds. Figure
\ref{fig:tdc1Vtdc2_unbiased.ps} presents the recorded TDC times for
unbiased events. Although the overwhelming majority of events
cluster at the expected time, we do observe occasional deviations
from the known expected time, indicative of either TDC noise or
`stray' hits.

\subsubsection{Comparison of times recorded by TDC's vs times 
derived from waveforms.}
\begin{figure}
\centerline{\includegraphics[width=9cm]{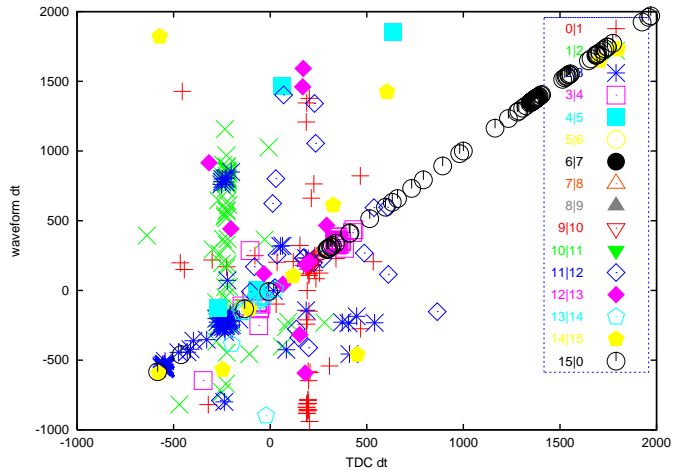}}
\caption{Channel i-channel j hit times,
extracted from either raw TDC's, or waveform times, 
general events.}
\label{fig:tdc-V-waveform.eps}
\end{figure}

A data acquisition system based entirely on fast TDC times would be
considerably less expensive, as well as less prone to incurring deadtime.
Such a system is effective only if the times extracted from the TDC's
are nearly as reliable as those extracted from waveforms. 
Unfortunately, it also deprives us of the possibility of imposing
matched filtering.
Figure
\ref{fig:tdc-V-waveform.eps} shows this correlation for a sample of data.
Although largely effective, the large number of off-diagonal entries indicate
that the TDC's are an inadequate proxy for full waveform information.

\subsection{Stability of Full Circuit Gain Calibration}
As detailed elsewhere\citep{rice03b}, we have calibrated the
total power gain of each receiver to within $\pm$3 dB.
In the final step of our amplitude calibration, the
antenna response to a calibration signal 
broadcast from an under-ice transmitter is measured {\it in situ}.
This test calibrates the combined effects of all cables, signal splitters,
amplifiers, etc. in the array.
A 1 milliwatt (0 dBm) 
continuous wave signal is broadcast through the transmit port of
an HP8713C NWA. The NWA scans through the frequency range
0$\to$1000 MHz in 1000 bins, 
at a slow enough sweep rate to ensure that the
in-ice amplifiers are not saturated. The signal 
is transmitted down through $\sim$1000 feet of coaxial cable to 
one of the five
under-ice dipole transmitting antennas. 
The transmitters subsequently broadcast this signal
to the under-ice receiver array, and the
return signal power from each of the receivers
(after amplification, passing upwards
through receiver cable and 
fed back into the return port of the NWA)
is then measured. Using laboratory measurements
made at the University of Kansas of: 
a) the effective height of the dipole antennas, as a function of
frequency (previously described), b) the dipole Tx/Rx
efficiency as a function of polar angle and
azimuth, c) cable losses and dispersive effects
(cables are observed
to be non-dispersive for the lengths of cable, and over the frequency
range used in this experiment),
d) the gain of the two stages of amplification as determined from RICE
data acquired {\it in situ} by
normalizing to thermal noise
$P_{noise}=kTB=\Sigma_{\omega}<V_{ant}^2(\omega)>/Z$, summing
over all frequency bins in the bandpass, and e) finally 
correcting for $1/r^2$ spherical spreading
of the signal power, one can model
the receiver array 
and calculate the expected signal strength returning
to the input port of the network analyzer.
This can then be directly compared with actual measurement.

\begin{figure*}[thpb]
\centerline{\includegraphics[width=11cm,angle=0]{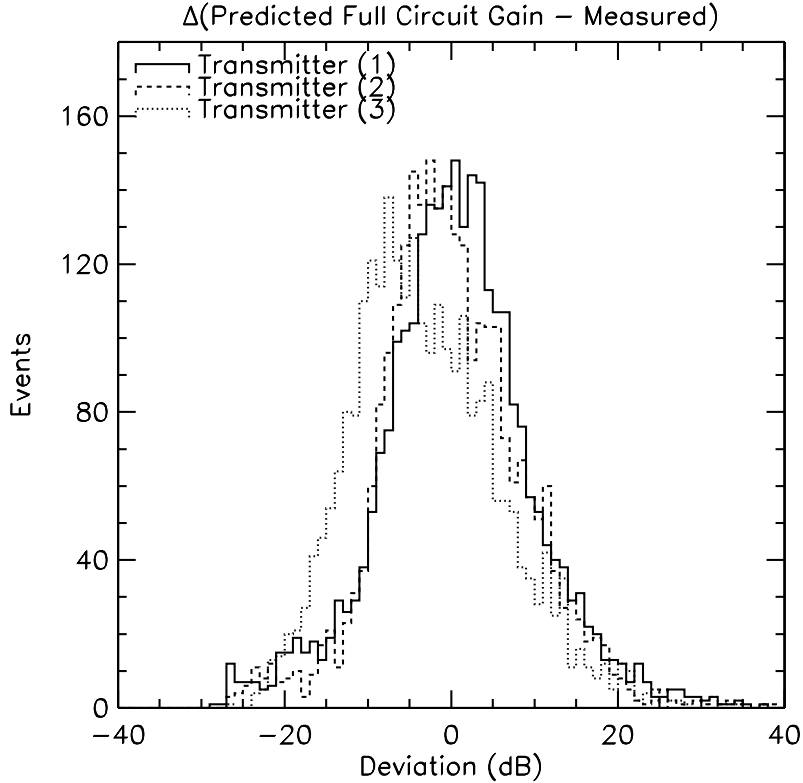}}
\caption{Deviation between expected vs. measured 
Tx$\to$Rx signal power for three transmitters broadcasting to
16 receivers.}
\label{fig:full_circuit_gain_dzb}
\end{figure*}

Such a comparison, as a function of
frequency, is shown in Figure 
\ref{fig:full_circuit_gain_dzb}; for each frequency bin, shown is the difference
between calculated vs. measured full-circuit gain.
Below 200 MHz, the attenuating effect of the high-pass filter is evident.
Figure \ref{fig:full_circuit_gain_dzb} thus shows the deviation between
the calculated gain minus the measured gain, for an ensemble of data runs.
Included in the Figure
are each of the 500 1-MHz bins between 200 MHz and 700 
MHz, for three transmitters. I.e., for each frequency bin, we determine
a gain deviation and enter that value in the Events vs. Deviation
histogram.
The average deviation between model 
and measurement over that frequency range
is calculated as
$\Delta(G_{calc}-G_{meas})\pm\sigma_{G_{calc}-G_{meas}}$, where
$\Delta(G_{calc}-G_{meas})$ is the mean of each of the
distributions shown in the Figure, and
$\sigma_{G_{calc}-G_{meas}}$ is the error in the mean, given
by the r.m.s. of the distribution itself divided by the
number of points in each distribution.
For the five currently functional
transmitters, the mean differences between the expected and
the measured gain
are $-0.6\pm0.6$, $-0.8\pm0.6$, $-2.3\pm0.5$, $-3.4\pm0.6$ and
$-2.8\pm0.6$ dB. 

Within the ``analysis''
frequency band of our experiment (200 MHz - 500 MHz), our
quoted level of
uncertainty in the total receiver 
circuit power is $\pm$6 dB; this
value is commensurate with the width of
the gain deviation distributions. On average, however,
the calculated gains quoted above are
well within these limits. 
Note that no correction for ice absorption has
been made, given the small scale of the array. Nor have 
corrections been made for possible AMANDA cable ``shadowing'' in
the same ice-hole, which may account for some of the undermeasurement
of signal relative to expectation for transmitter 97Tx4.
Given a typical transmitter-receiver separation distance of 100 m, an
electric field
attenuation length of 100 m would result in a shift of
$G_{calc}-G_{meas}\sim$8 dB. Although difficult to quantify,
our results are consistent with no observable attenuation.

\begin{figure}[htpb]
\centerline{\includegraphics[width=9cm]{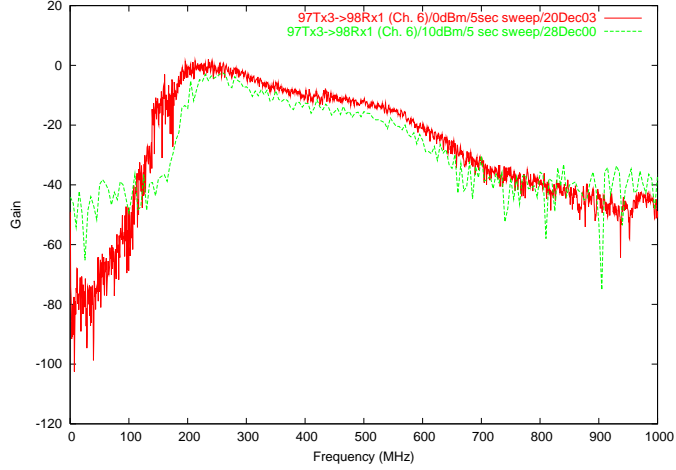}}
\caption{Comparison of network analyzer-derived full circuit gain for
$Tx\to Rx$, 2000 vs. 2003.}
\label{fig:97Tx3-X-98Rx1}
\end{figure}

We have investigated the stability of the gain over a three-year period.
Figure \ref{fig:97Tx3-X-98Rx1} shows the comparison of the 
``full-circuit'' gain, between 2000 and 2003. The later data are, in fact, 
slightly higher in amplitude than the older data in the primary
passband ($>$200 MHz), and also have a lower-frequency ``cut-off''
(due to the replacement of the NHP-250 highpass filters by SHP-200
highpass filters for the purposes of making radioglaciological measurements
in 2003-04).
To calculate eventual upper limits on the
neutrino flux, however, we continue to use the more conservative calculated
full-circuit gain, from the 2000 data.

\subsection{Hit Pattern Recognition}
\subsubsection{Timing Uncertainties}
\begin{figure}[htpb]\centerline{\includegraphics[width=10cm,angle=0]{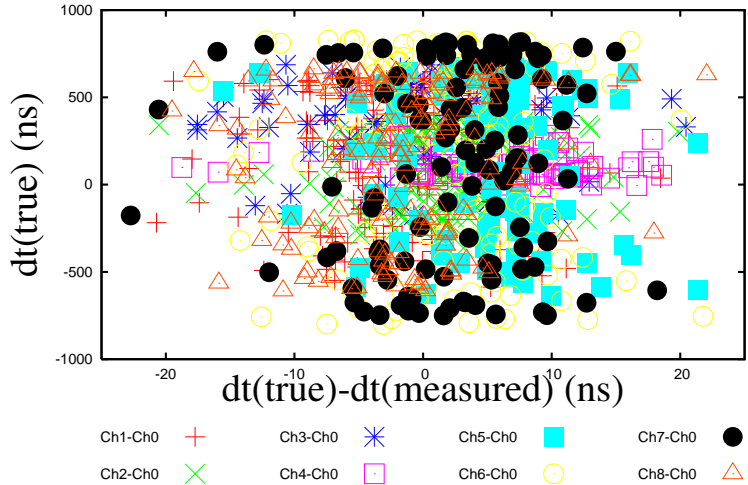}}
\caption{Scatter plot of the difference between the true (embedded) time between antenna hits minus the reconstructed hit time (after embedding) vs. the true time difference. The indicated timing resolution due to pattern recognition uncertainties is approximately 5 ns.}
\label{fig:dt-embedding}\end{figure}

\begin{figure}[htpb]\centerline{\includegraphics[width=10cm,angle=0]{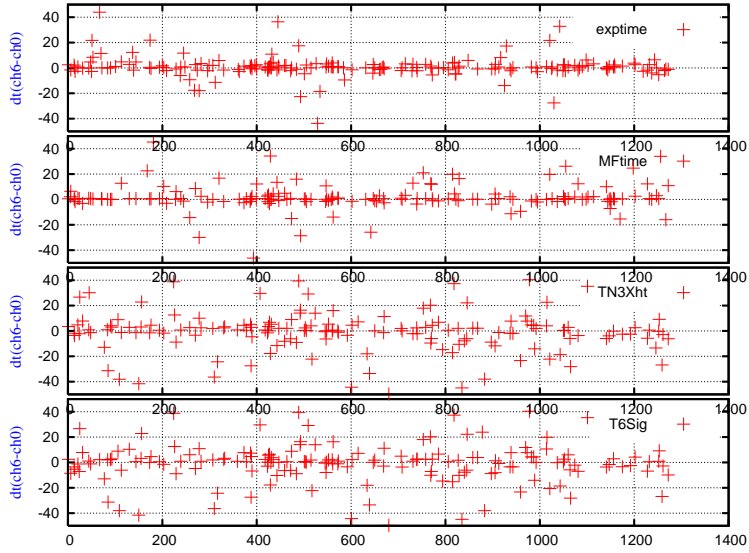}}
\caption{Comparison of the difference between the true (embedded) time between antenna hits minus the reconstructed hit time (after embedding) for 4 different hit time algorithms. The default RICE analysis uses the bottom criterion for hits; the 'exptime' algorithm, which simply looks for generalized exponential rings, obviously provides superior performance.}\label{fig:comp-hittimes.eps}\end{figure}

We have performed an embedding study to evaluate contributions to hit time
resolutions and the relative efficacy of various signal time estimation
algorithms. Monte Carlo simulations of neutrino-induced hits are
embedded into data unbiased events, and the extracted hit times then
compared with then known (true) embedded time.
Figures 
\ref{fig:dt-embedding} and \ref{fig:comp-hittimes.eps}
indicate that hit-recognition uncertainties
contribute approximately 5-10 ns to the overall timing resolution, per
channel.

\subsubsection{Signal Shape Uncertainties}
It is obviously essential that, in an experiment such as RICE, we accurately
be able to differentiate the hit corresponding to the signal induced in
an antenna by a neutrino-generated in-ice shower from
background. For a first approximation
to what these signals might look like, we used 'thermal noise hits' drawn
from the data itself.
Transient `hits'  were
extracted from a large number of ``unbiased'' trigger data events.
\message{(Aside: this argument has a long history. 
My original thought was that,
if you place an antenna in a 300 K infinite reservoir, but not in
contact with the walls of the reservoir, that the antenna would
eventually come into thermal equilibrium with the reservoir, so
that the temperature would be measured to be 300 K. If it 
were otherwise, then one could place two antennas into such a
reservoir, and, if one connected a cable, the one with the
higher bandwidth would be hotter. However, in retrospect, I 
now believe that that is, indeed, the case (that the higher-bandwidth
antenna is hotter). If one thinks about placing a 
perfect mirror inside
the same cavity, then, indeed, the mirror will not heat up. Or
even, on a more mundane level, considering the difference
between a black surface and a white surface on a hot day...).
More thoughts - if we conceptualize all the currents in the antenna
as due to the response of electromagnetic radiation in the medium
itself, then the transfer function should certainly be embedded in the
calculation of the noise at the antenna output, since the
transfer function is what is telling us the current induced in the
antenna given a unit input electric field. Note that ``thermal
contact'' is a mis-nomer - heat is transferred either through
conduction, convection or radiatively. For an isolated
antenna in-ice, there
is no conduction and no convection, so the transfer must be radiative.
However, if there is ``thermal contact'', then there is `conductive'
transfer of
thermal energy through collisions of electrons across the boundary.}
In the offline
analysis program, a 34 ns ``snippet'' of data is saved 
(10 ns before, and 24 ns after the time-of-maximum voltage, which
must exceed 6-sigma)
for short-duration
transient waveforms having small values of time-over-threshold.

\begin{figure}[htpb]
\centerline{\includegraphics[width=9cm,angle=0]{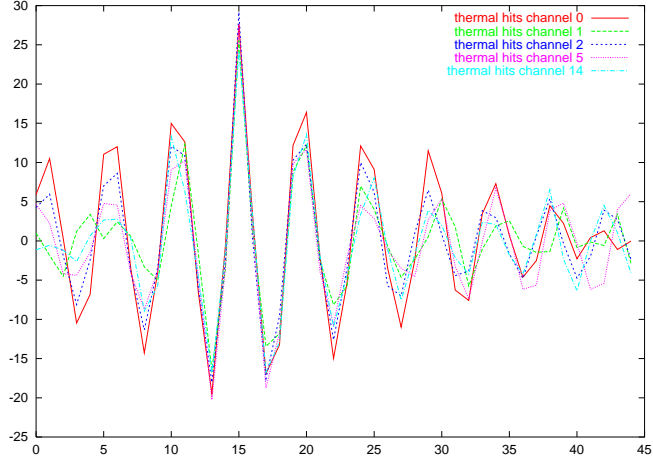}}
\caption{``Short-duration'' waveforms, taken from 2002 `noise' data, selecting
cases with fast impulsive responses (designated as
``thermal hits''), for the indicated channels.}
\label{fig:sigwvfm0}
\end{figure}

\begin{figure}[htpb]
\centerline{\includegraphics[width=7cm,angle=0]{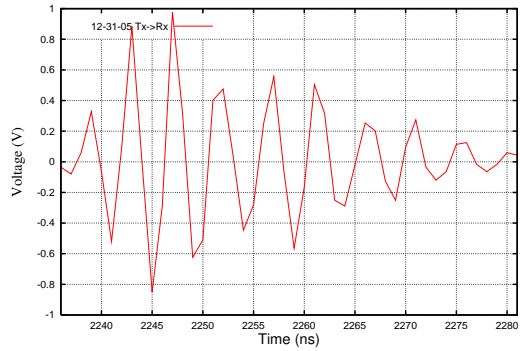}}
\vspace{0.5cm}
\caption{Transmitter data, channel 1, for comparison with 'thermal' hits.}
\label{fig:sigwvfm6}
\end{figure}

Figure \ref{fig:sigwvfm0} 
qualitatively indicates the reproducibility of such short-duration transient
responses from channel-to-channel (2002 data),
with transmitter signal shown in
\ref{fig:sigwvfm6} for comparison. These waveforms, channel-by-channel,
are used as one of our matched-filter, 
or `transfer function' templates in future analysis.

\begin{figure}[htpb]
\centerline{\includegraphics[width=9cm]{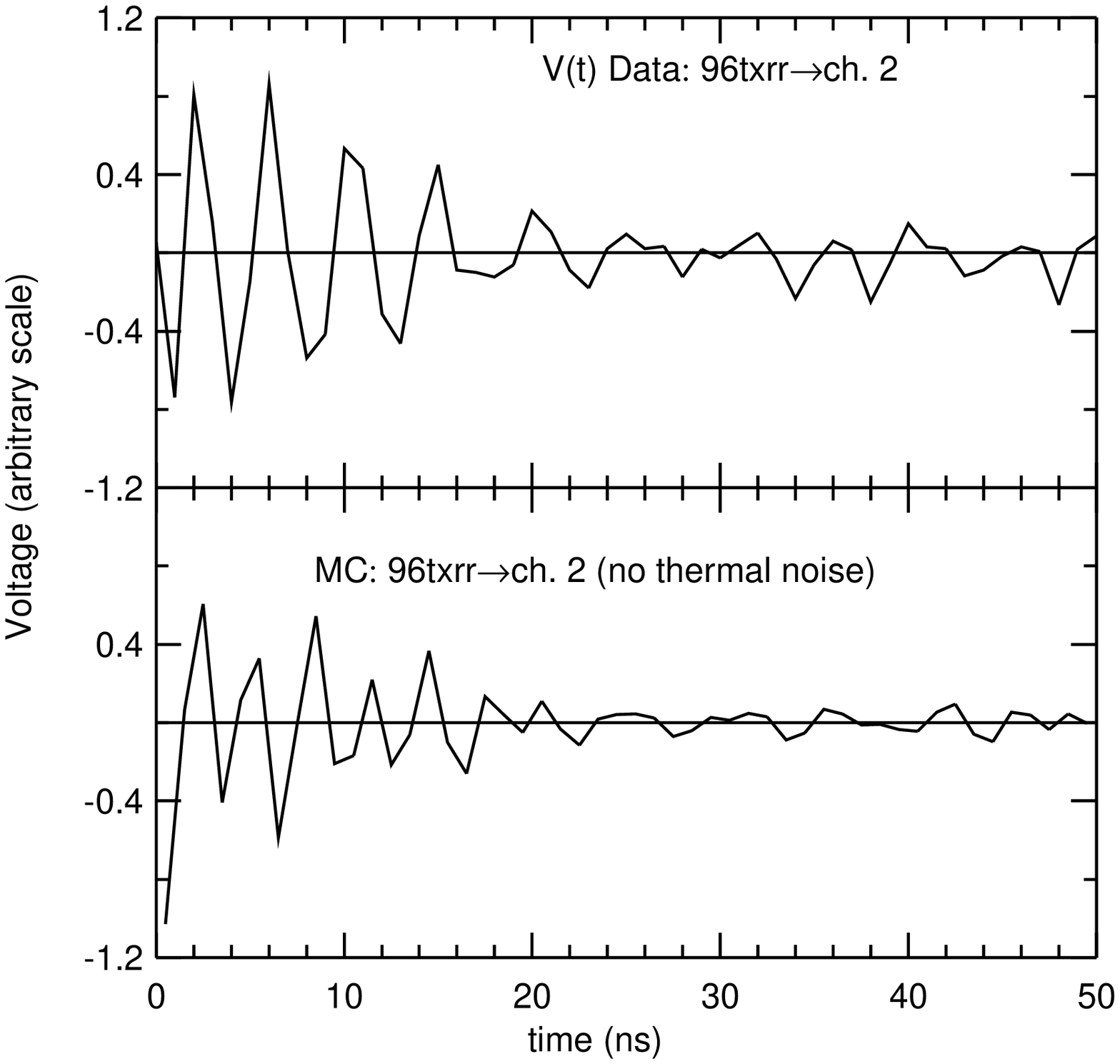}}
\caption{Receiver time-domain signal
resulting from a transmitter broadcasting in the ice, compared with 
predicted shape.}
\label{fig:MC-data-waveform-compare}
\end{figure}

An additional test of our derived filter is afforded by
comparing the signal we observe in transmitter data
with simulation of RICE transmitter$\to$receiver signals. Figure 
\ref{fig:MC-data-waveform-compare} shows the comparison between
true transmitter data (top, 96Tx1 data) with the expected signal
from 96Tx1 in a RICE receiver, using the
RICE `radiomc' Monte Carlo simulation code. 
Although $\omega_{MC}\sim1.2\omega_{data}$, the
general agreement between the two curves is reasonably good.

Qualitatively,
the decay time is related to the inverse of the bandwidth -- the
observed in-ice decay time of $\sim$5-8 ns corresponds to an expected
in-ice bandwidth of $\sim$150-200 MHz.
The number of cycles in the signal 
can be, as a general rule,
related to the fractional bandwidth: $\Delta f/f_0$, with
$f_0$ the center frequency of the antenna response and $\Delta f$ the 
bandwidth response (in MHz). For the RICE antennas, based on our
effective height curve (Figure \ref{fig:rice_dipole_H.eps}), we note that:
a) the bandwidth is $\sim$300 MHz (which gets down-shifted to $\sim$200 
MHz in ice), and b) the fractional bandwidth is about 300 MHz/800 MHz, or
0.3, corresponding to an expectation of 3 cycles in the signal. 
Cable attenuation reduces the bandwidth
somewhat, resulting in an expectation of 
$\sim$3--4 cycles in the signal. 

\begin{figure}[htpb]
\centerline{\includegraphics[width=9cm]{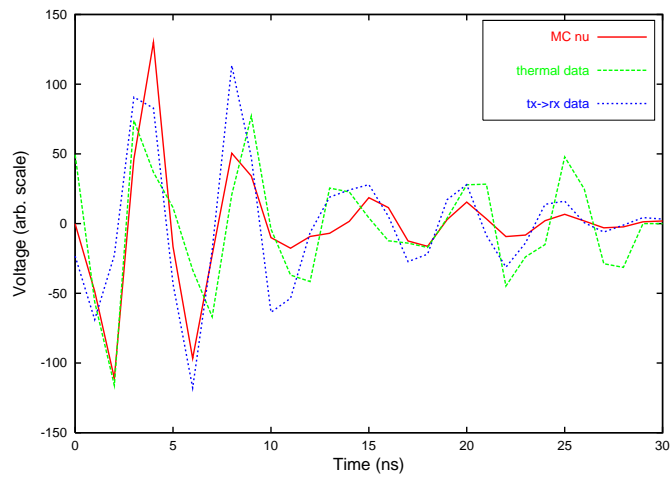}}\caption{Comparison of filters used in neutrino analysis (Channel 1, 2002 data). Red is
Monte Carlo neutrino simulations; green is ``thermal hit'' matched filter
derived from data and blue corresponds to matched filters derived from
transmitter data.}
\label{fig:filter-compare.eps} 
\end{figure}

A knowledge of the expected signal shape allows construction of a 
'matched filter', which provides, in principle, a more precise 
estimate of the time of a true signal than a simple threshold 
crossing. Calculation of a simple dot-product between the matched
filter vector and the waveform data vector gives a measure of the
correlation between putative and true signal shape.
A comparison of filters derived from RICE transmitter data, RICE Monte 
Carlo simulations of neutrino interactions, as well as short-duration
transients (``thermal'' filters) is shown in 
Figure \ref{fig:filter-compare.eps}. 

Numerically,
the filters are implemented by calculating a correlation
parameter at each point in the waveform, defined as:
${\bf C}=(\Sigma_iF_i\cdot(V_i-<V>))/\Sigma_i |F_i|^2)/\Sigma(F_i-(V_i-<V>))^2$, 
with
$F_i$ the $i^{th}$ element of the 34-element
filter vector, $<V>$ the average
voltage in the waveform ($\equiv$0 in the absence of baseline
drifts), and $V_i$ the
data voltage. The sum is carried out for the
34-elements of our filter arrays. For large
amplitude signals which match the shape of the filter, 
the numerator (generically designated as the
`Correlation Parameter' Corrparm, or the ``filter dot product'')
is large, and the denominator (generically designated as the
`$\chi^2$') small. The
particular value of the cut on this statistic
is set by determining the ${\bf C}$ distribution for
a large sample of unbiased events, and requiring that the probability of
a false positive be less than $10^{-3}$; i.e., that in our sample of
unbiased events, less than 0.1\% of the waveforms have ${\bf C}>C_{min}$.
For 20 waveforms per event, this corresponds to a fake rate of
order $<2$\%.

\begin{figure}[htpb]
\centerline{\includegraphics[width=9cm]{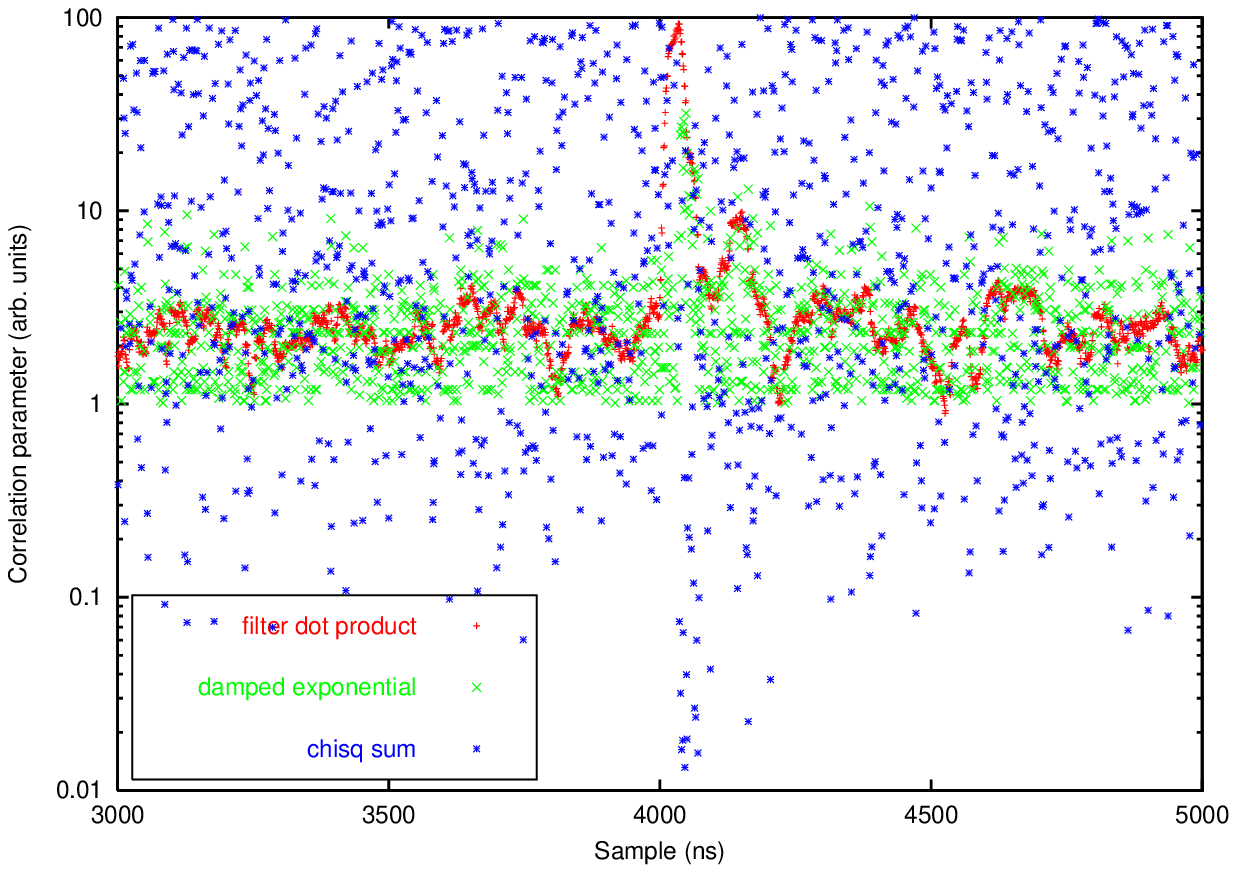}}
\caption{Typical correlation parameters, comparing the value of
Corrparm (red points) with the match of a waveform to a general damped
exponential form (green points). We have zoomed into the signal
region to show greater detail.
Also shown is the $\chi^2$ value (blue)
for the filter. In this event, there was a Monte Carlo signal embedded
in an unbiased event at time t=4000 ns in channel 0. The filter used here
was the thermal event filter.}
\label{fig:corrparmjpr.gnudat-6a.eps} 
\end{figure}

\begin{figure}[htpb]\centerline{\includegraphics[width=9cm]{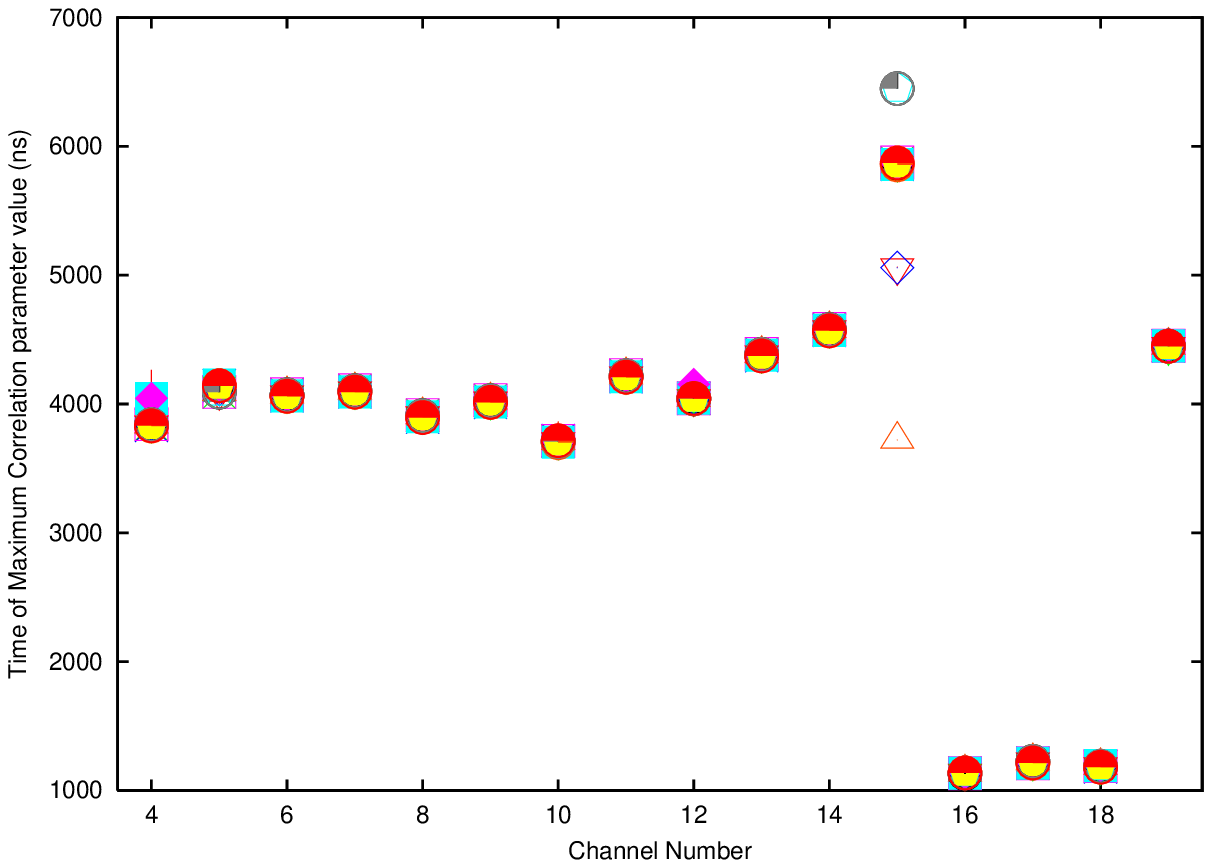}}\caption{Times of Maximum of correlation parameter for 8 different filters. The filters used are the `thermal' filters, the monte carlo `radiomc' filters, as well as filters derived from data using the five different transmitters broadcasting to the RICE array. We note general consistency in the returned times.}\label{fig:filter-0.eps} \end{figure}

\begin{figure}[htpb]\centerline{\includegraphics[width=9cm]{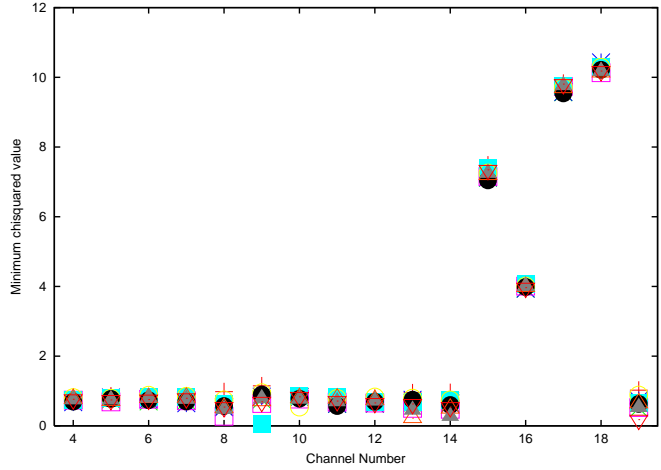}}\caption{Minimum $\chi^2$ at times corresponding to maximum of correlation parameter for 8 different filters.}\label{fig:filter-1.eps}\end{figure}

\begin{figure}[htpb]\centerline{\includegraphics[width=9cm]{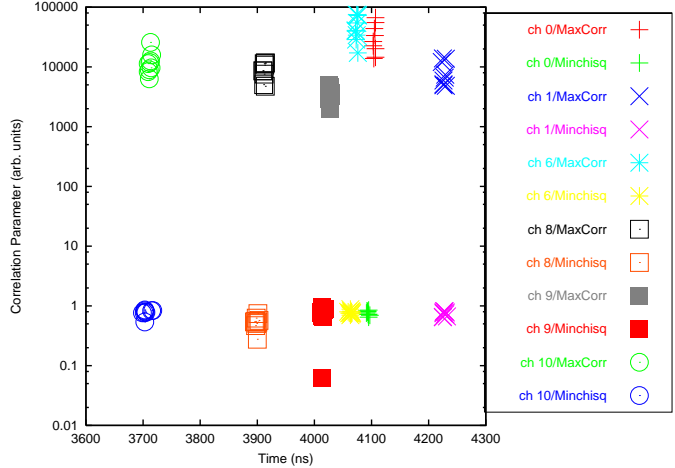}}\caption{Max Corrparm and Min $\chi^2$, and corresponding times in waveform.}\label{fig:filter0-chs0-10.eps}\end{figure}

Figure \ref{fig:corrparmjpr.gnudat-6a.eps} illustrates the
correlation parameter for a typical waveform. 
Figure \ref{fig:filter-0.eps} shows the agreement in the extracted hit time
for a set of general trigger events.
Figure \ref{fig:filter-1.eps} shows the agreement in the calculated
minimum $\chi^2$ and
Figure \ref{fig:filter0-chs0-10.eps} shows both the 
minimum in $\chi^2$ and the maximum in the overall correlation parameter
for 10 channels.

In general, the filters give comparable performance. We note that, in our
current data analysis, we have used only a threshold-crossing time to 
define the antenna hit time.

\section{Offline Data Filtering}
As data are taken, they are collected on the disk of the local PC. Once
a day, a script ftp's the most recent data to a PC (linux),
where the first stage of offline analysis occurs, as follows:  
\begin{enumerate}
\item For each event,
each waveform is scanned and we determine (among other things):
a) the maximum voltage in the waveform, as well as the
time of the maximum voltage, 
b) the time and voltage of the
first, and last 6$\sigma$ excursions in the waveform,
c) the rms voltage in each waveform.
To minimize the number of scans, the rms calculation
is based on the first 500 ns of data in the waveform, which should
be far from the trigger-time of $\sim$4$\mu$sec.
\item Once the rms-voltage in each waveform is determined, we can
further determine the number of samples
for which the waveform exceeds $6\sigma$ (aka ``Time-Over-Threshold'',
or, simply, ``TOT''), in a 2nd scan of the
waveform. Based on our knowledge of the transfer function (discussed
earlier),
this quantity is expected to typically
be less than 20 samples for real neutrino events.
\end{enumerate}
At this first stage of event selection,
we now require that: a) there be at least 4 channels
having $6\sigma$ excursions in the event, and that b) there be no channels
having TOT$>$25 ns (the number of total channels with TOT$>$25 ns is
designated ``NXTOT'').
Such cuts should anticipate any possible RICE analyses
(GRB's, wimps, monopoles, air showers, etc.) 
and be sufficiently general so as to have
$\sim$100\% efficiency for any such future analyses.\footnote{It is 
possible that air showers may produce 
events with many channels having 
large Time-Over-Threshold values. This has not, as yet, been fully modeled.} 
We emphasize, however, that all triggers are written to
tape and are available for future analysis.

To estimate what inefficiency the Time-Over-Threshold 
cut would incur for true neutrino
events, we consider what fraction of unbiased events would fail
the TOT cut, since such unbiased events are expected to constitute
the ``background'' on top of which a true neutrino event would be
superposed.
For August, 2000 data,
the fraction of unbiased events having NXTOT=0 is ($96.0\pm0.3$)\%.

\message{The 25 ns Time-over-Threshold cut has certainly not been tuned, although it is almost certainly a very loose cut, based on our estimates of the
signal shape thus far. Our observations of the receiver signal
shape in transmitter data, which must, if anything, be considerably longer
than the signal shape expected for a neutrino, also indicate that 25 ns is
a conservative cut.}

\subsection{Vertex Reconstruction}
\subsubsection{Hit-Definition and Gain Stability}
\begin{figure}[htpb]
\centerline{\includegraphics[width=9cm]{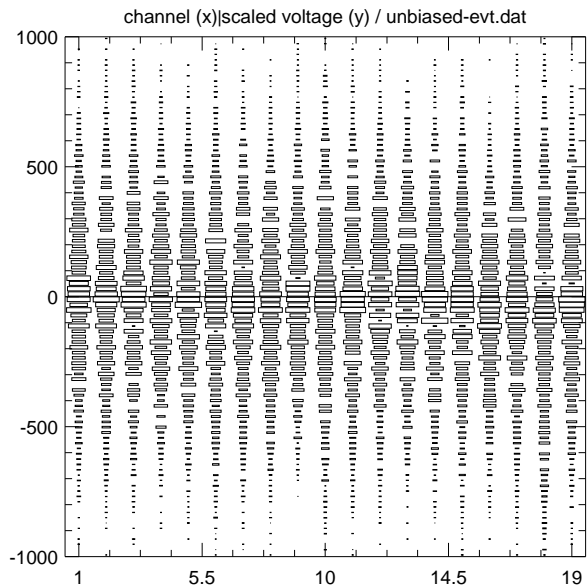}}
\caption{Distribution of raw voltages for the August, 2000 data, incrementing
by channel (along the x-axis). The absolute voltages have all been scaled
to a unit normalization, for display purposes.}
\label{fig:unbiased-rms-channel-dist}
\end{figure}

\begin{figure}[htpb]
\centerline{\includegraphics[width=9cm]{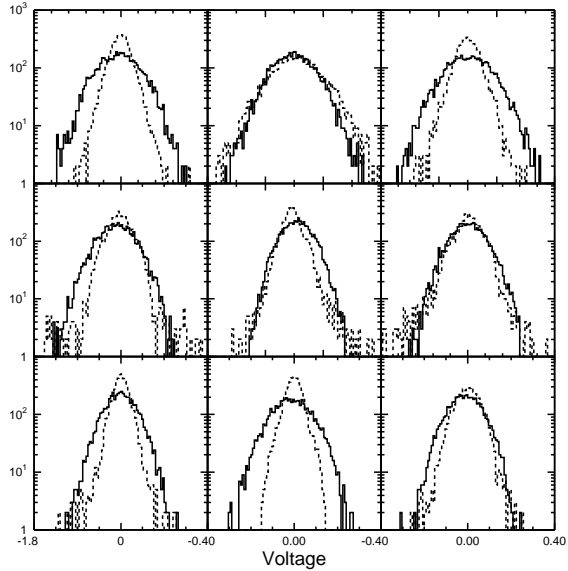}}
\caption{Distribution of rms voltages in nine channels 
for ``unbiased'' triggers,
taken from Jan., 2003 (solid), and compared with data taken from
June, 2003 (dashed). We note the presence of non-Gaussian ``tails''
in the latter data sample. The difference in widths, from channel-to-channel,
are primarily the result of different cable losses in the various channels.}
\label{fig:unbiased-voltages}
\end{figure}

As stated above, we require ``hits'' to exceed the measured rms
voltage in each channel at some minimum $N_\sigma$ level,
with $N\sim 5.5-6$.
The $N_\sigma$ requirement was selected on the following basis:
assuming Gaussian thermal noise, the probability of having an
$N\sigma$ excursion on any given sample is: $p_N=A\int_n^\infty exp(-n^2/2))$,
with the normalization constant $A=1/\sqrt{2\pi}$.
For $5.5\sigma$, this corresponds to $p_{5.5}=3.8\times 10^{-8}$. The 
probability of having at least one such excursion in one of 
16 waveforms, each waveform comprising 8192 samples is then
(approximately)
$p_{5.5}\times 8192\times 16\sim$0.5\%. For a 5-sigma
criterion, this probability is $\sim$7.5\%. 
In practice, there are non-Gaussian tails in the noise distribution.
Figure \ref{fig:unbiased-rms-channel-dist} shows
the two-dimensional
raw voltage distribution (from which the rms is calculated)
for eighteen channels (2002 data). 
Figure 
\ref{fig:unbiased-voltages} 
shows the calculated rms
voltages for nine RICE data channels taken from Jan., 2003 vs.
June, 2003. Although the former dataset is largely Gaussian, the latter
data set shows non-Gaussian tails. 

For August, 2000 data,
the fraction of the unbiased events having at least one
$4\sigma$ excursion is 100\%; for $5/6/7\sigma$, the corresponding
fractions are 65\%, 8.5\%, and 5.6\%, respectively. Taken at
face value, these numbers indicate that, in 8.5\% of all neutrino
events that might have occurred in august, 2000, there would have
been at least one spurious $6\sigma$ hit in the waveform.
This requires an event reconstruction algorithm which can 
robustly identify possibly spurious hits and reject them, based
on waveform information (discussed later in this document).

\subsection{Vertexing Studies and In-Ice Transmitter 
Reconstruction\label{sssect:timing}}
Beyond the initial filtering discussed in the document, vertex distributions
are perhaps the most direct discriminator of surface-generated ($z\sim 0$) vs.
non-surface (and therefore, candidates for more interesting processes) events.
Consistency between various source reconstruction algorithms is,
of course, desirable in giving confidence that the ``true'' source
has been located. 
\subsubsection{Application of Matched Filters to Vertexing}
\begin{figure}[htpb]
\vspace{1cm}
\centerline{\includegraphics[width=9cm]{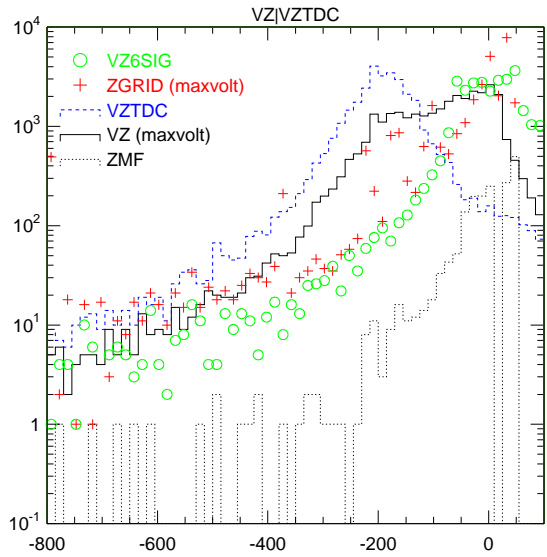}}
\caption{Reconstructed z-vertex for different source reconstruction
codes, and for different time-of-hit algorithms, for
80\% of the full 2000 dataset.}
\label{fig:zcomp}
\end{figure}

Figure \ref{fig:zcomp} shows the z-distribution
reconstructed using various codes: top) depth of source vertex obtained
using grid algorithm (zgrid), using time of
``maximum voltage'' in a waveform 
as the hit time, for all channels; middle) 4-hit source reconstruction
algorithm, using same maxVolt criterion as in top plot; bottom) grid-based
algorithm, using as the hit time the time
when the waveform 
best matches a ``ringing exponential''. We note general agreement
between the three depth distributions.
The smaller statistics for the last matched filter distribution
is a consequence of the smaller fraction of times the filter
finds 4 hits.

\begin{figure}[htpb]
\centerline{\includegraphics[width=9cm]{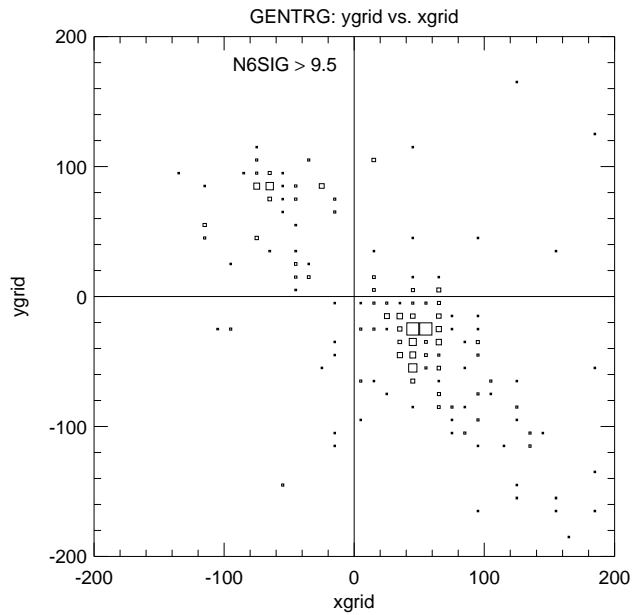}}
\caption{Distribution of xgrid vs. ygrid, August, 2000 general events.
In making this distribution, we have required a high hit multiplicity
(NHIT$\ge$10).}
\label{fig:ZxVXy}
\end{figure}

Figure \ref{fig:ZxVXy} shows the 
x-y projection of found vertices (grid-vertex-finder) for general
triggers in the August, 2000 data, for which there are 10 or more receivers
hit in the event. The main ``cluster'' is consistent with MAPO. The 
`reflection' in the upper left results from the case where one of the
shallow receivers in the northwest corner of the grid registers
a spurious, early hit, which pulls the vertex in that direction. This
cluster can be suppressed with increasingly stringent hit definitions.

\subsubsection{Time Residual Studies}
An additional discriminant of ``well-reconstructed'' vs. 
``poorly-reconstructed'' sources is provided by calculating the average
time residual per hit. This is done by: 1) finding the most-likely vertex
for the event, 2) calculating the expected recorded hit time, for each
channel, assuming that vertex. This is a combination of ice-propagation
time plus cable delays plus electronics propagation delays at the surface.
3) calculating the difference between the expected time minus the 
actual, measured time, for that reconstructed vertex (this is, in fact, the
same parameter which, when minimized, defines the reconstructed vertex, using
the grid-based algorithm). That time difference is defined as the ``time
residual'' (as defined for fits to the helical
trajectory expected for a charged track traversing a multi-layer drift 
chamber). 

On a more fundamental level, it is necessary that the 
overall channel-to-channel
timing calibration be performed satisfactorily, and that the
timing delays channel-to-channel, within the DAQ, are known, if we
are to confidently perform source vertexing.
The initial timing calibration is done with simple TDR 
(Time Domain Reflectometry) measurements at the Pole, for which
a signal is sent to each antenna, and the
time delay between the SEND signal and the reception of the RETURN
signal divided by two gives the one-way transit time delay. 
If all the system
delays are known, then (as mentioned previously), time residuals should be
zero for perfectly fit 
vertices.\footnote{Unfortunately, TDR measurements do not probe possible
surveying errors of the locations of the buried receivers. Those must be determined using transmitter data and looking at channel-by-channel spatial residuals.}

\begin{figure}[htpb]
\centerline{\includegraphics[width=9cm]{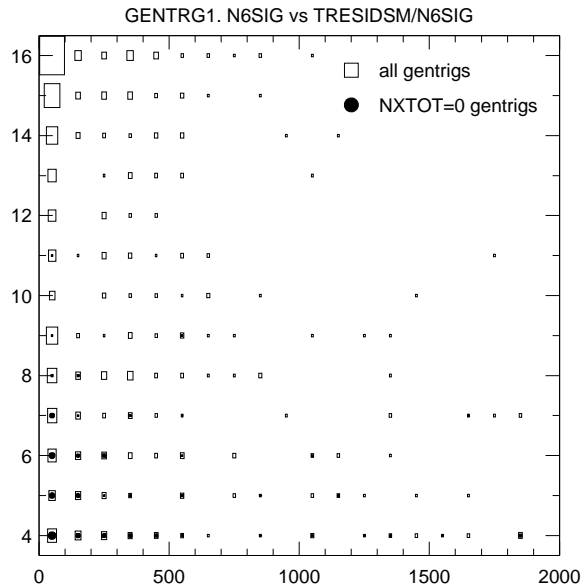}}
\caption{Time-residual per hit antenna (in units of ns, horizontal) vs.
the total number of 6-sigma excursions (vertical). The open squares show
the distribution for all general triggers; the filled in circles show the
distribution for general triggers having NXTOT=0.}
\label{fig:n6sigVtresidsm}
\end{figure}

Figure \ref{fig:n6sigVtresidsm} shows the scatter plot of 
the average time residual, per channel, vs. the number of 6-sigma excursions
in an event, comparing all general triggers (August, 2000 data) with the
subset of those triggers having NXTOT=0. We note two features of this
scatter plot: 1) as expected, the events which have NXTOT=0 favor lower
hit multiplicities - i.e., the more channels that are ON, the more chances
to have one channel fail the $TOT<$25 ns cut, 2) as N6SIG increases, the
time residual per hit does, in fact, decrease, indicating a 
better-reconstructed vertex.

\begin{figure}[htpb]
\centerline{\includegraphics[width=9cm]{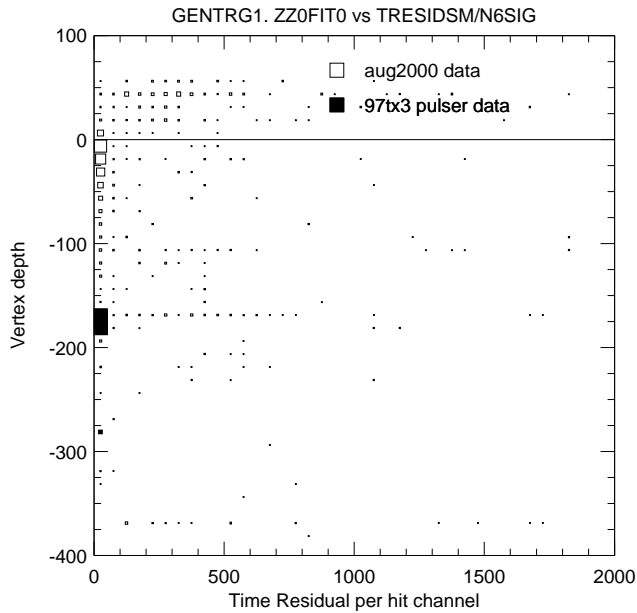}}
\caption{Average time residual (``$<\delta t>$'') per hit 
(horizontally, in units of ns) vs. location of reconstructed
vertex (vertical, units are meters) for August, 2000 data vs.
97Tx3 pulser data (from 2001).}
\label{fig:ZzVtresidsm}
\end{figure}

To corroborate that conclusion, Figure \ref{fig:ZzVtresidsm} shows
the distribution of time residual per hit channel (ns) vs. the
depth of the reconstructed vertex (vertical). Open squares are general
triggers, Aug. 2000 data. Filled-in squares correspond to 
97Tx3 pulser data, taken January, 2001. We note that: a) the
time-residual per channel for pulser data clusters at very small
values, b) there is very little scatter in the reconstructed depth for 
the 97Tx3 pulser data, c) the cluster of events corresponding to
$zgrid\to$0 occurs predominantly for small values of 
$<\delta t>$. Such a goodness-of-vertex cut
can be used to select 
well-fit vertices.
Following Figure \ref{fig:ZzVtresidsm},
Figure \ref{fig:aug2000-tresid} 
presents channel-by-channel time
residuals for August, 2000 general events (black), compared to 
time residuals for 97Tx3 pulser data (the same data that were used for
comparison purposes in Figure \ref{fig:ZzVtresidsm}). Typical 
channel-to-channel time residuals are of order $<$10 ns for the 97Tx3
pulser data (shown in red).
\begin{figure}[htpb]
\centerline{\includegraphics[width=9cm]{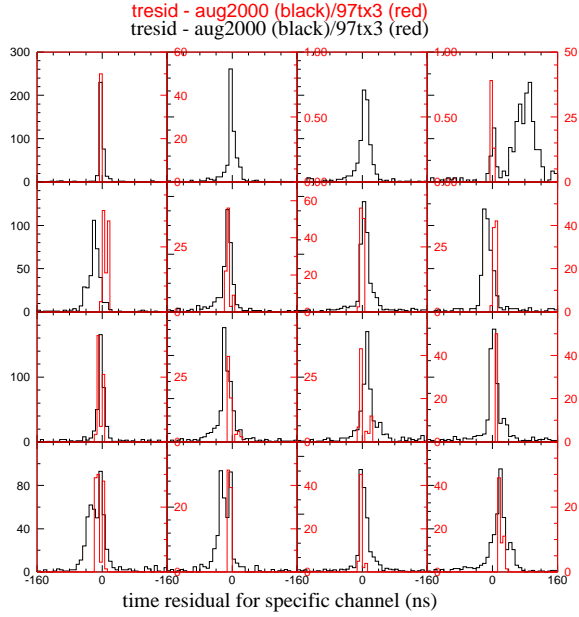}}
\caption{Time residual, channel-by-channel, for general triggers,
August, 2000 data (shown in black) compared to pulser data
(97Tx3). Note that the 97Tx3 transmitter is in the 
same hole as the channel 3 receiver; transmitter
receiver coupling effects `contaminate' the waveform 
and therefore bias the reconstructed
hit time. Horizontal axis is time residual, in units of
nanoseconds. Top row channels are, left to right: 
0, 1, 2, 3. Second row: 
4, 5, 6, 7 (etc.)}
\label{fig:aug2000-tresid}
\end{figure}
By contrast, for the August, 2000 general events (shown in black),
distributions are considerably wider.

\begin{figure}[htpb]
\centerline{\includegraphics[width=9cm]{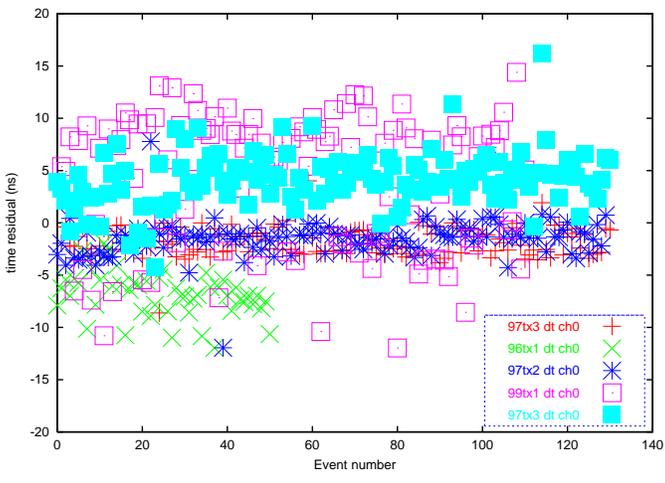}}
\caption{Channel 0 time residuals, for five different transmitters, showing
typical channel timing calibration and response.}
\label{fig:ch0-tresiduals.eps}
\end{figure}

\begin{figure}[htpb]
\centerline{\includegraphics[width=9cm]{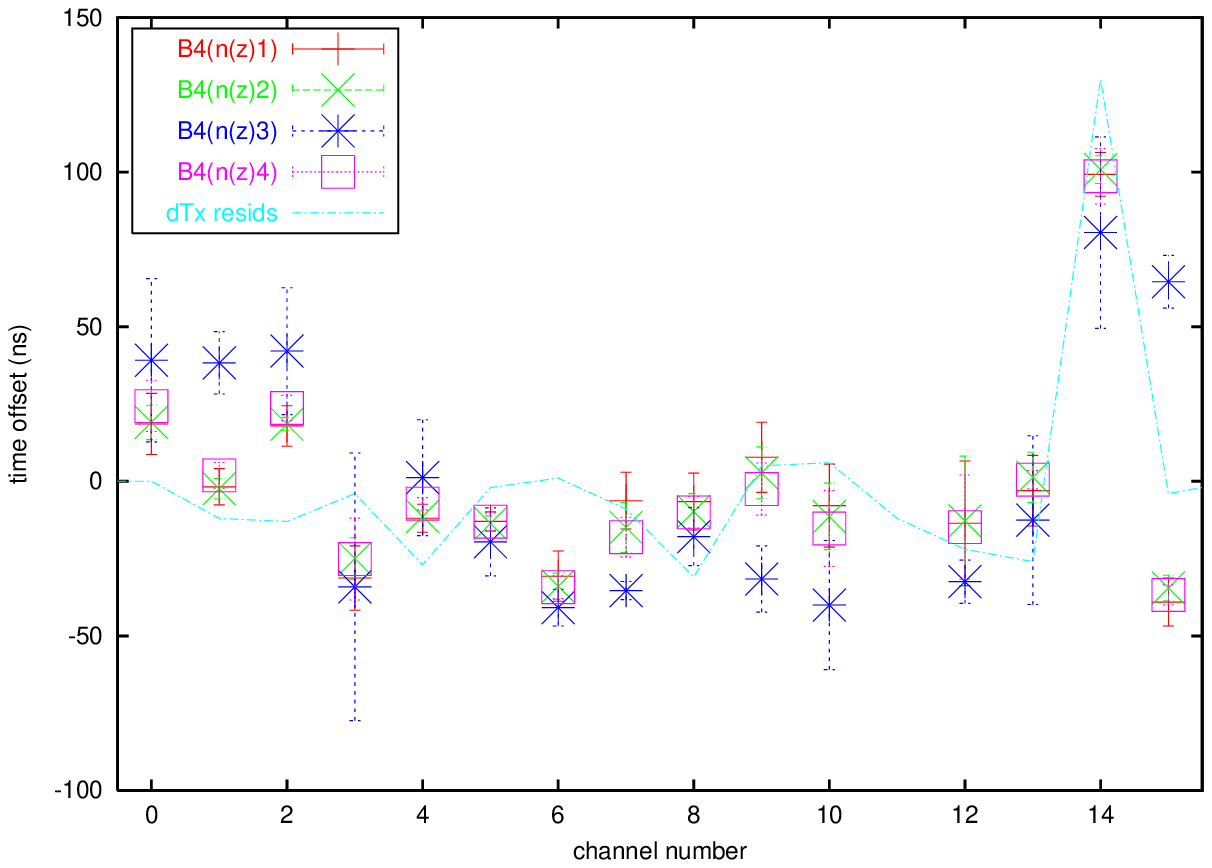}}
\caption{Timing offsets derived from B4 data vs. tx data.}
\label{fig:B4-n-z-t0cal.eps}
\end{figure}

For a perfectly fit vertex, the time residuals are all zero.
Additional 
transmitter data are shown in Figure \ref{fig:ch0-tresiduals.eps}, for
samples of transmitter data taken in January, 2003.
Figures such as these 
can be used to determine an average timing correction, channel-by-channel,
based
on information from all transmitters.
Based on the observed channel-per-channel average time residuals,
we can
derive a 2nd-order correction to the channel-by-channel time
delays derived from time-domain-reflectometer measurements.
These 2nd-order timing offsets are shown in Figure \ref{fig:B4-n-z-t0cal.eps}.
The points are derived from the same data set used to measure
n(z) in a previous publication (\citep{KravchenkoRFre}). 

\begin{figure}[htpb]
\centerline{\includegraphics[width=9cm, angle=-90]{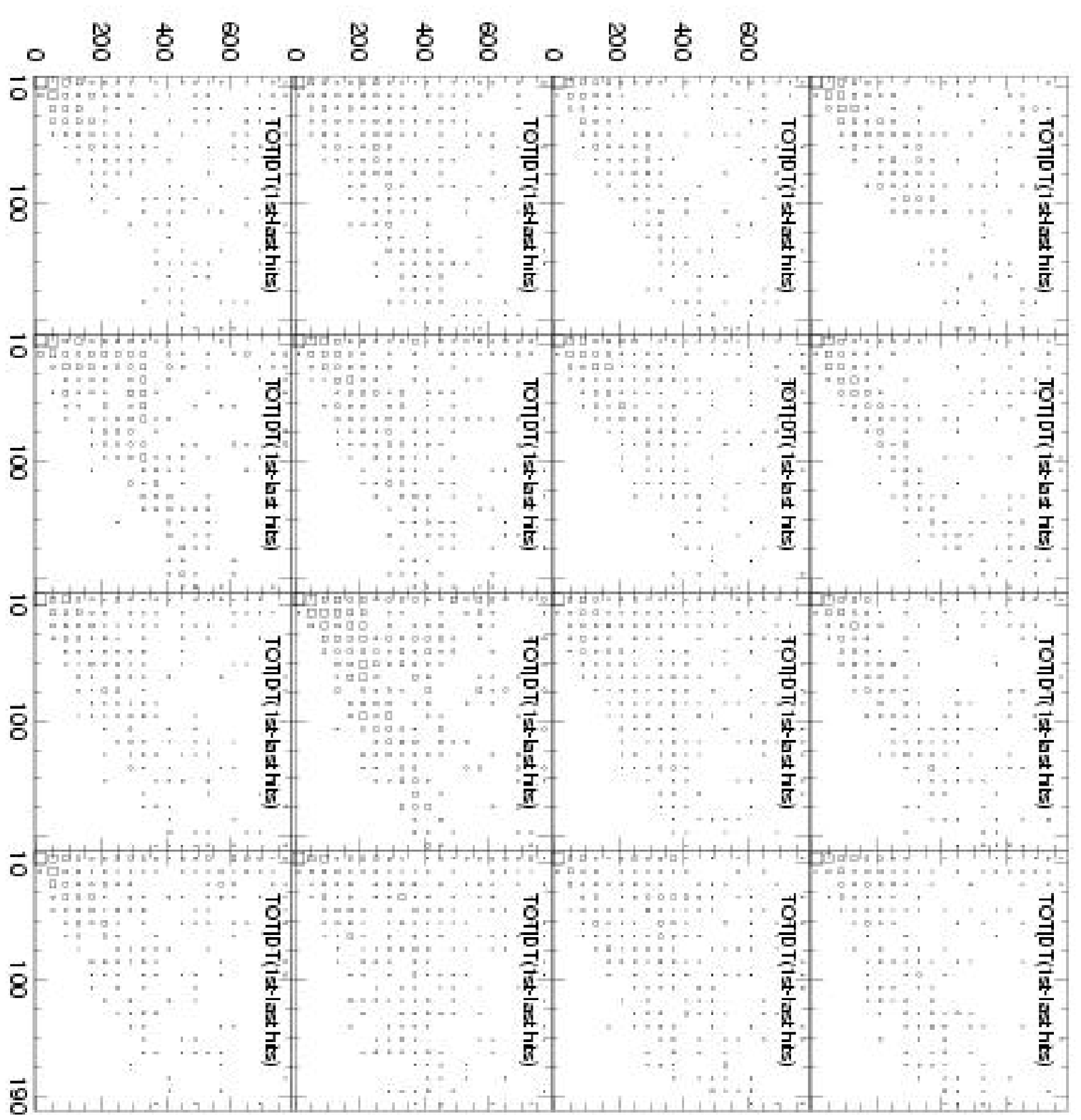}}
\caption{Chi-squared contour, using grid-based
vertexing algorithm, of a typical ``general'' trigger; data taken
from 2000 sample (Day 88, event 103). Shown is the $\chi^2$, in
units of $ns^2$, around the selected vertex. Displayed is only 
the vertex in x-y.}
\label{fig:day88_event103.ps}
\end{figure}

\begin{figure}[htpb]
\centerline{\includegraphics[width=9cm, angle=-90]{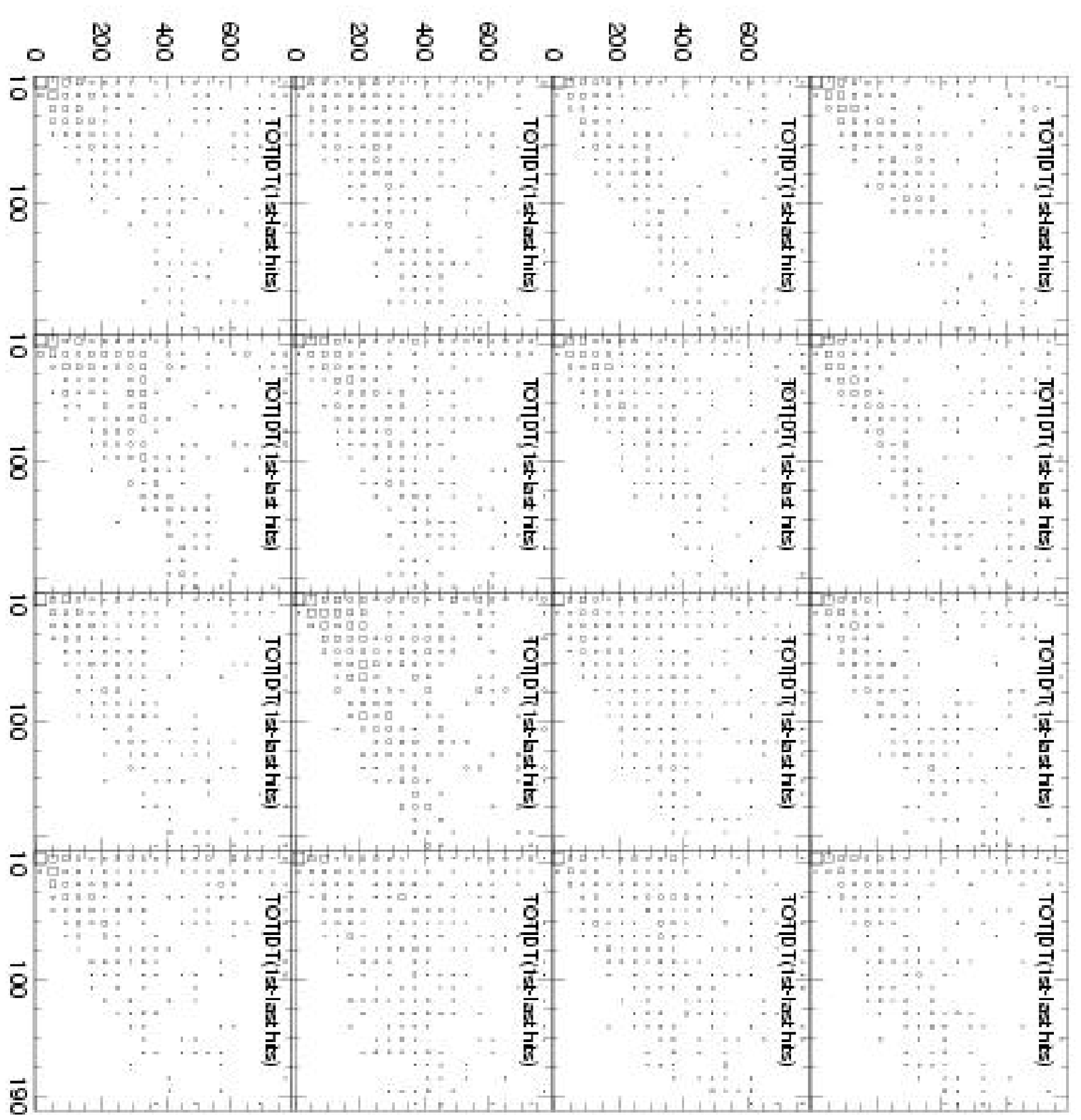}}
\caption{Chi-squared contour, using grid-based
vertexing algorithm, of a typical transmitter event; data taken
from 2001 sample (Day 2, event 1). Shown is the $\chi^2$, in
units of $ns^2$, around the selected vertex. Displayed is only 
the vertex in x-y.}
\label{fig:day2_event1.ps}
\end{figure}

The sum of the timing residuals, for any putative event vertex, gives
us a value of chi-squared, which, when minimized, defines the 'grid'-vertex.
The shallowness of the chi-squared
distribution, indicating the typical uncertainty
in the reconstructed vertex, is shown for two typical
events in
Figures \ref{fig:day88_event103.ps} and \ref{fig:day2_event1.ps}. The
minimum is observed to be fairly broad, corresponding to a typical
uncertainty (statistical) within the center of the
array of order tens of meters.

\subsubsection{Spatial Residuals}
\begin{figure}[htpb]
\centerline{\includegraphics[width=9cm]{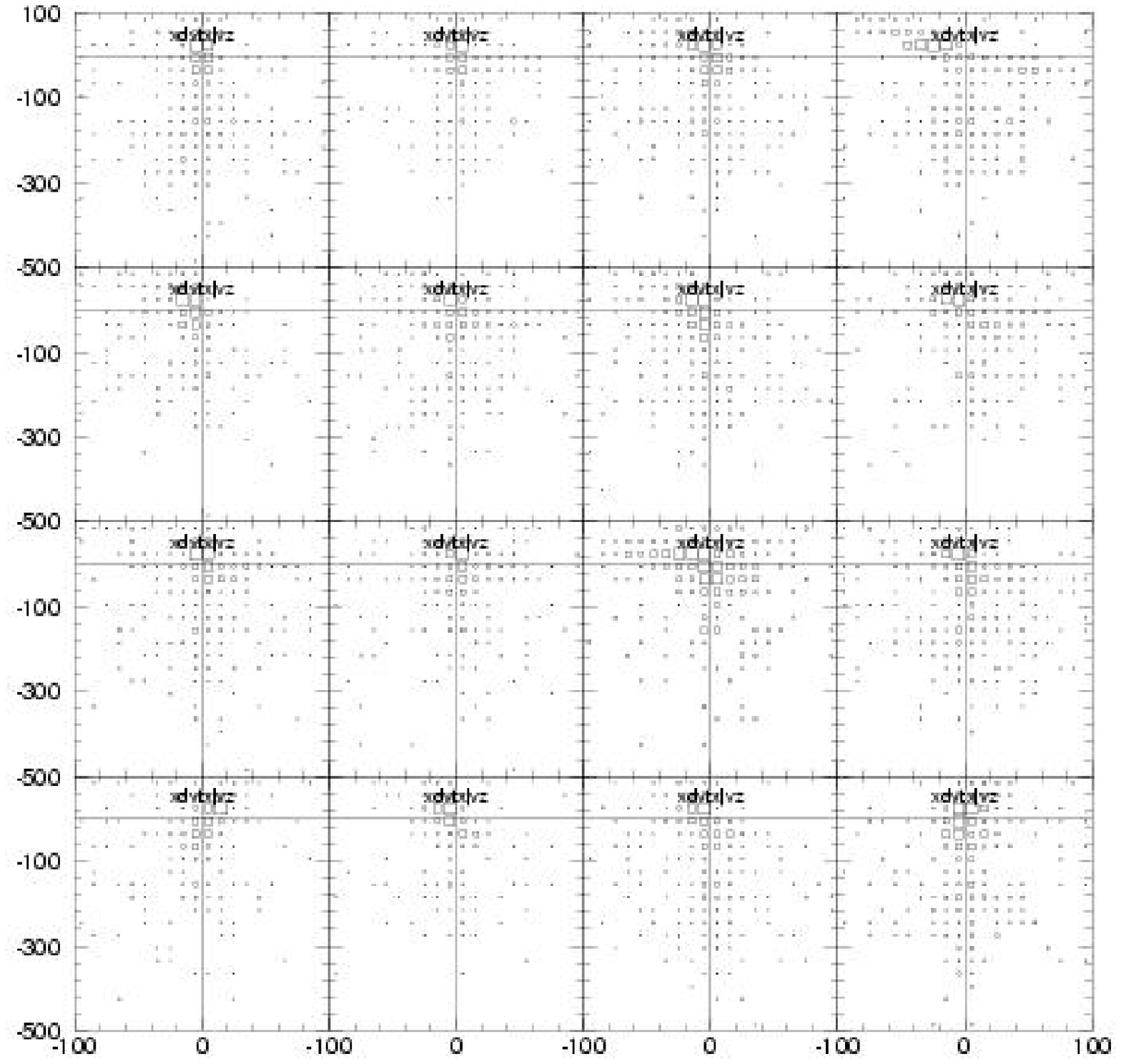}}
\caption{
Difference between
x-coordinate of the vertex calculated including the indicated
channel minus x-vertex calculated excluding the indicated channel (horizontal),
obtained from the 4-hit algorithm,
vs. reconstructed vertex depth (vertical), for August, 2000 data.
Units are meters.
}
\label{fig:vzVdxvtx}
\end{figure}

\begin{figure}[htpb]
\centerline{\includegraphics[width=9cm]{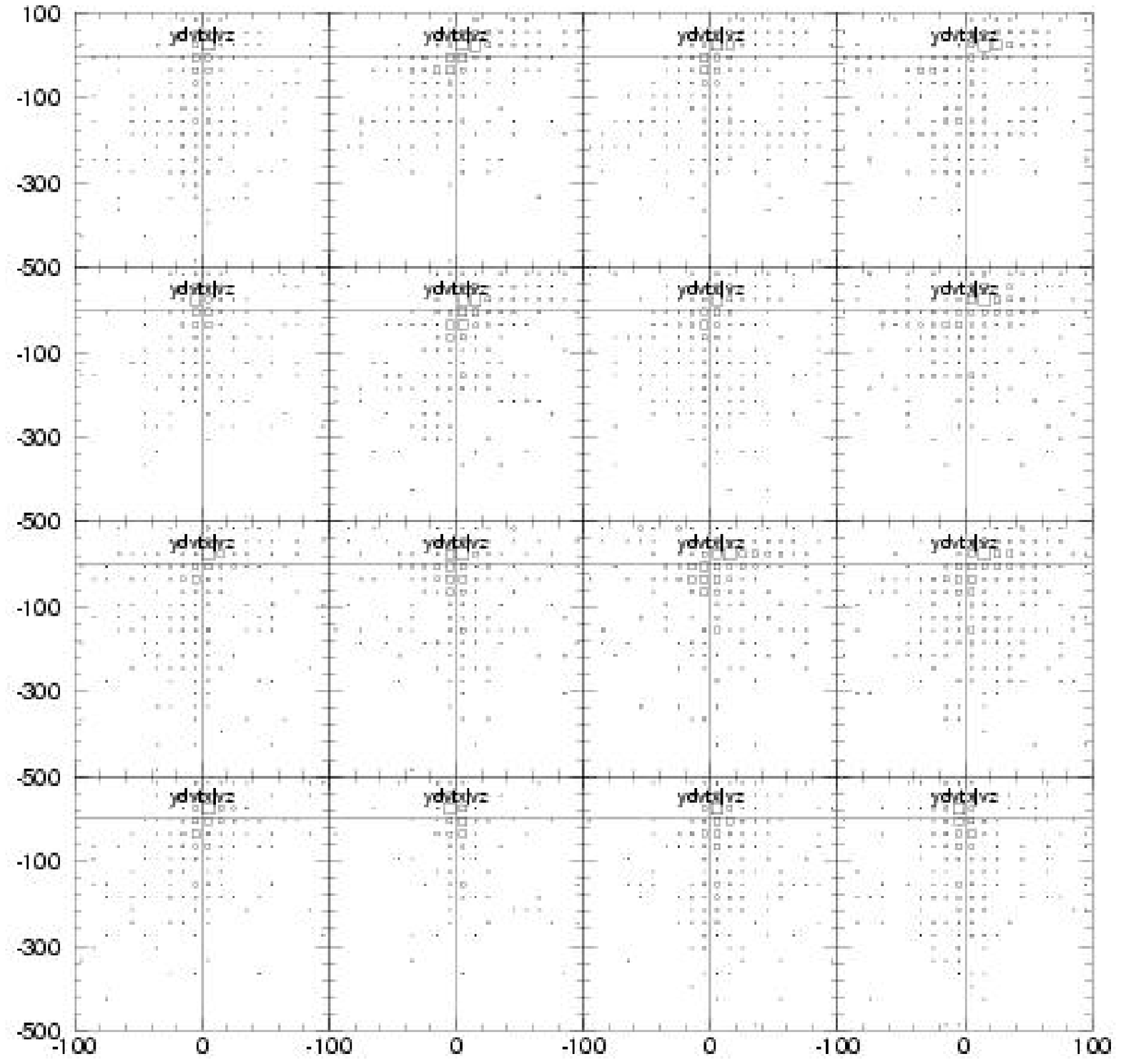}}
\caption{Difference between
y-coordinate of the vertex calculated including the indicated
channel minus y-vertex calculated excluding the indicated channel (horizontal),
obtained from the 4-hit algorithm,
vs. reconstructed vertex depth (vertical), for August, 2000 data.}
\label{fig:vzVdyvtx}
\end{figure}

\begin{figure}[htpb]
\centerline{\includegraphics[width=9cm]{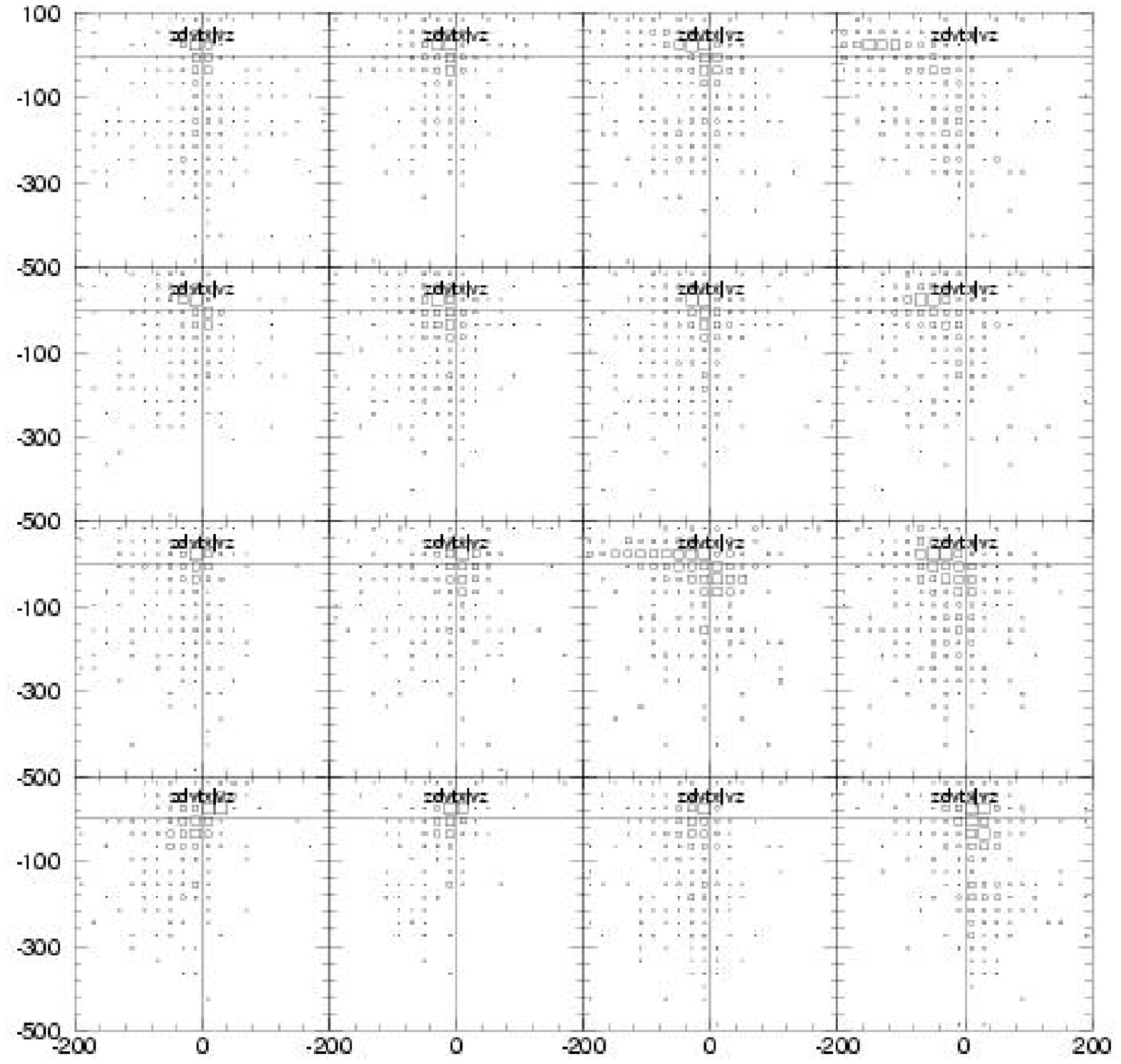}}
\caption{
Difference between
z-coordinate of the vertex calculated including the indicated
channel minus z-vertex calculated excluding the indicated channel (horizontal),
obtained from the 4-hit algorithm,
vs. reconstructed vertex depth (vertical), for August, 2000 data.
}
\label{fig:vzVdzvtx}
\end{figure}

The spatial residuals are correlated with, but also provide somewhat
independent information relative to the time residuals.
To investigate the
possibility of incorrect spatial pulls of any one channel on the
event reconstruction, Figures \ref{fig:vzVdxvtx},
\ref{fig:vzVdyvtx}, and \ref{fig:vzVdzvtx} are derived from the
4-hit `analytic' fitting algorithm, as follows: for each event, having 
hit multiplicity $N$, for which $N>$4, one makes 4-hit subsets
of the total $N$ hits and vertexes using those 4-hit subsets.
At the end of calculating all the possible vertices, one constructs
an ``average'', or ``best''-vertex, which is then reported back
to the user. One can compare the vertex calculated from
those subsets including channel $i$ with the vertex calculated 
from those subsets excluding channel $i$. If the timing
calibration has been performed properly, and if the 
surveyed locations of all the receivers have also been
recorded properly, the two sets of calculated vertices
(the set based on 4-fold
combinations including channel $i$ vs. the set based on
4-fold combinations excluding channel $i$) should coincide with
each other. We show the difference between those two
vertices, in the x-coordinate (Figure \ref{fig:vzVdxvtx}), y-coordinate
(Figure \ref{fig:vzVdyvtx}), and z-coordinate (Figure \ref{fig:vzVdzvtx}),
respectively, scatter plotted against the reconstructed depth
of the vertex (the vertical coordinate in each of these plots)
for the August, 2000 data. 
Calibration constants giving corrections to the surveyed locations
of the receivers (in particular,
for channel 3 and channel 10) are based on distributions
such as these.

\begin{figure}[htpb]
\centerline{\includegraphics[width=9cm]{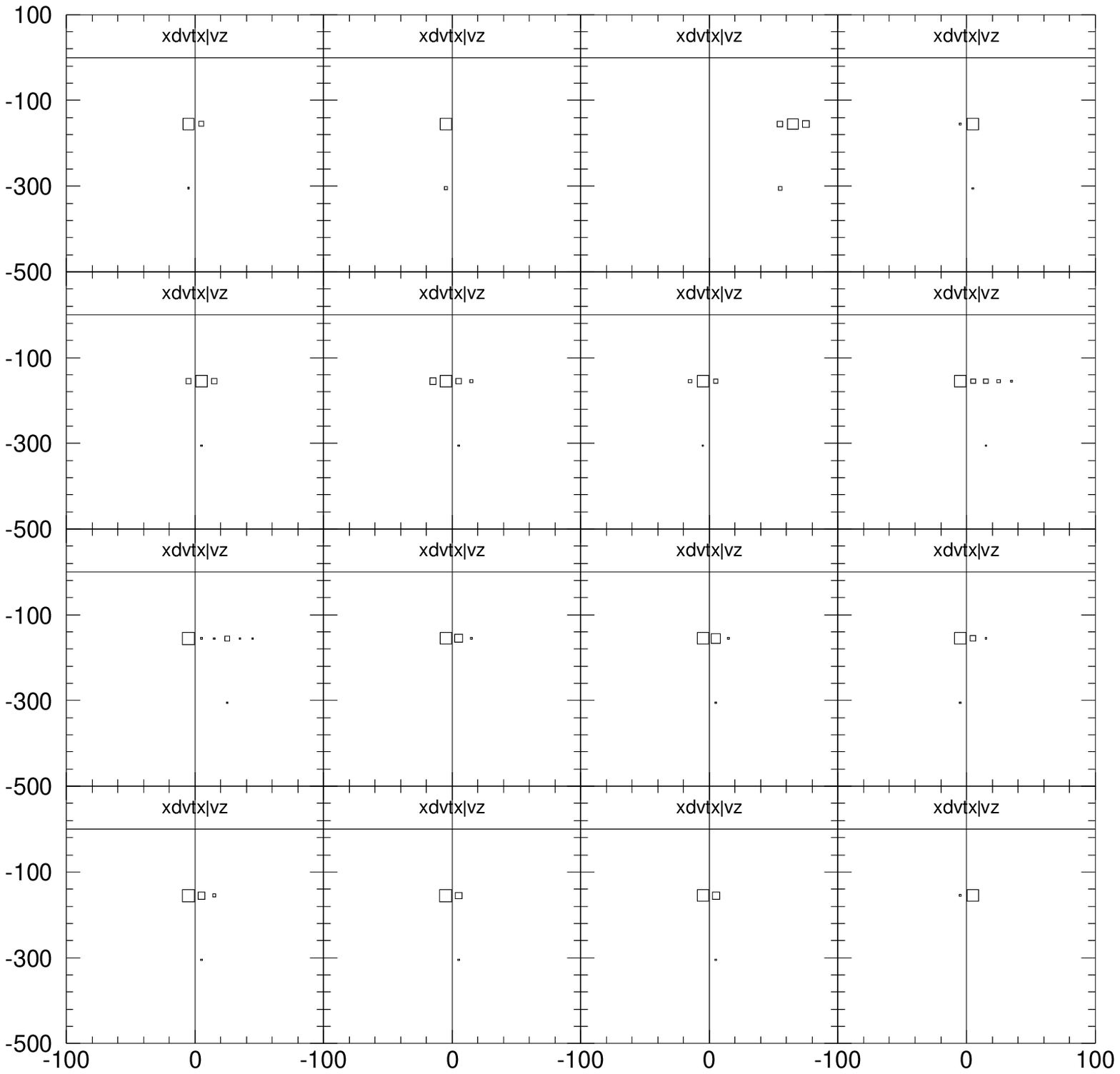}}
\caption{
x-spatial residual: Difference between
x-coordinate of the vertex calculated including the indicated
channel minus x-vertex calculated excluding the indicated channel (horizontal),
obtained from the 4-hit algorithm,
vs. reconstructed vertex depth (vertical), for January, 2001 pulser data.
The large apparent error in the
Channel 2 receiver (top row, 3rd column) results from the transmitter
being pulsed being in the same hole as that receiver.
}
\label{fig:pulser_dxvtx}
\end{figure}

\begin{figure}[htpb]
\centerline{\includegraphics[width=9cm]{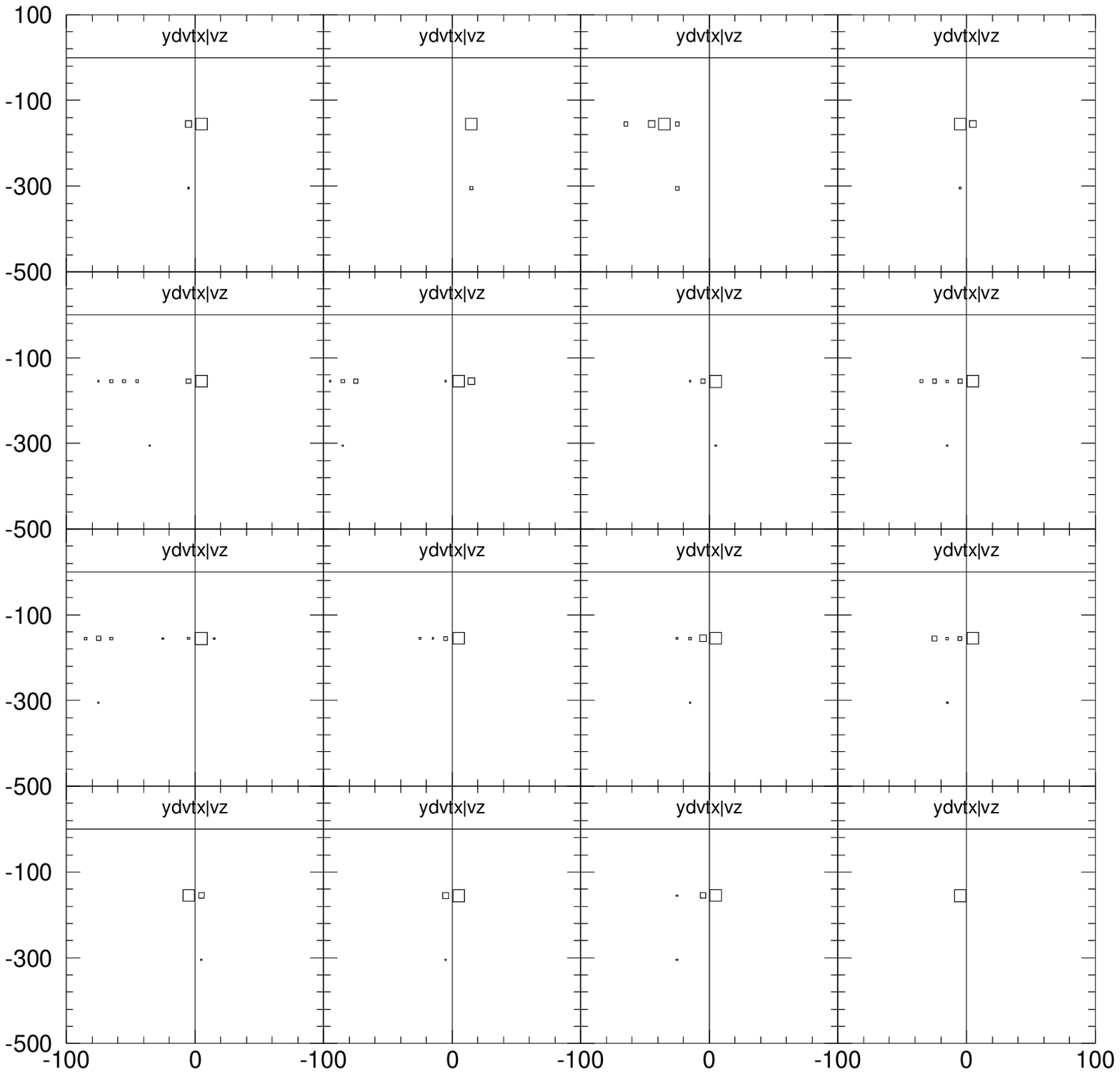}}
\caption{
y-spatial residual: Difference between
y-coordinate of the vertex calculated including the indicated
channel minus y-vertex calculated excluding the indicated channel (horizontal),
obtained from the 4-hit algorithm,
vs. reconstructed vertex depth (vertical), for January, 2001 pulser data.
The large apparent error in the
Channel 2 (top row, 3rd column) receiver results from the transmitter
being pulsed being in the same hole as that receiver.
}
\label{fig:pulser_dyvtx}
\end{figure}

\begin{figure}[htpb]
\centerline{\includegraphics[width=9cm]{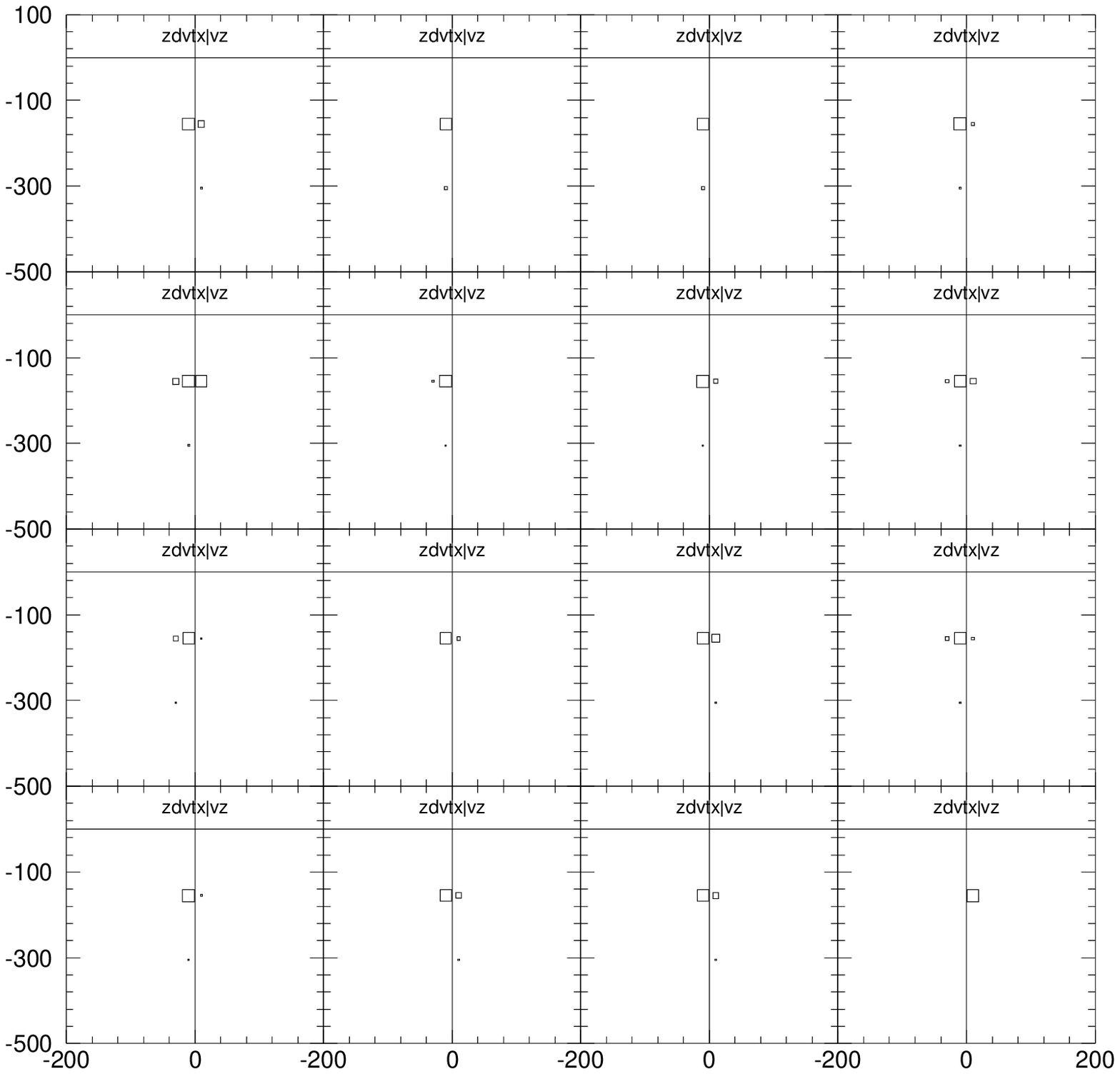}}
\caption{
z-spatial residual: Difference between
z-coordinate of the vertex calculated including the indicated
channel minus z-vertex calculated excluding the indicated channel (horizontal),
obtained from the 4-hit algorithm,
vs. reconstructed vertex depth (vertical), for January, 2001 pulser data.
}
\label{fig:pulser_dzvtx}
\end{figure}

Figures \ref{fig:pulser_dxvtx}, \ref{fig:pulser_dyvtx},
and \ref{fig:pulser_dzvtx}, which show the corresponding plots for 
97Tx3 data, are obtained using the timing-derived calibration constants. 
We note that, for surface source locations,
ray tracing must be performed through the firn to the surface. 
Even after application of ray-tracing corrections, 
we note that a large fraction
of the solid angle above the surface is folded into caustics around the 
total internal reflection angle, which makes unique identification of
a surface source difficult if it is not directly over the array.
For sources such as 97Tx3 (and, in general, for $z<-150$m, which 
corresponds to the neutrino search region), 
such corrections are less important, although still non-zero for
shallow receivers.

\section{Further consideration of physics backgrounds}
\subsection{Detectability of cosmic ray muons in RICE}
Several processes can, in principle, 
contribute to the actual muon event rate expected in RICE. These include:
contributions from the Coulomb field of
muons passing close to a RICE radio receiver. 
For vertical muons passing near two
vertically displaced receivers, the signature is, in principle, clean -
two hit times separated by t=d/c rather than
t=d/(c/n). 
Pictorally, the transverse field increases by $\gamma$, whereas the
longitudinal field (to which we are sensitive for downcoming muons) is
invariant.
(In frequency space, the vertical extent of a typical RICE dipole sets the scale of the bandwidth response. Although the field lines are Lorentz-contracted, the Lorentz free space contracted time scale is $\sim\gamma$ smaller, and the bandwidth is $\sim\gamma$ larger, so $\gamma_{net}\sim$1.)
Therefore, there is a maximal mismatch of geometries -- 
for the predominantly vertical
muon flux, the vertically-oriented 
dipoles are most sensitive to the (unboosted)
longitudinal component of the Coulomb field. 

\begin{figure}[htpb]
\centerline{\includegraphics[width=9cm]{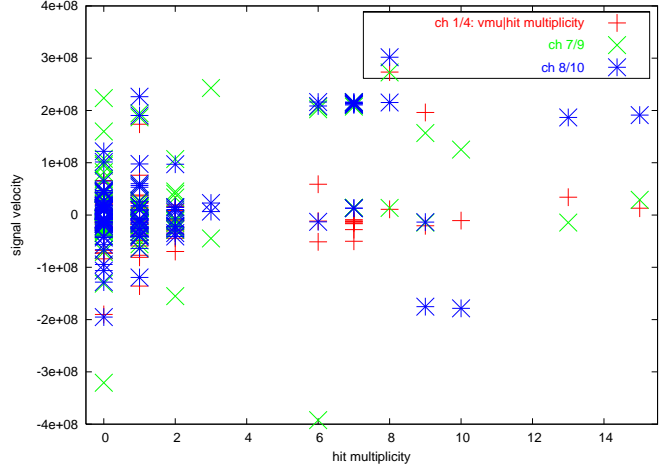}}
\caption{Propagation velocity for signals registered in
two receivers in same icehole, for non-general triggers,
data taken 10Dec03. No obvious band at v=c is observed.}
\label{fig:cCR.eps}
\end{figure}

Nevertheless, we can estimate the field strengths, for, e.g., 
vertically-incident flux.
Neglecting effects due to the Lorentz contraction of the field lines, 
zt a distance
of one meter from an antenna, $V\sim (q/(4\pi\epsilon_0r^2))\times h_{eff}$, or
approximately 0.144 nV -- roughly 5 orders of magnitude smaller than
the mean rms thermal noise. We therefore require $\gamma\sim 10^5$, or
$E_\mu\sim 10^{13}$ eV, with the greatest sensitivity to horizontal 
muons (unfortunately, the muon flux $dN/d\theta\sim sec\theta$).
We can search for events by looking for signals in multiple channels having
a transit time between channels corresponding to motion of an 
ultra-relativistic charged particle.
Figure \ref{fig:cCR.eps} shows no obvious band corresponding to a particle
moving through the ice at v=c.

\subsubsection{Muon Bremstrahlung}
\begin{figure}[htpb]
\centerline{\includegraphics[width=9cm]{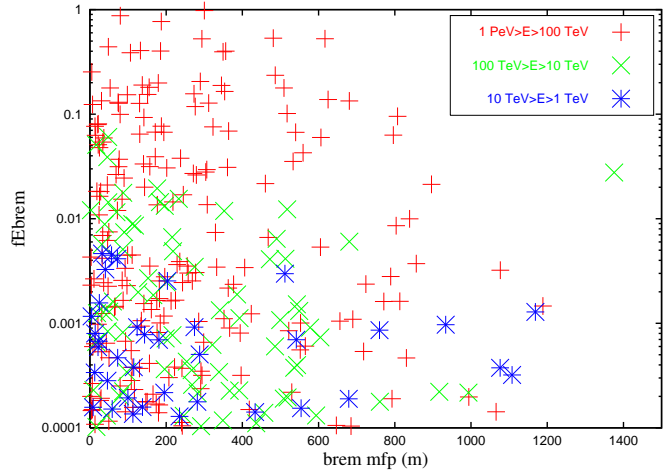}}\caption
{Fraction of muon energy lost to bremstrahlung, as a function of muon energy, using GEANT simulations.}
\label{fig:mubrem1p100.eps}
\end{figure}

We have written a crude simulation to estimate the expected signal
from possible muon bremstrahlung.
GEANT simulations of muons passing through ice allow
us to investigate the expected increase in the bremstrahlung cross-section
with muon energy. (Note that the GEANT estimates do not include the
enhancements in the muon observability due to photonuclear interactions, which
become dominant at energies greater than 100 EeV\citep{spencer04}.)
Figure \ref{fig:mubrem1p100.eps} shows the correlation
of mean-free-path (distance between successive bremstrahlungs) and
fraction of energy lost in a bremstrahlung for three different
muon energy ranges.
At each bremstrahlung, we calculate the expected signal voltage 
induced in a receiver antenna, assuming $V_{eff}\sim 0.1$m, and 
using the original
ZHS\citep{ZHS92} estimate of the signal strength on the Cherenkov cone. 
This is then compared with the rms thermal noise voltage $V_{thermal}=kTB$.
For this simulation, we generate the muon flux on the ground,
as given by Gelmini, Gondolo and Varieschi\citep{ggv}
which includes both
muons from high energy primaries, as well as muons from charm
produced by cosmic ray interactions in the atmosphere. We model
these two contributions using the prescription in the Review of Particle
Properties\citep{RPP06}: $dN_\mu/dE$(conventional)$\sim(0.14E^{-2.7})$*((1/(1+((1.1E)/115))) + 
(0.054/(1+((1.1E)/850));  (E in GeV, flux in $GeV^{-1}s^{-1}sr^{-1}$).
The charm flux, as presribed by GGR, is modeled as a 
simple power law, with a vertical component which is approximately
$E^{0.5}$ harder than the conventional flux, and crosses the 
conventional flux at approximately 1 PeV.

\begin{figure}\centerline{\includegraphics[width=9cm,angle=0]{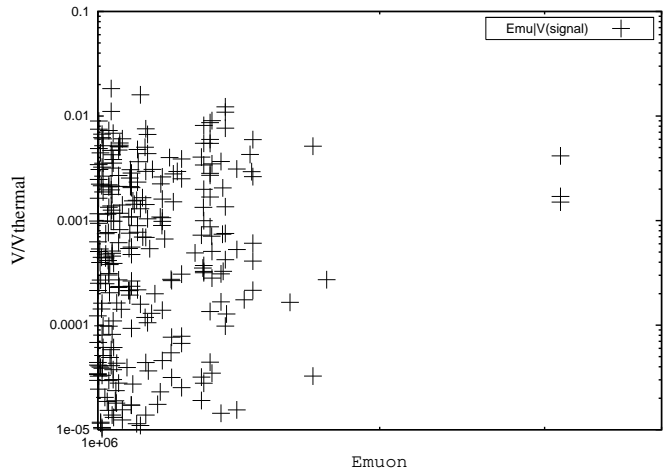}}\caption{Signal distribution expected from muon bremstrahlungs in one year.}\label{fig:muons-out.eps}\end{figure}



\begin{figure}[htpb]
\centerline{\includegraphics[width=9cm,angle=0]{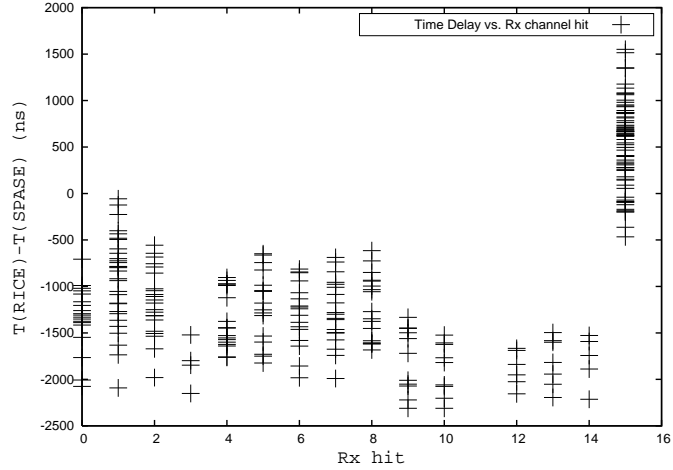}}
\caption{Expected timing correlations between RICE and SPASE hits due to
single muons.}
\label{fig:RICE-SPASE-timing-corr.eps}
\end{figure}

The expected signals induced by bremstrahlungs, in one year, are shown
in Figure \ref{fig:muons-out.eps}. (Multiple y-values for the same
x-value correspond to multiple receivers hit for the same muon). 
We note that there are no signals
which exceed a value of S:N$\ge$1. 
The trigger efficiency of such signals is, in principle, enhanced by 
writing to disk all events for which there is a coincidence
between a high-multiplicity
SPASE (South Pole Air Shower Experiment) event and a hit RICE receiver.
The timing delay between the expected
RICE signal and the expected SPASE trigger time received at the RICE DAQ
is shown in Figure \ref{fig:RICE-SPASE-timing-corr.eps}, taking into
account known cable delays, etc. Given
the timing coincidence of 1.25$\mu$sec, we should have adequate efficiency
for detection of bremstrahlungs, in the event that the signal were 
sufficiently large. However, as shown in the Figure, our timing delays
are, in fact, not tuned to match the expected delay between the SPASE and
RICE triggers.

We point out that this timing correlation estimate
is, admittedly, extremely crude -- in a realistic scenario, an air shower
would probably produce multiple muons, one of which would trigger the
DAQ early, and a later one of which would trigger RICE (we also
have only very crudely estimated the details of the SPASE trigger, with
which we have very limited familiarity).

\begin{figure}[htpb]
\centerline{\includegraphics[width=9cm]{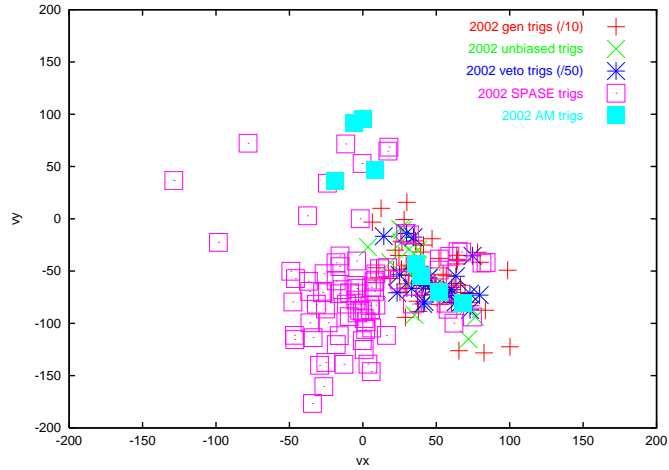}}
\caption{Vertex distributions for various trigger types, xy-projection. 
``Veto'' triggers are triggers identified as likely of surface origin, based
on fast timing (TDC) information. ``General'' triggers are multiplicity 4
events which have not been identified as ``veto'' events. ``Unbiased'' 
triggers correspond to data captures taken at random times, ``AM'' and
``SPASE'' refer to coincidences between AMANDA-B triggers or SPASE
triggers, to within 1.25 microseconds.}
\label{fig:trig-compare-vx-vy.eps}
\end{figure}

We have separated our events by trigger type, and reconstructed
corresponding event vertices. As shown in Figure
\ref{fig:trig-compare-vx-vy.eps}, the xy-location of the
reconstructed vertex is, in fact, different for the various trigger
types - perhaps most interestingly, the SPASE coincidences show a
marked shift, in the direction of the SPASE array, relative to 
`general' triggers (which are consistent with 
the sub-sample of veto triggers that are saved to disk).
A full analysis of these triggers has, however, not yet been conducted. 

\begin{figure}[htpb]
\centerline{\includegraphics[width=9cm]{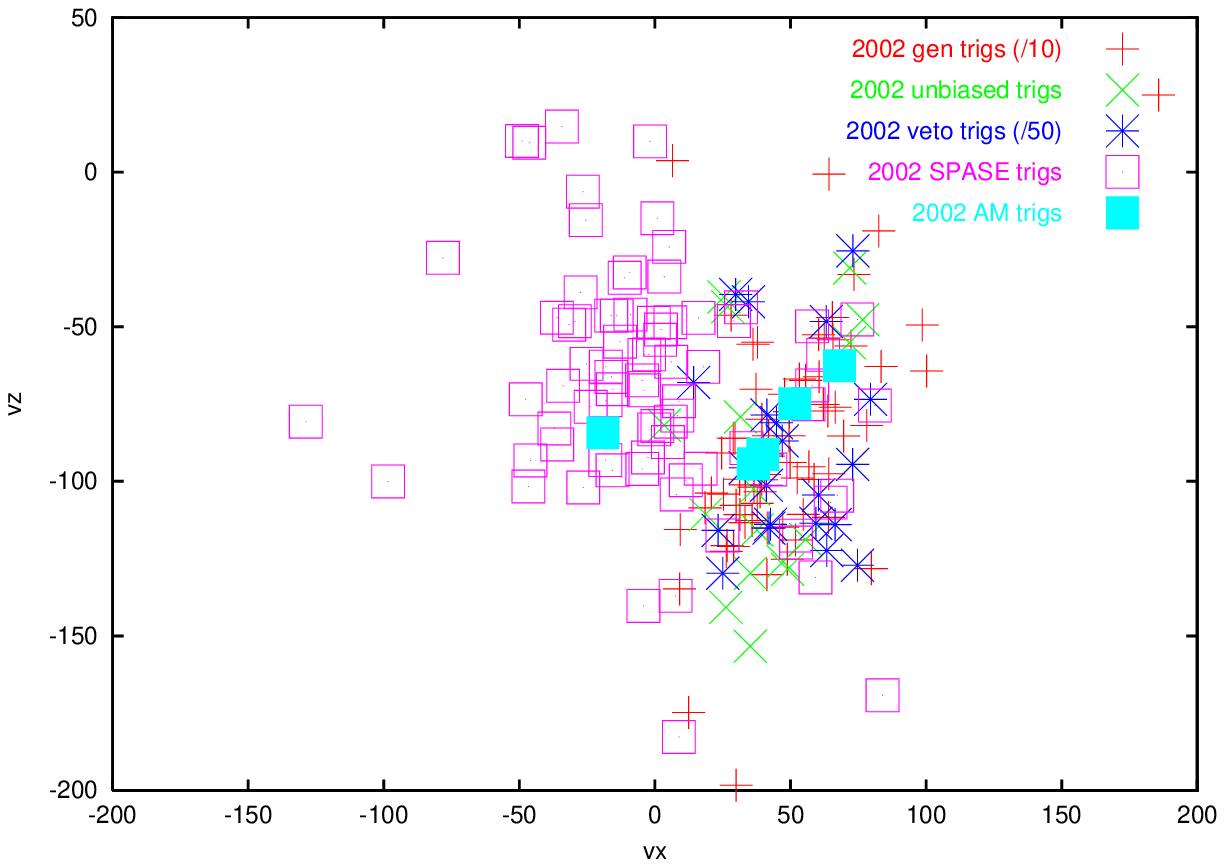}}
\caption{Vertex distributions for various trigger types, yz-projection.
Scheme is the same as in the previous Figure.}
\label{fig:trig-compare-vx-vz.eps}
\end{figure}

In principle, timing correlations between receivers, in the case of a ``muon bundle'', e.g., might be used to accentuate the possible signal -- in this case, one takes advantage of the fact that the muon bundle is propagating at the velocity of light while Cherenkov radiation, e.g., will propagate at $v\sim 0.6c$. This has not yet been fully investigated. 

\subsection{Air Shower Backgrounds}
\begin{figure}[h]
\begin{center}
\begin{minipage}{32pc}
\includegraphics[width=18pc,angle=0]{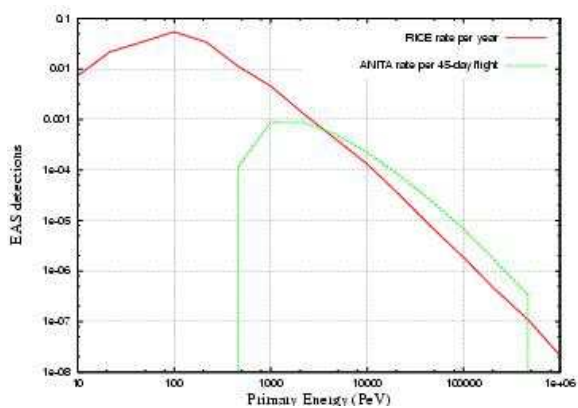}
\caption{\label{fig:EAS}Estimated EAS detections per year for RICE
and ANITA.}
\end{minipage}
\end{center}
\end{figure}

Direct radiofrequency
signals from extensive air showers (EAS)
which propagate into the ice, as measured by the CODALEMA
and LOPES experiments, give signals which
peak in the tens of MHz regime. There should be a signal
resulting from the impact of the shower core with the ice,
producing the same kind of Cherenkov radiation signal that
RICE seeks to measure. Nevertheless (Fig. \ref{fig:EAS}), the
expected EAS signal rate should be almost immeasurably small.

\subsection{Solar RF Backgrounds}
\begin{figure}[htpb]
\centerline{\includegraphics[width=9cm]{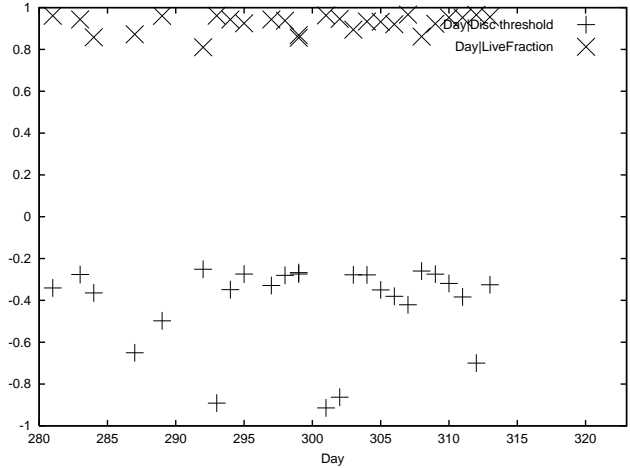}}
\caption{LiveTime (crosses) and also discriminator thresholds (in 
Volts) for 
days 280 -- 310 of 2003. Solar flares were registered on Oct. 31
(registered at Earth at 9:54 am UT) and
Nov. 6 (7:20 am UT), corresponding to Julian days 304/305 and 310/311. }
\label{fig:D1LiveFrac2003.eps}
\end{figure}

Radio frequency noise correlated with solar activity has been
the subject of extensive investigation. Auroral discharges have
been continuously monitored in Antarctica in the tens of MHz frequency range,
over the last decade. In 2003, there were high-intensity solar 
flares recorded between Oct. 19, 2003 to Nov. 4, 2003; typically, these
result in electrical disturbances at Earth some 24-48 hours 
later.\footnote{Other notable flares were recorded on 
July 14, 2000.}
We have looked for correlations, during this time period, with 
high backgrounds as registered by RICE (based on low livetime, or,
alternately, high discriminator thresholds needed to maintain 
reasonable livetimes). Figure \ref{fig:D1LiveFrac2003.eps} shows
these quantites, as monitored through this time period. No
obvious correlation is observed. 


\message{FOR AIR SHOWERS: A)
 CAN WE SKIM EVENTS WHICH ARE AT THE EDGE OF THE C-CONE
CAUSTIC, POINTING UP, AND INCONSISTENT WITH STUFF DIRECTLY OVERHEAD OR
STUFF POINTING TOWARDS SPASE OR THE DOME? B) IS THERE A MAGNIFICATION EFFECT
WHEREBY ALL SIGNAL CLOSE TO THE SURFACE IS FUNNELED INTO RICE??

more thoughts on air showers - all signal outside 45-degree angle folds into
45-degree angle. How much signal is opposite MAPO? Alternately, how many
times do we see signal sweeping in from channel 0?!

NEED TO LOOK AT HIGH-MULTIPLICITY SPASE TRIGGERS (TRIGCODE=3)! WHY SHOULD
THERE BE ANY HIGH-MULTIPLICITY EVENTS DUE TO BACKGROUND IF WE REQUIRE ONLY
A 1.2 MICROSECOND WINDOW FOR A 1-HIT COINCIDENCE. ALSO NEED TO KNOW THE
RAW SPASE TRIGGER RATE FOR THEIR 30-FOLD COINCIDENCE TO FIRE!

SPASE.gnu - why do SPASE triggers have high-multiplicity???

INPUT NEW TEXT FROM JENNI/SURUJ!}

\subsection{Correlation with Machine Activity at South Pole 
Station}
We have searched for correlations with aircraft activity at Pole, but
find no obvious correlation. Nevertheless, operating machinery
(in particular, the IceCube drill) creates large amplitude backgrounds
that render our data-taking during the austral summer largely
ineffective. Fortunately, conditions during the austral winter
tend to be generally radio-quiet.

\section{Future Running}
The current RICE data-taking is expected to continue for the
next 3-4 years, yielding an incremental improvement in sensitivity
largely proportional to the increase in livetime.
A second-generation experiment would address:
1) the limited
bandwidth of the experiment, resulting from cable losses at high
frequencies, the need to highpass filter above 250 MHz in order to
reduce low-frequency noise from the AMANDA (and later, IceCube) phototubes,
and the 500 MHz bandwidth of the digital oscilloscopes, and 2) the 
high thresholds resulting from a simple one-tier trigger system and 
the inability to effectively reduce surface backgrounds in hardware.
We are currently developing a successor to the present RICE array
(``AURA''), featuring
in-ice custom 
digitizer boards, and a local coincidence multiplicity
trigger, which only considers ``hits'' for which a local antenna cluster
(consisting of 4 antenna) itself satisfies a local trigger coincidence 
inconsistent with down-coming signals.

\begin{figure}[htpb]
\centerline{\includegraphics[width=10cm,angle=0]{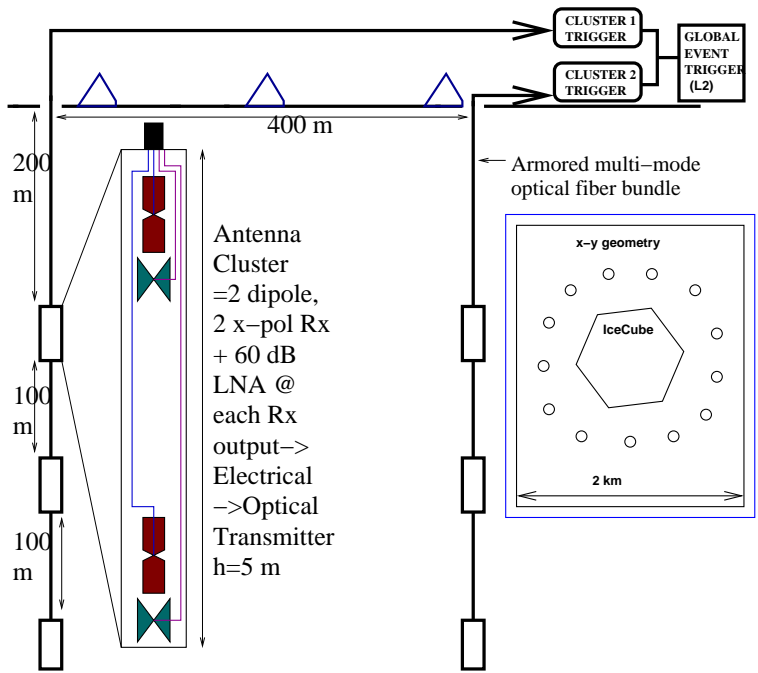}}
\caption{RICE-II schematic.}
\label{fig:RICEII.xfig.eps}
\end{figure}

\begin{figure}[htpb]
\centerline{\includegraphics[width=10cm,angle=0]{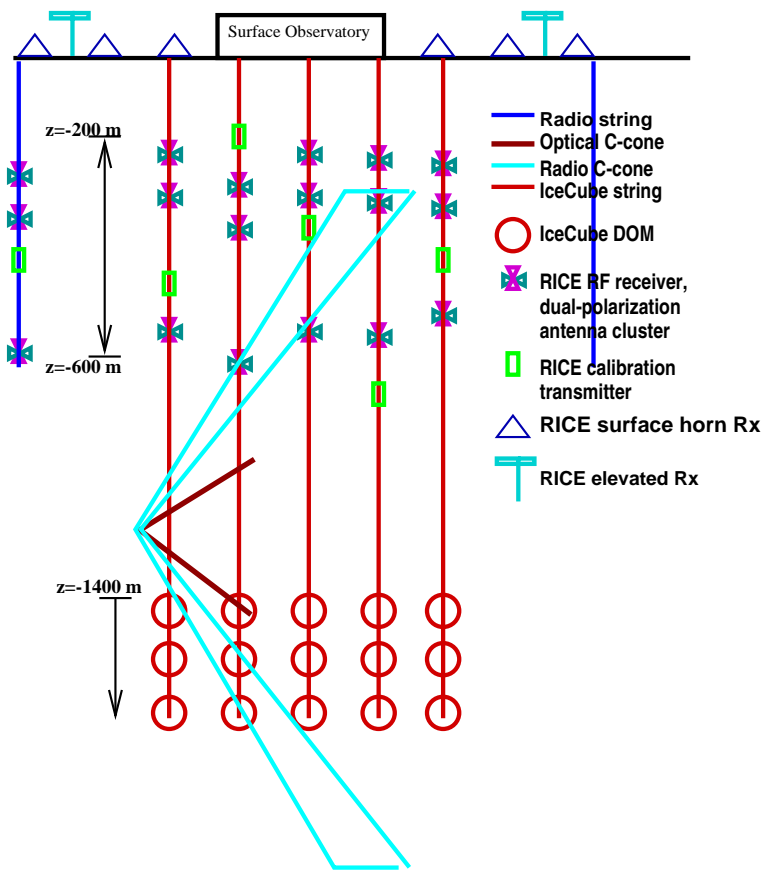}}
\caption{Geometry of ICECUBE - RICE event.}
\label{fig:IceCubeXRICE.xfig}
\end{figure}

In the IceCube era, we hope to expand the capabilities of RICE with
improvements to both the receiver modules as well as the data acquisition
system. A sketch of the future array is outlined in Figure
\ref{fig:RICEII.xfig.eps}. We anticipate in-ice sampling, and transmission of
digital signals to the central Counting House, where the event trigger
is formed and digitization occurs. 
Vertically, there are two ``close'' clusters, both below the firn, and
one deep cluster. Assuming no dedicated 
ice-hole drilling and only co-deployment with IceCube, we show the 
horizontal footprint expected in the ``IceCube-Plus''
scenario. The in-ice antennas are complemented by downward-looking
surface horn antennas, which view only upcoming signals (and
obviate in-ice deployment issues) and also are necessary for ensuring the
maximal rejection of down-coming noise.
Local ``clusters'' allow suppression
of spurious randoms and also ensure that recorded signals are up-coming
rather than down-going.
Figure
\ref{fig:IceCubeXRICE.xfig} illustrates the geometry of a
possible RICE-IceCube overlap event.

\begin{figure}[htpb]
\centerline{\includegraphics[width=9cm,angle=0]{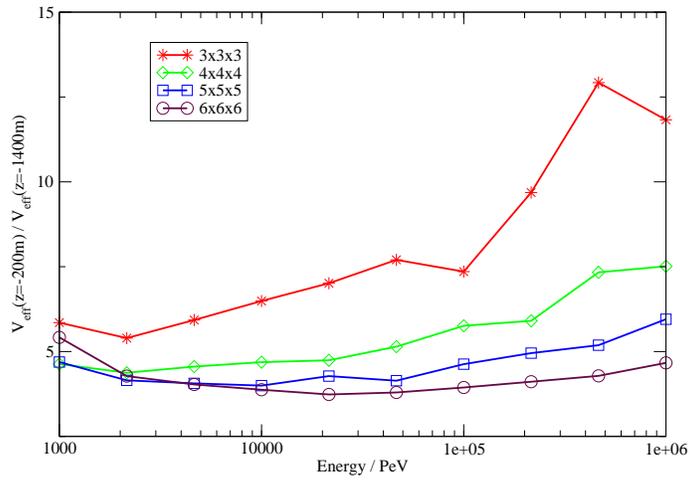}}
\caption{Neutrino Effective volume, as a function of depth of array.}
\label{fig:200m-1400m-comp.eps}
\end{figure}

\begin{figure}[htpb]
\centerline{\includegraphics[width=9.cm,angle=0]{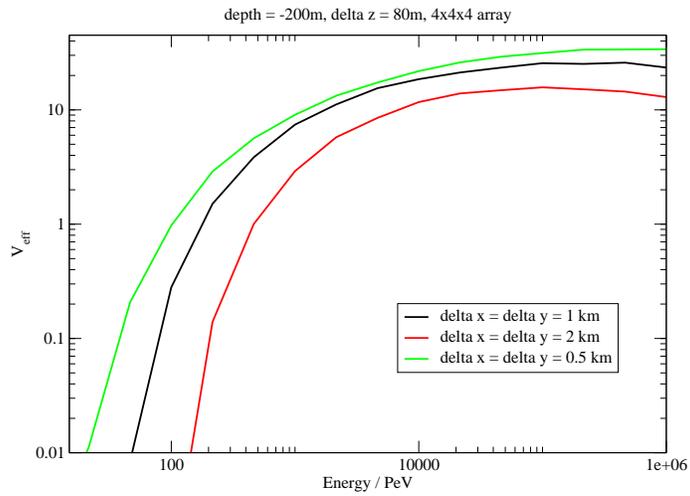}}
\caption{Neutrino Effective volume, as a function of the surface extent
of the array.}
\label{fig:dxdy.eps}
\end{figure}

\begin{figure}[htpb]
\centerline{\includegraphics[width=9.cm,angle=0]{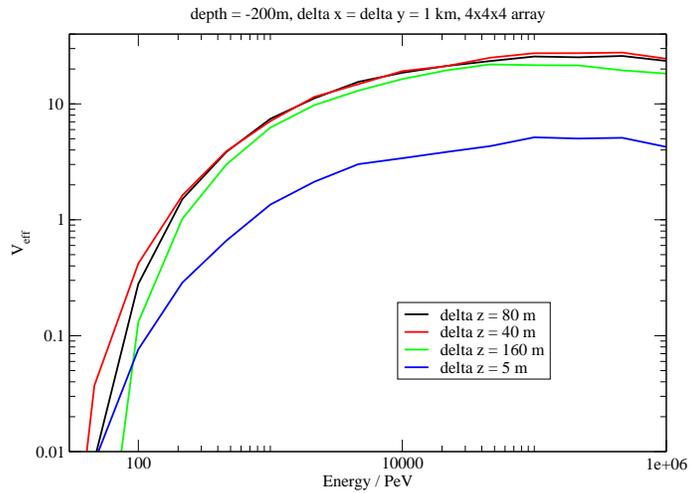}}
\caption{Neutrino Effective volume, as a function of vertical
spacing of the array.}
\label{fig:dz.eps}
\end{figure}

Monte Carlo array optimization studies are in progress.
Figures \ref{fig:200m-1400m-comp.eps}, \ref{fig:dxdy.eps},
and \ref{fig:dz.eps}
show the
expected sensitive volume of a neutrino array, and its dependence
on the array geometry and spacing between array elements. In general,
the most efficient array studied is a shallow array (filling the
depth region between --200 m and --600 m) with typical horizontal
spacing of $\sim$1 km and typical vertical spacing of $\sim$100 m,
reflecting the typical scale of the attenuation length, and the width
of the Cherenkov cone for a typical neutrino interaction, respectively.

\begin{figure}[htpb]
\centerline{\includegraphics[width=9cm,angle=0]{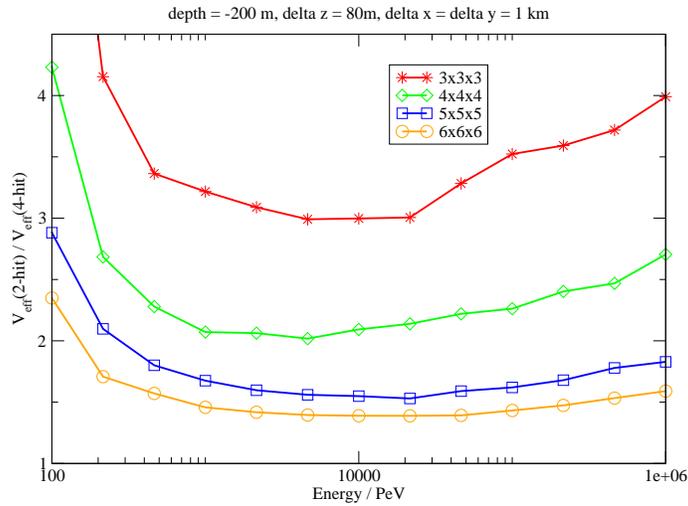}}
\caption{Neutrino Effective volume, as a function of minimum
``event'' multiplicity required.}
\label{fig:2-4hit2.eps}
\end{figure}

Expected performance is also being evaluated. Figure
\ref{fig:2-4hit2.eps} displays the effective volume dependence on
the minimum required event multiplicity.
\vspace{0.4cm}

\begin{figure}[htpb]
\centerline{\includegraphics[width=10cm,angle=0]{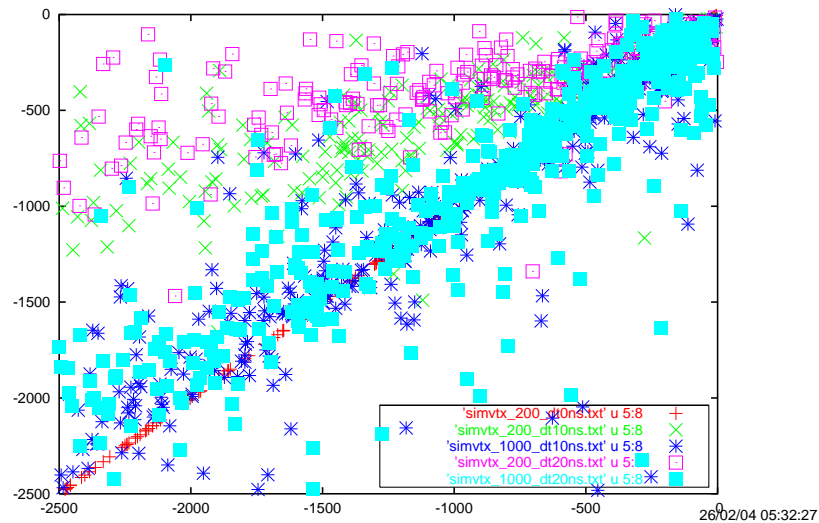}}
\caption{Generated (horizontal) vs. reconstructed source depth, for
various configurations of antennas and timing resolutions. Perfect
source reconstruction results
in a narrow band along the diagonal.}
\label{fig:simvtx-arraysize-tsmear}
\end{figure}

Figure \ref{fig:simvtx-arraysize-tsmear} illustrates that
relatively large timing uncertainties are tolerable if the array
is distributed over a large area.

One of the most important experimental factors determining signal
detection rates is the center frequency of the antenna, and how that
relates to the bandwidth of noise (and the S:N for true neutrinos);
the S:N rate is what then dictates the appropriate
operating discrimator threshold.
\begin{figure}[htpb]\centerline{\includegraphics[width=10cm,angle=0]{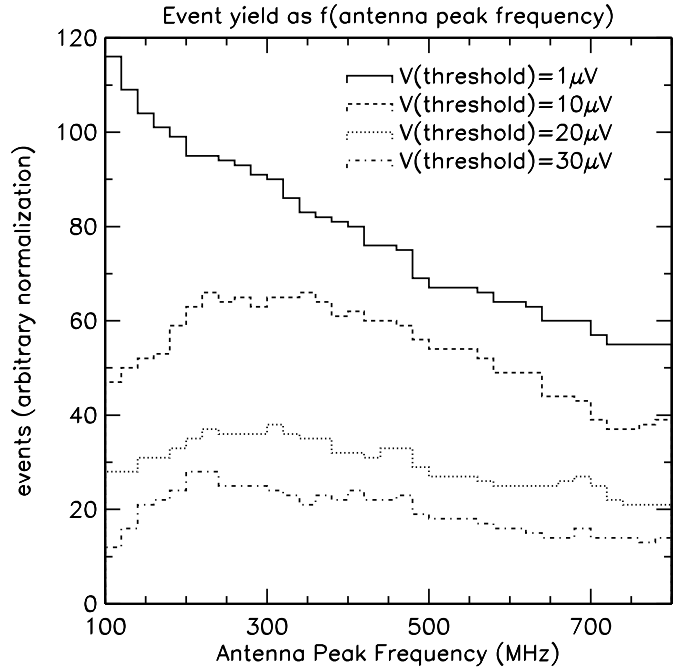}}\caption{Relative signal yield as a function of discriminator threshold and center frequency of antennas.}\label{fig:Signal_f_frequency_1PeV}\end{figure}

A multiparameter optimization and expected event rates for various 
experimental configurations has been ongoing, and will continue as our
understanding of such things as ice properties improve
and practical limitations such as drilling costs and overhead become
better-defined.

\section{Acknowledgments}
The RICE experiment is
supported by NSF Office of Polar Programs Award No. 0338219,
the University of Kansas, and the Research Corporation.
We thank our RICE colleagues for making the experiment possible.

\end{document}